\documentclass[letterpaper]{JHEP3}
\usepackage{epsfig,mathrsfs}

\def\ba{\begin{eqnarray}}
\def\ea{\end{eqnarray}}
\def\be{\begin{equation}}
\def\ee{\end{equation}}
\def\nn{\nonumber}
\def\exd{{\rm d}}
\def\pd{\partial}
\makeatletter
\def\x@arrow{\DOTSB\Relbar}
\def\xlongequalsignfill@{\arrowfill@\x@arrow\Relbar\x@arrow}
\newcommand{\xlongequal}[2]{%
    \ext@arrow 0099\xlongequalsignfill@{#1}{#2}}
\makeatother

\newcommand{\roughly}[1]{\mathrel{\raise.3ex\hbox{$#1$\kern-0.85em
\lower1ex\hbox{$\sim$}}}}

\def\endignore{}
\def\ignore #1\endignore{} 

\def\be{\begin{equation}}
\def\beq\begin{equation}
\def\ee{\end{equation}}
\def\bea{\begin{eqnarray}}
\def\eea{\end{eqnarray}}

\def\eqa{\begin{eqnarray}}
\def\eeqa{\end{eqnarray}}
\def\eq{\begin{equation}}
\def\eeq{\end{equation}}

\def\nn{\nonumber}

\def\pref#1{(\ref{#1})}

\def\ol#1{\overline{#1}}

\def\exd{{\rm d}}
\def\nn{\nonumber}
\def\pref#1{(\ref{#1})}
\def\be{\begin{equation}}
\def\ee{\end{equation}}

\def\gR{g_\ssR}
\def\gB{g_\ssB}

\def\beq{\begin{equation}}
\def\eeq{\end{equation}}
\def\beqa{\begin{eqnarray}}
\def\eeqa{\end{eqnarray}}

\def\fL{\mathscr{L}} 
\def\fM{\mathscr{M}} 

\def\cA{{\cal A}}
\def\cB{{\cal B}}
\def\cC{{\cal C}}

\def\cF{{\cal F}}
\def\cG{{\cal G}}
\def\cH{{\cal H}}

\def\cL{{\cal L}}

\def\cN{{\cal N}}
\def\cO{{\cal O}}

\def\cR{{\cal R}}

\def\cU{{\cal U}}
\def\cV{{\cal V}}
\def\cW{{\cal W}}

\def\ssA{{\scriptscriptstyle A}}
\def\ssB{{\scriptscriptstyle B}}

\def\ssE{{\scriptscriptstyle E}}
\def\ssF{{\scriptscriptstyle F}}
\def\ssG{{\scriptscriptstyle G}}
\def\ssH{{\scriptscriptstyle H}}
\def\ssI{{\scriptscriptstyle I}}
\def\ssJ{{\scriptscriptstyle J}}
\def\ssK{{\scriptscriptstyle K}}
\def\ssL{{\scriptscriptstyle L}}
\def\ssM{{\scriptscriptstyle M}}
\def\ssN{{\scriptscriptstyle N}}
\def\ssP{{\scriptscriptstyle P}}
\def\ssQ{{\scriptscriptstyle Q}}
\def\ssR{{\scriptscriptstyle R}}

\def\ssT{{\scriptscriptstyle T}}

\def\UV{{\scriptscriptstyle UV}}

\def\Vone{\cV_{\rm 1-loop}}

\newcommand{\bmat}{\left(\begin{array}}
\newcommand{\emat}{\end{array}\right)}

\def\-{\hphantom{-}}

\def\s2{\frac{1}{2}}

\def\tr{{\rm tr \,}}
\def\Tr{{\rm Tr \,}}

\def\IF{\relax{\rm I\kern-.18em F}}
\def\II{\relax{\rm I\kern-.18em I}}
\def\IP{\relax{\rm I\kern-.18em P}}
\def\IC{\relax{\rm I\kern-.48em C}}
\def\IR{\relax{\rm I\kern-.18em R}}
\def\IK{\relax{\rm I\kern-.20em K}}
\def\IM{\relax{\rm I\kern-.25em M}}

\def\nott#1{\setbox0=\hbox{$#1$}                
   \dimen0=\wd0                                 
   \setbox1=\hbox{/} \dimen1=\wd1               
   \ifdim\dimen0>\dimen1                        
      \rlap{\hbox to \dimen0{\hfil/\hfil}}      
      #1                                        
   \else                                        
      \rlap{\hbox to \dimen1{\hfil$#1$\hfil}}   
      /                                         
   \fi}                                         %

\def\y2{Y_{\ssM\ssN} Y^{\ssM\ssN}}
\def\Riem2{R_{\ssA\ssB\ssM\ssN} R^{\ssA\ssB\ssM\ssN}}
\def\Ricci2{R_{\ssM\ssN} R^{\ssM\ssN}}

\def\f2{F^{a}_{\ssM\ssN} F^{\ssM\ssN}_a}

\def\Dsl{\,\raise.15ex\hbox{/}\mkern-13.5mu D}
\def\ksl{\,\raise.15ex\hbox{/}\mkern-10.5mu k}
\def \one{\relax{\rm 1\kern-.26em I}}

\def\exd{{\rm d}}

\def\nn{\nonumber}

\def\({\left(}
\def\){\right)}

\preprint{FTUAM-13-125, IFT-UAM/CSIC-13-007}
\title{Accidental SUSY: Enhanced Bulk\\
Supersymmetry from Brane Back-reaction}

\author{
C.P.~Burgess,${}^{1,2}$
L.~van Nierop,${}^1$
S.~Parameswaran,${}^3$
A.~Salvio${}^{4,5}$ and
M.~Williams${}^{1}$\\

$^1$ Department of Physics \& Astronomy, McMaster University,
 Hamilton ON, Canada\\
$^2$   Perimeter Institute for Theoretical Physics,
 Waterloo ON, Canada\\
$^3$   Institute for Theoretical Physics, Leibniz University, 
Hannover, Germany\\
$^4$ Scuola Normale Superiore and INFN, 
Pisa, Italy\\
$^5$ Departamento de F\'isica Te\'orica, Universidad Aut\'onoma de Madrid and \\ 
Instituto de F\'isica Te\'orica IFT-UAM/CSIC, Cantoblanco, 28049 Madrid, Spain
}

\date{}

\abstract { We compute how bulk loops renormalize both bulk and brane effective interactions for codimension-two branes in 6D gauged chiral supergravity, as functions of the brane tension and brane-localized flux. We do so by explicitly integrating out hyper- and gauge-multiplets in 6D gauged chiral supergravity compactified to 4D on a flux-stabilized 2D rugby-ball geometry, specializing the results of a companion paper, {\tt arXiv:1210.3753}, to the supersymmetric case. While the brane back-reaction generically breaks supersymmetry, we show that the bulk supersymmetry can be preserved if the amount of brane-localized flux is related in a specific BPS-like way to the brane tension, and verify that the loop corrections to the brane curvature vanish in this special case. In these systems it is the brane-bulk couplings that fix the size of the extra dimensions, and we show that in some circumstances the bulk geometry dynamically adjusts to ensure the supersymmetric BPS-like condition is automatically satisfied. 
We investigate the robustness of this residual supersymmetry to loops of non-supersymmetric matter on the branes, and show that supersymmetry-breaking effects can enter only through effective brane-bulk interactions involving at least two derivatives. We comment on the relevance of this calculation to proposed applications of codimension-two 6D models to solutions of the hierarchy and cosmological constant problems.
}

\begin{document}
\section{Introduction}
\label{sec:introduction}

If teflon theories are those to which lack of experimental support does not stick, then supersymmetry is their poster child. Indeed, supersymmetry continues to play a central role in particle theory --- and has done so for more than 3 decades --- despite its so-far disappointing prediction: the perpetually imminent discovery of superpartners for all Standard Model particles.

Its longevity in the teeth of such disappointment has many reasons, but an important one is its good ultraviolet properties. Supersymmetry is one of the few symmetries (another is scale invariance) that can suppress both scalar masses and vacuum energies when unbroken, and so potentially might be useful for the hierarchy and the cosmological-constant problems. The challenge is to enable this suppression to survive the symmetry breaking required to explain the experimental absence of superpartners. Moreover, supersymmetry arises organically in string theory, which remains our best candidate for physics at the highest energies.

These observations suggest the utility of re-thinking the (apparently signature) prediction of Standard-Model superpartners, since it is the absence of evidence for these that so far provides the best evidence for absence of supersymmetry. The key assumption that underlies the prediction of superpartners (and so, more broadly, of the supersymmetric Standard Model, minimal or otherwise) is the assumption that supersymmetry is linearly realized. After all, nonlinear realization does not require superpartners, because nonlinearly realized supersymmetry acts on a single-particle state (say, the electron) to give a two-particle state (an electron plus a goldstino) \cite{SUSYNLR, WessBagg} rather than the single-particle state (a selectron) required by linear realization.

Nonlinearly realized supersymmetry also arises organically in string theory when supersymmetry is broken by the presence of branes \cite{TASIBrane}. $D$-branes often break half of the supersymmetries present in the bulk, and by so doing provide counter-examples \cite{PartialBreaking} to previously conjectured no-go theorems \cite{PartialNoGo} precluding partial supersymmetry breaking. In general, a configuration of branes can break all or only part of the supersymmetries present in the bulk. This observation has spawned a variety of studies of brane-induced partial supersymmetry breaking within both string and brane-world models \cite{Sequest, SUSYBUV}.

Physically, nonlinear realization is appropriate if the symmetry-breaking scale, $M_s$, is larger than the UV scale, $M_\UV$, above which the theory's UV completion intervenes \cite{CCWZ, GBEFT}. In this case symmetry multiplets can be split by more than $M_\UV$, and so the low-energy theory need not contain the particle content required to linearly realize the symmetry. For $D$-branes the UV completion is string theory itself, so the brane spectrum need never linearly realize supersymmetry in the field theory limit below the string scale.

When supersymmetry breaks on a brane it is often true that the bulk sector is more supersymmetric than the brane sector, since the bulk must pay the price of a (possibly weak) bulk-brane coupling before it learns that supersymmetry is broken. As a result, unlike for the branes, the bulk spectrum has equal numbers of bosons and fermions, whose masses could be split by as little as the Kaluza Klein scale. It therefore has the field content to linearly realize supersymmetry, and so can have much milder UV properties than would be expected for the branes.

\subsection*{The gravity of SUSY}

All of this suggests a somewhat unorthodox picture of how low-energy supersymmetry might be realized despite the apparent experimental absence of superpartners \cite{MSLED}. If Standard-Model particles were localized on a supersymmetry-breaking brane sitting within a more supersymmetric bulk, then Standard-Model superpartners would be avoided and the low-energy world would have a gravity sector that is much more supersymmetric than is the Standard-Model sector to which accelerators have access.

Supersymmetric signals would be much harder to find in such a world, and would depend somewhat on the number of degrees of freedom present in the gravity sector \cite{Bouchart:2011va}. Although each mode is gravitationally coupled, observable energy loss rates into the gravitational sector can be possible (such as in the specific realizations involving supersymmetric large extra dimensions \cite{Towards, MSLED, SUSYADD}). In such scenarios the enormous phase space can compensate the small gravitational couplings, just as one obtains for gravitons in ordinary large extra dimensions \cite{ADD, ADDpheno}.

Can the good UV properties of supersymmetry still be useful within this kind of picture? A hint that they can comes from the observation that both the hierarchy problem --- `Why is the weak scale so far below the Planck scale?' --- and the cosmological constant problem --- `Why does the vacuum energy gravitate so weakly?' --- involve gravity in their formulation. Perhaps they might be ameliorated by the same physics if the gravity sector were very supersymmetric.

But there is no substitute for testing these ideas with an explicit calculation of the size of loop effects. A well-developed, fairly simple and concrete framework within which to do so is to describe the supersymmetric bulk using the field equations of 6D chiral, gauged supergravity \cite{NS}, which has long been known to allow (marginally) stable compactifications to 4D on a sphere \cite{SS, ABPQ}. All but a single modulus, $\varphi$, of this supergravity is stabilized by the presence of a Maxwell flux that threads the sphere in a monopole configuration. It is also known how to embed SUSY-breaking branes into this system including their back-reaction onto the extra-dimensional geometry, which (for two branes) deforms from a sphere into a rugby ball\footnote{North American readers should think `football' here, but we use `rugby' to avoid cultural disagreements about the shape of a football.} (with the branes located at the tips) \cite{Towards} or into something even more distorted \cite{GGP, OtherConical}. (
See also \cite{preSLED} for similar flux-stabilized rugby-ball constructions within a non-supersymmetric context.)

In this paper we test the UV properties of this kind of framework by explicitly computing the contribution of bulk loops to the 1PI quantum action (as well as to the vacuum energy), including the supersymmetry-breaking influence of the branes. We do so by adapting to the supersymmetric case a general calculation of bulk loops on rugby-ball geometries \cite{Companion}. For technical reasons these calculations are only for low-spin bulk fields --- {\em i.e.} spins zero, half and one --- but work is in progress to extend our present results to higher spins. By combining with earlier results for brane loops \cite{TNc2B}, we can piece together how the complete one-loop result depends on brane and bulk properties.

In particular, because our interest is in the low-energy effects of UV modes, we track how short-wavelength bulk loops renormalize the local effective interactions both on the brane and in the bulk. In particular we ask how they depend on the single bulk modulus, $\varphi$, as well as on the two main brane properties relevant at low energies: their tension, $T_b$, and the amount of stabilizing Maxwell flux, $\Phi_b$, that is localized on the branes.\footnote{Maxwell (or gauge) flux can be localized on a 3-brane in 6 dimensions in the same way that magnetic flux can be localized on a string (or vortex) in 4 dimensions.} Although $\Phi_b$ may seem unfamiliar, its presence is in general required in order for the full brane-bulk system to have low-energy deformations that can satisfy flux-quantization constraints that relate $\Phi_b$ to $T_b$ \cite{localizedflux}. Both $T_b$ and $\Phi_b$ correspond to the coefficients of the first two terms in a generic derivative expansion of the brane action:\footnote{Although 
our introductory discussion motivated brane supersymmetry breaking on known properties of $D$-branes, notice that this is not a $D$-brane action. Because our focus is ultimately on very low-energy properties we treat the brane phenomenologically, as a generic localized object whose microscopic structure is not resolved in detail. A resolution of this more detailed structure would be required in any embedding of our discussion into a UV completion. It is not yet known what the full string provenance is of the 6D chiral supergravity considered here (see, however, \cite{UVcompl} for steps in this direction).}
\be \label{E:TPhidefs}
 S_b = - \int_\cW \exd^4x \sqrt{- \gamma} \;\; T_b + \frac{2 \pi }{\tilde g^2} \int_\cW \Phi_b \, e^{-\phi}\, {}^\star F + \cdots \,,
\ee
where ${}^\star F$ is the 6D Hodge dual of the background 6D Maxwell flux, $F_{\ssM \ssN}$, (whose gauge coupling is $\tilde g$); $\cW$ denotes the 4D world-surface of the brane and $\gamma_{ab} = g_{\ssM \ssN} \partial_a x^\ssM \partial_b x^\ssN$ is the induced metric on $\cW$. The ellipses in eq.~\pref{E:TPhidefs} correspond to terms involving at least two derivatives of the bulk fields.

In general, both $T_b$ and $\Phi_b$ can depend on the various bulk scalar fields --- in particular on the bulk dilaton, $\phi$, that appears in the 6D gravity supermultiplet --- and generically this dependence breaks the classical scaling symmetry whose presence is responsible for the bulk geometry's one classical modulus, $\varphi$. Because of this the brane-bulk backreaction combines with flux quantization to fix this last remaining modulus. This is why quantities like the 1PI action can depend on $\varphi$ once branes are present, even at the classical level.

\subsection*{Accidental SUSY}

Remarkably, we find for the simplest situation --- two identical branes that do not couple at all to the 6D dilaton $\phi$, situated at opposite ends of a rugby-ball geometry \cite{Towards}  --- the one-loop vacuum energy precisely vanishes. On closer inspection it does so because all bulk Kaluza-Klein (KK) modes come in degenerate bose-fermi pairs. In retrospect this happens because once the bulk modulus, $\varphi$, adjusts to relate $\Phi_b$ to $T_b$ as dictated by flux quantization, the boundary conditions at the brane allow a Killing spinor to exist in the bulk. That is, the branes unexpectedly leave unbroken the single `accidental' 4D subset of 6D supersymmetry that is also left unbroken by the bulk \cite{SS}. This residual supersymmetry was not noticed earlier because its existence requires $\Phi_b$ to be nonzero. Consequently it is not present for the `pure-tension' branes that are the usual fare of brane-world calculations.

This unbroken supersymmetry is accidental in the sense that it arises automatically for two identical branes, provided these are described only up to one-derivative level ({\em i.e.} by eq.~\pref{E:TPhidefs}), assuming only that $T_b$ and $\Phi_b$ do not depend on $\phi$. It is in general broken once higher-derivative effective brane-bulk interactions are also included, since these modify the boundary conditions of bulk fields in such a way as to preclude there being a Killing spinor. What is remarkable is how generic this supersymmetry is, since it depends on only to two requirements: ($i$) that the branes not couple to the bulk dilaton, $\phi$; and ($ii$) that both branes are identical\footnote{Even though supersymmetry breaks when ($ii$) is not satisfied, it is known \cite{GGP, OtherConical} that the bulk geometry obtained is still flat in the on-brane directions, provided only that ($i$) is satisfied.} (such as might be enforced by a $Z_2$ symmetry).

Because of this accidental supersymmetry, the bulk contribution to the vacuum energy should vanish to all orders in the absence of brane-localized fields and of two-derivative (and higher) interactions on the brane. We explicitly verify that this is true at one-loop order, by generalizing results derived earlier for the non-supersymmetric case \cite{Companion}.

More generally, non-supersymmetric configurations can also be explored for which the localized flux differs on the two branes. We find that integrating out massive bulk supermultiplets at one-loop gives a low-energy vacuum energy contribution that is generically of order
\be \label{E:Coverrsq}
 \Lambda \simeq \frac{C}{(4 \pi r^2)^2} \,,
\ee
where $C$ is an order-unity constant obtained by summing the contributions of all fields in the problem (and to which bosons and fermions contribute with opposite signs). Generically $C$ is proportional to whatever quantities break supersymmetry, for instance giving $C \propto (\Delta \Phi)^2$ for branes with unequal fluxes: $\Delta \Phi = \Phi_+ - \Phi_-$. In the supersymmetric case of identical branes $C = 0$.

Ultimately, the surprisingly small size of \pref{E:Coverrsq} has two sources. It can be partially traced to the supersymmetry of the bulk geometry, since 6D supersymmetry strongly restricts how the bulk action is renormalized by short-wavelength UV modes. In particular, one-loop renormalizations of the bulk action (and its higher-derivative corrections) vanish once summed over a 6D supermultiplet for supersymmetric rugby balls, independent of what the brane properties are. This generalizes (for low-spin fields) to rugby balls an earlier result for Ricci-flat geometries \cite{RicciFlatUV}.

The second important ingredient underlying \pref{E:Coverrsq} is classical scale invariance, which ensures the bulk action can be written in the form
\be
 S_\ssB = \int \exd^6 x \sqrt{- \hat g} \; e^{-2\phi} \, L_\ssB (\hat g_{\ssM\ssN}, \partial_\ssM \phi, \cdots ) \,,
\ee
where $L_\ssB$ does not depend on $\phi$ undifferentiated, and the Jordan-frame metric, $\hat g_{\ssM\ssN}$ is related to the Einstein-frame metric in 6D by $\hat g_{\ssM \ssN} = e^\phi \, g_{\ssM \ssN}$. This guarantees that a factor of $e^{2\phi}$ accompanies each loop in the Jordan frame, and so provides the bulk theory's loop-counting parameter. $e^\phi$ turns out to be very small for large rugby balls because flux stabilization dictates that $e^{\phi} \sim 1/(M_6 r)^2$, where $M_6$ is of order the 6D Planck scale (more about which below). Consequently each bulk loop contributes a factor proportional to $1/r^4$, making the one-loop vacuum energy naturally of order the KK scale.

In the 6D Einstein frame these same factors of $e^\phi$ are also easily understood, since there they arise because Einstein-frame masses, $m$, are related to Jordan-frame masses, $M$, by $m^2 = M^2 e^\phi$. Consequently $m \simeq 1/r$ even if $M \simeq M_6$. To obtain $m \simeq M_6$ would require $M \gg M_6$, for which a proper treatment requires understanding the UV completion above $M_6$, likely a string theory. It is here that bulk supersymmetry is likely to play an even more important role.

Finally, we use the results of the one-loop calculation to estimate the size of higher loops. In particular, we explore the size of two-loop contributions in the supersymmetric case for which the one-loop result vanishes. Here we find the most dangerous contributions involve both a bulk and a brane loop, and in some circumstances these can contribute $\Lambda \propto \mu^2 m^2 /(4\pi)^4 \propto \mu^2/(16\pi^2r)^2$, where $\mu$ is a brane mass and $m^2 \simeq M^2 e^\phi$ is the bulk mass encountered above. When present such contributions dominate, and we explore when this obtains.

The rest of this paper is organized as follows. First, \S\ref{sec:bulksugra} describes the bulk supergravity of interest, its rugby ball solutions and their supersymmetry properties. Then \S\ref{sec:ModeSums} briefly recaps the results of ref.~\cite{Companion} for the one-loop 1PI action as computed for spins zero, half and one propagating within the rugby-ball geometry, with a focus on how short-wavelength modes renormalize the bulk and brane actions. Next, \S\ref{sec:SUSYmult} assembles these renormalization results for individual particles into a result for several 6D supermultiplets. Then \S\ref{sec:4Dvaceng} computes how to get from the 1PI action to the 4D vacuum energy, tracking how the bulk back-reacts to the loop-changed brane energy densities, contributing an amount comparable to the direct loop-generated changes themselves. Finally, a brief summary of our conclusions, and the estimate of higher-loop bulk-brane effects can be found in \S\ref{sec:Conclusions}.

\section{Bulk field theory and background solution}
\label{sec:bulksugra}

We begin by summarizing the field content and dynamics of the bulk field theory of interest: six-dimensional gauged, chiral supergravity \cite{NS,MS,SS} coupled to a number of 6D gauge- and hyper- supermultiplets.

\subsection{6D gauged, chiral supergravity}
\label{S:6DSS}

The field content of the supergravity sector of the theory consists of the minimal supergravity multiplet plus a single chiral Kalb-Ramond tensor multiplet; that is, a metric ($g_{\ssM\ssN}$), antisymmetric Kalb-Ramond field ($B_{\ssM\ssN}$), dilaton ($\phi$), gravitino ($\psi_\ssM$) and dilatino ($\chi$). The theory has a lagrangian formulation\footnote{In general 6D supergravities need not \cite{MS}, when self-dual or anti-self dual Kalb-Ramond fields are present.} because the Kalb-Ramond field has both self-dual and anti-self-dual parts (one comes from the gravity multiplet and the other from the tensor multiplet) and this is the purpose of including the single chiral tensor multiplet. The supergravity is chiral because the fermions are all complex 6D Weyl spinors -- satisfying $\gamma_7 \, \psi_M = \psi_M$ and $\gamma_7 \, \chi = - \chi$.

This gravity multiplet can also couple to matter supermultiplets, of which we consider two types: gauge multiplets -- containing a gauge potential ($A^a_\ssM$) and a chiral gaugino ($\gamma_7 \, \lambda^a = \lambda^a$); or hyper-multiplets --- comprising two complex scalars ($\Phi^\ssI$) and their chiral hyperini ($\gamma_7 \, \Psi^\ssI = -\Psi^\ssI$). 6D supersymmetry requires the scalars within the hypermultiplets to take values in a quaternionic manifold, and precludes them from appearing in the gauge kinetic terms or in the kinetic term for the dilaton field $\phi$ \cite{janber}.

The supergravity is called `gauged' because the 6D supersymmetry algebra has an abelian $U(1)_\ssR$ symmetry that does not commute with supersymmetry and is gauged by one of the gauge multiplets. The fermion fields $\psi_\ssM$, $\chi$ and $\lambda^a$ all transform under the $U(1)_\ssR$ gauge symmetry, as do the hyper-scalars, $\Phi^\ssI$ (but not the hyperini, $\Psi^\ssI$). For instance, the gravitino covariant derivative is
\begin{equation} \label{E:covderiv}
    D_{\ssM}\psi_{\ssN} = \left(\partial_{\ssM} -
    \frac14 \, {\omega_{\ssM}}^{\ssA\ssB}\,\Gamma_{\ssA\ssB}  - iA_{\ssM}\right)\psi_{\ssN} - \Gamma^\ssL_{\ssM\ssN} \psi_\ssL \,,
\end{equation}
where ${\omega_{\ssM}}^{\ssA\ssB}$ denotes the spin connection, $\Gamma^\ssL_{\ssM\ssN}$ the metric's Christoffel symbol, $\Gamma_{\ssA\ssB} := \frac12 \left[ \Gamma_\ssA , \Gamma_\ssB \right]$ is the commutator of two 6D Dirac matrices and the gauge field $A_\ssM$ gauges the 6D $U(1)_\ssR$ symmetry.

\subsubsection*{Anomaly cancellation}

Because the fermions are chiral there are gauge and gravitational anomalies, which must be cancelled using a version of Green-Schwarz anomaly cancellation \cite{GSAC, AGW}. In 6D this is not possible for generic anomalous theories, but under some circumstances can be done. In particular, Green-Schwarz anomaly cancellation requires: a Kalb-Ramond field which shifts under the anomalous gauge symmetry (and so whose field strength contains a Chern-Simons term for this symmetry), and some restrictions on the gauge groups and number of chiral matter fields present \cite{6DAC}. In particular, the number of gauge- and hyper-multiplets, $n_\ssG$ and $n_\ssH$, must satisfy \cite{6DAC}
\be \label{eq:nHvsnV}
 n_\ssH = n_\ssG + 244 \,.
\ee
We see from this that anomaly-freedom ensures there are literally hundreds of matter multiplets.

For the theory of interest here the Kalb-Ramond field required by anomaly cancellation is simply $B_{\ssM\ssN}$ of the supergravity multiplet, whose field strength, $G_{\ssM \ssN \ssP}$, is required by supersymmetry to contain Chern-Simons contributions. For instance, at lowest order
\begin{equation}
    \label{E:CSterm}
    G_{\ssM\ssN\ssP} = \partial_{\ssM} B_{\ssN\ssP} + \frac{\kappa}{\gR^2} \, F_{\ssM\ssN} A_{\ssP} + \hbox{(cyclic
    permutations)} \, ,
\end{equation}
where $F_{\ssM\ssN} = \partial_\ssM A_\ssN - \partial_\ssN A_\ssM$ is the abelian gauge field strength for the $U(1)_\ssR$ gauge symmetry, and $\gR$ is its coupling constant. More generally, at higher orders anomaly cancellation also requires $G_{\ssM\ssN\ssP}$ to contain gravitational Chern-Simons terms.

\subsubsection*{Bulk action and field equations}

The bosonic part of the classical 6D supergravity action
is:\footnote{Our metric is `mostly plus' and we follow Weinberg's curvature conventions \cite{GandC}, which differ from those of MTW \cite{MTW} only by an overall sign in the definition of the Riemann tensor. To keep the same notation as \cite{Companion} we adopt here a convention for gauge fields that differs in normalization by a factor of the relevant gauge coupling, $g_a$, compared with our earlier papers on 6D supergravity.}
\bea
\label{E:BactionSF}
    \frac{{\cal L}_\ssB}{\sqrt{- \hat g}} &=& e^{-2 \phi} \left[ -\, \frac{1}{2\kappa^2} \, \hat g^{\ssM \ssN} \Bigl( \hat R_{\ssM \ssN} +
    \partial_{\ssM} \phi \, \partial_\ssN \phi \Bigr)  -
    \frac{1}{12} \; G_{\ssM \ssN\ssP} \,
    G^{\hat\ssM \hat\ssN \hat\ssP} \right. \nn\\
    && \qquad\qquad \left. - \frac{1}{4 g^2_a} \; F^a_{\ssM\ssN} F_a^{\hat\ssM \hat\ssN}- \frac12 \, \cG_{\ssI \ssJ} (\Phi) \, \hat g^{\ssM \ssN} D_\ssM \Phi^\ssI D_\ssN \Phi^\ssJ - \frac{2 \gR^2}{\kappa^4} \, \cU(\Phi) \right] \,,
\eea
where carets indicate curvatures, determinants or raised indices that are computed using the metric, $\hat{g}_{\ssM \ssN}$. Here the sum over gauge fields includes, in particular, the abelian factor that gauges the $U(1)_\ssR$ symmetry --- whose gauge coupling, $\gR$, appears in the scalar potential on the right-hand side. $\cG_{\ssI\ssJ}(\Phi)$ is the metric of the quaternionic coset space, $M = G/H$, in which the $\Phi^\ssI$ take their values.

Eq.~\pref{E:BactionSF} can be rewritten in the 6D Einstein frame by rescaling $\hat g_{\ssM \ssN} = e^\phi \, g_{\ssM \ssN}$, to give
\bea
\label{E:Baction}
    \frac{{\cal L}_\ssB}{\sqrt{- g}} &=& -\, \frac{1}{2\kappa^2} \Bigl( R +
    \partial_{\ssM} \phi \, \partial^\ssM \phi \Bigr)  -
    \frac{e^{-2\phi}}{12} \; G_{\ssM\ssN\ssP} \,
    G^{\ssM\ssN\ssP}  \nn\\
    && \qquad\qquad - \frac{e^{-\phi}}{4g_a^2} \; F^a_{\ssM\ssN} F_a^{\ssM\ssN}- \frac12 \, \cG_{\ssI \ssJ} (\Phi)\, g^{\ssM \ssN} D_\ssM \Phi^\ssI D_\ssN \Phi^\ssJ - \frac{2 \gR^2}{\kappa^4} \,  e^\phi \, \cU(\Phi) \,.
\eea
The potential, $\cU(\Phi)$, is nontrivial and depends on the gauge group and other details but in the cases for which it is known \cite{RandjbarDaemi:2004qr}  it is extremized for $\Phi^\ssI = 0$, near which
\be
  \cU = 1 + \frac{\kappa^2} 2 \, \cG_{\ssI\ssJ}(0) \Phi^\ssI \Phi^\ssJ + \dots \,.
\ee
In particular $\Phi^\ssI = 0$ is consistent with the full equations of motion.

The presence of $e^{-2\phi}$ as an overall prefactor in eq.~\pref{E:BactionSF} reveals $e^{2\phi}$ as the loop-counting parameter, and this action neglects higher-order corrections that are suppressed relative to the ones shown by powers of $e^{2\phi}$ and/or higher derivatives. Among these are interactions that are related by supersymmetry to anomaly canceling terms, such as one-loop corrections to the gauge kinetic function, $\sqrt{- \hat g} \; F^a_{\ssM\ssN} F_a^{\hat\ssM \hat\ssN} = \sqrt{-g} \; e^\phi \, F^a_{\ssM\ssN} F_a^{\ssM \ssN}$ \cite{6DSUSY, HigherTerms}.

The equations of motion for the bosonic fields which follow from the action, eq.~\pref{E:Baction}, after using $\Phi^\ssI = 0$ are:
\eqa \label{E:Beom}
 &&\Box \, \phi + \frac{\kappa^2}6 \, e^{-2 \phi} \,
 G_{\ssM\ssN\ssP} \, G^{\ssM\ssN\ssP} + \frac{\kappa^2}{4g_a^2} \, e^{-\phi} \; F_{\ssM\ssN}^a F^{\ssM\ssN}_a - \frac{2 \gR^2}{\kappa^2} \, e^\phi = 0 \nn\\
 &&D_\ssM \Bigl( e^{-2\phi} \, G^{\ssM\ssN\ssP} \Bigr) = 0  \,,
 \qquad\qquad D_\ssM \Bigl( e^{-\phi} \, F_a^{\ssM\ssN} \Bigr)  = 0 \\
 &&D_\ssM \Bigl( e^{-\phi} \, F^{\ssM\ssN} \Bigr) + \kappa \, e^{-2\phi} \, G^{\ssM\ssN\ssP} \,
 F_{\ssM\ssP} = 0 \nn \\
 &&R_{\ssM\ssN} + \partial_\ssM\phi \, \partial_\ssN \phi + \frac{\kappa^2}2 \,
 e^{-2\phi} \, G_{\ssM\ssP\ssQ} \, {G_\ssN}^{\ssP\ssQ} + \frac{\kappa^2 e^{-\phi}}{g_a^2} \, F_{\ssM\ssP}^a {F_a\ssN}^{\ssP}
 + \frac12 \,  (\Box \phi )\, g_{\ssM\ssN} = 0 \,, \nn
\eeqa
where the second-last equation is for the $U(1)_\ssR$ gauge potential whose Chern-Simons term appears in the field strength $G_{\ssM \ssN \ssP}$, as in eq.~\pref{E:CSterm}.

\subsection*{Massive supermultiplets}

Both the gauge- and hyper- supermultiplets described above furnish representations of 6D supersymmetry for massless particles. By contrast, the particle content for a massive 6D matter multiplet consists of a massive gauge particle, a massive Dirac fermion and three scalars - a total of 8 bosonic and 8 fermionic states.

Since this is also the combined field content of a gauge- plus a hyper-multiplet, one expects to be able to form a massive multiplet by having the gauge boson from a gauge multiplet `eat' one of the scalars of a hypermultiplet through the Higgs mechanism. For ungauged supergravity, with vanishing scalar potential, this is indeed what happens in general as the hyperscalars can take arbitrary constant values in the vacuum. This picture is also consistent with the observation that massive states should not alter the anomaly cancellation conditions since the condition, eq.~\pref{eq:nHvsnV}, is not modified when equal numbers of gauge and hypermultiplets are added to the system.

If $w$ denotes the v.e.v. of the field that breaks the relevant gauge symmetry, we expect the common mass of all elements of the massive supermultiplet to be of order $m^2 \sim e^{\phi} w^2$ (in the 6D Einstein frame\footnote{A frame-independent way to write this is $\kappa m^2 \sim e^\phi \kappa w^2$.}). This dependence of $m^2$ on $\phi$ can be seen in several ways: for the gauge fields it arises because of the presence of $e^{-\phi}$ in the gauge kinetic term. Alternatively, the proportionality $m^2 \propto e^{\phi}$ can also be seen from the overall factor of $e^\phi$ in the hypermultiplet scalar potential, $U = 2 \, \gR^2 e^\phi \, \cU(\Phi)$. These factors of $e^{\phi}$ play an important role in the overall size of the effects found later from loops of massive fields.

\subsection{Rugby-ball compactifications}

The simplest compactified solutions \cite{SS, Towards} to the field equations \pref{E:Beom} are found using the Freund-Rubin {\em ansatz} \cite{FR} in which: $\phi =$ constant and
\eq \label{E:FRansatz}
    {g}_{\ssM\ssN} = \pmatrix{
    {g}_{\mu\nu}(x) & 0 \cr 0 & {g}_{mn}(y) \cr}
    \qquad \hbox{and} \qquad
    {F}_{\ssM\ssN} = \pmatrix{0 & 0 \cr 0 &
    f\, {\epsilon}_{mn}(y) \cr}  \,.
\eeq
Here ${g}_{\mu\nu}$ is a maximally-symmetric Lorentzian metric --- {\it i.e.} de Sitter, anti-de Sitter or flat space --- while ${g}_{mn}$ and ${\epsilon}_{mn}$ are the metric and volume form on the two-sphere, $S_2$. The Bianchi identity requires the quantity $f$ appearing in the background gauge field --- which could be any one of the gauge fields present in the theory --- is a constant. All other fields vanish.

As is easily verified, the above ansatz solves the field equations provided that the following three conditions are satisfied: ${R}_{\mu\nu} = 0$,
\be
 \frac{1}{\gB^2} \, {F}_{mn} {F}^{mn} = \frac{2f^2}{\gB^2} = \frac{8 \, \gR^2 }{ \kappa^{4}} \; e^{2{\phi}} \quad \hbox{and} \quad
 {R}_{mn} = - \frac{\kappa^2 }{ \gB^{2}} \; e^{-{\phi}} \, {F}_{mp} \, {{F}_n}^p = - \frac{ f^2 \, \kappa^2 }{ \gB^{2}} \;  e^{-{\phi}} \, {g}_{mn} \,,
\ee
where\footnote{The coupling $\gB$ as defined here is $\phi$-independent, and so is related to the coupling $\tilde g(\phi)$ used in \cite{Companion} by $\tilde g^2(\phi) = \gB^2 \, e^\phi$.} $\gB$ is the gauge coupling, $g_a$, for the specific gauge generator whose potential is nonzero in the background. This in general differs from the gauge coupling, $\gR$, of the abelian $R$-symmetry, $U(1)_\ssR$ (that enters through its appearance in the scalar potential). These imply the four dimensional spacetime is flat, plus the two conditions
\eq \label{E:phircond}
 e^{\phi}  = {\kappa^2 \over 4 \gR^2\, {r}^2}
 \quad \hbox{and} \quad
 f = \pm \frac{\gB}{2 \,\gR \, r^2} = \pm \frac{2 \,\gB \gR \, e^\phi}{\kappa^2}  \, .
\eeq
Notice that these expressions determine the values of $f$ and $\phi$ in terms of the size of the extra dimensions, implying in particular that $e^\phi$ becomes very small when $r$ is very large.

The gauge potential, ${A}_m$, that gives rise to the field strength $F_{mn}$ is the potential of a magnetic monopole. As such, it is subject to the condition that the total magnetic flux through the sphere is quantized:\footnote{This expression assumes that all charged matter fields couple to the background gauge potential with strength $\gB$, and so differs from the corresponding one in \cite{Companion} which allows the coupling strength to be $q \gB$. Although $\gB$ can be defined so that $q = 1$ for any particular matter field, this cannot be done for more than one field at a time. See \cite{Companion} for the expressions with general $q$.}
\be
  \int_{S_2} F = 4 \pi r^2   f = 2 \pi N \qquad \qquad \hbox{(sphere with no branes)}\,,
\ee
with $N = 0, \pm 1, ..$. This requires the normalization constant, $f$, to satisfy:
\eq \label{E:fquant}
    f = {N \over 2 \, {r}^2}  \qquad \qquad \qquad \qquad\qquad
    \hbox{(sphere with no branes)}
\eeq
where ${r}$ is the radius of the sphere. Comparing eqs.~\pref{E:phircond} and eq.~\pref{E:fquant} then implies $N = \pm \, \gB/g$, which is only possible if $\gB$ is an integer multiple of $g$.

Of particular interest in what follows is the special case where the bulk background flux lies in the $U(1)_\ssR$ direction, in which case
\be \label{eq:SScase}
 \gB = \gR \quad \hbox{and so} \quad N = \pm 1
 \qquad\qquad (\hbox{Salam-Sezgin solution})\,.
\ee
This solution turns out to preserve precisely one 4D supersymmetry \cite{SS}, and in later sections we seek to identify the size of supersymmetry breaking effects due to the back-reaction of the source branes.

Notice, however, that the value of $r$ itself is {\em not} determined by the field equations, indicating the existence of a (classical) flat direction. Because of eq.~\pref{E:phircond} this flat direction can be parameterized either by $r$ or $\phi$, and its existence is a general consequence of the following rigid classical scaling symmetry of the supergravity field equations:
\be \label{eq:scalingsym}
 \phi \to \phi + c \quad \hbox{and} \quad
 g_{\ssM\ssN} \to e^{-c} \, g_{\ssM\ssN} \,,
\ee
(and so $\hat g_{\ssM \ssN}$ is fixed). Since this is only a symmetry of the classical bulk equations, the flat direction can be lifted, even classically, once the bulk is coupled to brane sources that break this symmetry. Alternatively, it is also generically lifted by quantum effects, with $\ell$-loop corrections to the action proportional to $e^{2(\ell - 1)\phi}$ when expressed in terms of the scale-invariant metric, $\hat g_{\ssM\ssN}$.

\subsubsection*{Brane sources}

The solutions as outlined so far describe an extra-dimensional 2-sphere supported by flux, with metric
\be
 \exd s^2 = r^2 \Bigl( \exd \theta^2 + \sin^2 \theta \, \exd \varphi^2 \Bigr) \,,
\ee
without the need for brane sources \cite{SS}. However brane sources can be introduced into this supergravity solution \cite{Towards} simply by allowing the angular coordinate to be periodic with period $\varphi \simeq \varphi + 2 \pi \alpha$ with $\alpha$ not equal to unity. Geometrically, this corresponds to removing a wedge from the sphere along two lines of longitude and identifying points on opposite sides of the wedge \cite{preSLED}. This introduces a conical singularity at both the north and south poles, with defect angle $\delta = 2\pi (1 - \alpha)$, a geometry called the {\em rugby ball}.

Physically, this geometry describes the gravitational field of two identical brane sources, one situated at each of the two poles, with Einstein's equations relating the defect angle to the properties of the branes. Concretely, take the action of the brane to be\footnote{A more covariant way of writing the term linear in $F_{mn}$ is as the integral of the 6D Hodge dual, ${}^\star F$, over the 4-dimensional brane world-sheet \cite{localizedflux}.}
\bea \label{eq:genbraneaction}
 S_b &=& - \int \exd^4x \, \sqrt{-\gamma} \; L_b \nn\\
 \hbox{with} \quad L_b &=& T_b - \frac{ \cA_b}{2 \gB^2} \, \epsilon^{mn} F_{mn} 
 + \cdots \,,
\eea
with $\gamma_{ab} := g_{\ssM\ssN} \, \partial_a x^\ssM \, \partial_b x^\ssN$ being the brane's induced metric and ellipses denoting terms involving two or more derivatives. In general the coefficients $T_b$ and $\cA_b$ and so on could depend on any of the 6D scalars, $\phi$ or $\Phi^\ssI$.

The back-reaction of such a brane onto the extra-dimensional geometry is governed by the near-brane boundary condition the brane induces on all bulk fields. This boundary condition relates the radial derivative of the field to the brane action, for instance implying for the hyperscalars \cite{uvcaps}
\be \label{E:hsbranebc}
 \lim_{\rho \to 0} \Bigl[ \cG_{\ssI\ssJ}(\Phi) \, \rho \, \pd_\rho\Phi^\ssJ \Bigr] = \frac{\kappa^2}{2\pi} \left( \frac{\delta S_b}{\delta\Phi^{\ssI}} \right) \,,
\ee
where $\rho$ denotes proper distance from the brane. In general, a bulk field having a nonzero derivative near a brane diverges at the brane positions, leading to curvature singularity there. But it turns out that if the coefficients $T_b$, $\cA_b$ {\em etc.} are all independent of the bulk scalars, then the singularity is fairly mild: a conical defect such as found in the above rugby-ball geometries. In this case the near-brane boundary conditions degenerate to a formula \cite{Vil, localizedflux} for the defect angle at the brane's position:
\be \label{E:deltavsL}
 \delta_b = \kappa^2 L_b \,.
\ee
In the special case of a rugby-ball solution, since the defect angle is the same at both poles the same must be true of $L_b$ for the corresponding branes at each pole,\footnote{See \cite{GGP, OtherConical, laterconical} for solutions with conical singularities that can differ at the two poles.} with
\be
 2\pi(1 - \alpha) = \kappa^2 L_\pm = 8\pi G_6 L_\pm \,,
\ee
where $b = \pm$ labels the two poles.

The presence of the brane sources complicates the flux quantization condition in two important ways. The first complication arises because the resulting defect angle changes the volume of the sphere, which appears in the flux-quantization condition when integrating over the bulk magnetic field,
\be
 \int_{S_2(\alpha)} F = 4\pi\alpha \, r^2 f \,.
\ee
The second complication arises because the branes themselves can carry a localized flux, given by
\be
 2 \pi \Phi_b = \cA_b \, e^\phi \,.
\ee
This can be seen by asking how $\cA_b$ changes the boundary conditions for the bulk gauge field, and tracking these through the flux-quantization condition, which becomes \cite{localizedflux}
\be \label{eq:BLFquantization}
 2 \pi N = 2\pi \,\Phi + \int_{S_2(\alpha)} F
 = 2\pi\,\Phi + 4 \pi \alpha\, r^2 f \,,
\ee
where $\Phi := \sum_b \Phi_b$.

Solving this for $f$ and comparing with the bulk field equation, eq.~\pref{E:phircond}, we find that eq.~\pref{E:fquant} generalizes to
\be
 f= \frac{\cN}{2\,r^2} = \pm \frac{\gB}{2 \, \gR \, r^2} \,,
\ee
where $\cN:=\omega(N-\Phi) = \pm \gB/\gR$ and we follow \cite{Companion} by defining (for later convenience)
\be
 \omega:=\frac{1}{\alpha} \,.
\ee

Notice that if $\cA_b \propto e^{-\phi}$ then $2\pi \Phi = \sum_b \cA_b e^{\phi}$ is independent of $\phi$, and so also independent of the flat direction guaranteed by eq.~\pref{eq:scalingsym} (which can be parameterized by $\phi$). However, if $\cA_b$ has any other $\phi$-dependence (and in particular if it is $\phi$-independent) then $\Phi$ varies with $\phi$, and eq.~\pref{eq:BLFquantization} lifts the degeneracy of the flat direction. It then should be regarded as an equation to be solved for $\Phi$ (and so also for $\phi$), to give
\be
 \label{eq:fluxstabilizationcondition}
 \Phi = N - \alpha \, \cN = N \mp \frac{\alpha \, \gB}{\gR} \,.
\ee
For instance, if $\gB = \gR$ and $N = \pm 1$ (as in the Salam-Sezgin solution), then using $\alpha = 1 - \delta/2\pi$ implies
\be \label{E:SSqntznresult}
 \Phi = \pm ( 1 - \alpha ) = \pm \frac{\delta}{2\pi}
 \qquad\qquad \hbox{(if $\gB = \gR$ and $N = \pm 1$)}\,.
\ee
This can be regarded as a dynamical adjustment of $\Phi$ to track the defect angle (and so also the brane tensions) so long as $\Phi$ depends on the flat direction, $\phi$ ({\em i.e.} so long as $\sum_b \cA_b$ is not proportional to $e^{-\phi}$).

Notice that because eq.~\pref{E:deltavsL} gives the defect angle as a function of tension and brane flux, once the brane-localized flux adjusts to track the brane tension the defect angle is completely determined by the brane tensions alone. However the presence of the flux acts to change the size of the defect angle produced by a particular tension, $T$, relative to its naive value. That is, for a `pure tension' brane --- {\em i.e.} in the absence of higher-derivative brane interactions (including brane-localized flux) --- each brane's contribution to the defect angle would be controlled by its tension
\be \label{E:naivedeltavsT}
 2\pi (1 - \alpha) = \kappa^2 T = 8\pi G_6 T
 \qquad\qquad \hbox{(no brane-localized flux)}\,.
\ee
But in the presence of brane-localizing flux the brane lagrangian instead evaluates to
\be \label{E:LbvsTPhi}
 L_b = T_b - \frac{\cA_b \, f}{\gB^2} + \cdots
 = T_b - \frac{2\pi \Phi_b \, e^{-\phi} f}{\gB^2} + \cdots
 = T_b \mp \frac{ 4\pi \, \gR \, \Phi_b}{\gB \kappa^2}+ \cdots \,,
\ee
where the ellipses represent terms like $R$ that involve at least two derivatives (and so are down by at least $1/r^2$). We see that for the rugby ball with equal fluxes and tensions ($T_+ = T_- = T$ and $\Phi_+ = \Phi_- = \frac12 \, \Phi$) combining this with eq.~\pref{eq:fluxstabilizationcondition} gives the relation between defect angle and tension as
\be \label{E:deltavsTgN}
 \delta = 2\pi( 1 - \alpha) = 4\pi G_6 T + \pi \left( 1 \mp \frac{\gR N }{\gB} \right)  \,.
\ee
For instance, in the Salam-Sezgin case --- where $\gB = \gR$ and $N = \pm 1$ --- the presence of $\Phi$ makes the defect angle $\delta = 2\pi(1 - \alpha)$ half as large as it would have been -- {\em i.e.} eq.~\pref{E:naivedeltavsT} -- if $\Phi$ had vanished:
\be \label{E:deltavsalphaSUSY}
 \delta = 2\pi(1 - \alpha) = \frac{\kappa^2 T}{2} = 4\pi \,G_6 T \qquad \hbox{(with brane-localized flux)} \,.
\ee

Having a single-derivative and no-derivative term compete in this way might raise concerns for the validity of the derivative expansion for the brane action. However the brane-localized flux can be larger than the other terms in a derivative expansion for two reasons: its dependence on the zero mode and its participation in the flux-quantization condition. On one hand the dependence on the (otherwise undetermined) zero mode makes its coefficient free to adjust to satisfy flux quantization. And on the other hand, flux quantization makes it compete with the bulk flux and so drives its coefficient out to a volume-enhanced value. The same is not true for other terms in the derivative expansion of the brane action.

\subsection{Supersymmetry of the solutions}

It was famously shown by Salam and Sezgin \cite{SS} that the spherical solution (no defect angle) using $N = \pm 1$ unit of $U(1)_\ssR$ flux preserves a single 4-dimensional supersymmetry. We here reproduce their argument to identify how back-reaction in the presence of branes changes this conclusion.\footnote{See ref.~\cite{6D4DSUSY} for a precursor to the argument we present here.} We find that pure tension branes always break all of the bulk supersymmetry, but supersymmetry can be preserved if both tension and brane-localized flux are present. In particular, we find that the condition for unbroken supersymmetry is precisely the same condition as is imposed by flux quantization, as found earlier (eq.~\pref{E:SSqntznresult}).

A background configuration does not break supersymmetry if the supersymmetry transformations all vanish when evaluated at the background solution. Since the variations of bosonic fields all vanish trivially (because all fermions vanish in the background), it suffices only to evaluate the fermionic variations. For the 6D supergravity of interest, with background $U(1)_\ssR$ flux and vanishing hyperscalars, this requires all of
\ba
 \delta\lambda &=& \frac1{2\sqrt 2 \, \gR} \, e^{-\phi/2} F_{\ssM\ssN} \Gamma^{\ssM\ssN} \epsilon - \frac {i\sqrt 2 \, \gR}{\kappa^2} \, e^{\phi/2} \epsilon \nn\\
 \delta\chi &=& \frac1{\kappa\sqrt 2} (\pd_\ssM\phi) \Gamma^\ssM \epsilon + \frac1{12} \, e^{-\phi} G_{\ssM\ssN\ssP} \Gamma^{\ssM\ssN\ssP} \epsilon \nn\\
 \delta\psi_\ssM &=& \frac{\sqrt 2}\kappa \, D_\ssM \epsilon + \frac1{24} \, e^{-\phi} \, G_{\ssP\ssQ\ssR} \Gamma^{\ssP\ssQ\ssR} \Gamma_\ssM \epsilon
\ea
to vanish.

First consider the variation of the dilatino, $\chi$. Since 4D maximal symmetry and 2 internal dimensions require vanishing $G_{\ssM \ssN \ssP}$, the condition $\delta \chi = 0$ implies the dilaton must be a constant: $\partial_\ssM \phi = 0$. Since back-reaction relates $\delta S_b/\delta \phi$ to the near-brane limit of $\rho \, \partial_\rho \phi$, the requirement that $\phi$ be a constant implies all brane actions must be stationary with respect to dilaton variations when evaluated at the background. A sufficient condition for this to be so is to have all of the coefficient functions, $T_b$, $\cA_b$ {\em etc.}, be completely independent of the dilaton.

Next, the condition $\delta \lambda = 0$ can be written as
\be
 0 = \sqrt 2 \; e^{-\phi/2} \left( \frac1{4 \gR} \, F_{\ssM\ssN} \Gamma^{\ssM\ssN} - \frac{i\gR e^\phi}{\kappa^2} \right) \epsilon
 = \frac{\sqrt 2 \; e^{-\phi/2}}{4\gR r^2} \left( \pm\frac1{2} \epsilon_{mn} \Gamma^{mn} -i\right) \epsilon \,,
\ee
when evaluated with $F_{mn} = \pm \epsilon_{mn}/(2 r^2)$ and $e^\phi = \kappa^2/ (4\gR^2 r^2)$. Using the following representation of 6D Gamma matrices:\footnote{In what follows, we follow the conventions in Appendix C of \cite{Companion}.}
\be
\Gamma^\mu = \left(
\begin{array}{cc}
0 & \gamma^\mu \\
\gamma^\mu & 0
\end{array}
\right) \,,\quad
\Gamma^4 = \left(
\begin{array}{cc}
0 & \gamma_5 \\
\gamma_5 & 0
\end{array}
\right) \,,\quad
\Gamma^5 = \left(
\begin{array}{cc}
0 & -i \one_4 \\
i \one_4 & 0
\end{array}
\right)
\ee
where $\gamma^\mu$ are the usual 4D Dirac matrices and $\gamma_5 = -i \gamma^0 \gamma^1 \gamma^2 \gamma^3$, we have
\be
 \epsilon_{mn} \Gamma^{mn} = 2i \left(
 \begin{array}{cc}
 \gamma_5 & 0 \\
 0 & -\gamma_5
 \end{array}
 \right) \,,
\ee
and so the condition $\delta \lambda = 0$ implies the 6D Weyl spinor $\epsilon$ satisfies
\be \label{E:6DWeylspinor}
 \epsilon = \left(
 \begin{array}{c}
 \varepsilon_{4\pm} \\  0
 \end{array}
 \right) \,,
\ee
where $\varepsilon_{4\pm}$ is a 4D spinor that satisfies the 4D Weyl condition $\gamma_5 \varepsilon_{4\pm} = \pm \varepsilon_{4\pm}$, with the sign correlated with that of $\cN = 2 r^2 f = \pm \gB/\gR = \pm 1$.

Finally the condition $\delta \psi_\ssM = 0$ boils down to the existence of a covariantly constant (Killing) spinor:
\be
 D_\ssM\epsilon= \left(\pd_\ssM - \frac{i}4 \Gamma_{\ssA\ssB} \, \omega_\ssM^{\ssA\ssB} - i  A_\ssM \right)   \epsilon  = 0 \,,
\ee
where the covariant derivative of $\epsilon$ depends on $A_\ssM$ because the corresponding symmetry is an $R$ symmetry (and so does not commute with supersymmetry). The integrability condition for such a spinor states $[D_\ssM , D_\ssN] \epsilon = -i  \left( \frac12 \, R_{\ssM \ssN \ssP \ssQ} \Gamma^{\ssP \ssQ} + F_{\ssM \ssN} \right) \epsilon = 0$, which for the rugby-ball background becomes
\be
 \frac{i}{2 r^2} \Bigl( \Gamma_{mn} - \cN \epsilon_{mn} \Bigr) \epsilon
 =0 \,.
\ee
This is automatically satisfied by eq.~\pref{E:6DWeylspinor} together with $\gamma_5 \varepsilon_{4\pm} = \cN \varepsilon = \pm \varepsilon_{4\pm}$.

To find the Killing spinor we take two coordinate patches, centered about the North and South poles (labeled by $b = \pm$), and use the frame fields
\be
 {e_a}^m =
 \frac1r \left(
  \begin{array}{cc}
 \cos\varphi & \frac{-b\,\sin\varphi}{\alpha \sin\theta} \\
 b\,\sin\varphi & \frac{\cos\varphi}{\alpha \sin\theta}
  \end{array}
 \right) \,,\,\, {e_\alpha}^\mu = \delta_\alpha^\mu \,,\,\, {e_\alpha}^m = 0 \,,
\ee
to compute the following non-zero components for the spin connection $\omega_\ssM^{\ssA\ssB}$:
\be
 \omega_\varphi^{45} = \alpha\cos\theta-b = -\omega_\varphi^{54} \,.
\ee
The background gauge potential satisfying the near-brane boundary conditions dictated by back-reaction \cite{localizedflux} is similarly given by
\be
  A_\varphi = -\frac{\cN\alpha}2 (\cos\theta-b) + b \, \Phi_b \,,
\ee
where $\cN = \pm 1$. The non-trivial component of the covariant derivative becomes
\be
 D_\varphi \epsilon = \left[\pd_\varphi -\frac{i}2 \left(
 \begin{array}{cc}
 \gamma_5 & 0 \\
 0 & -\gamma_5
 \end{array}
 \right) (\alpha \cos\theta -b) + \frac{i\cN\alpha}2 (\cos\theta-b) - ib \, \Phi_b  \right] \epsilon =0\,,
\ee
and so $\varepsilon_{4\pm}$ must satisfy
\be
 \left\{ \partial_\varphi + ib \left[ \pm \frac12 (1 - \alpha) - \Phi_b \right] \right\} \varepsilon_{4\pm} = 0 \,.
\ee
Equivalently,
\be
 \Bigl[ \partial_\varphi - i (\Phi_+ - \Phi_-) \Bigr] \varepsilon_{4\pm} = 0
 \quad \hbox{and} \quad
 \left[ \pm (1 - \alpha) - \Phi \right] \varepsilon_{4\pm} = 0 \,,
\ee
where $\Phi := \Phi_+ + \Phi_-$, and so solutions exist (and are constants) when the branes satisfy
\be \label{E:SUSYcond}
 \Phi_+ = \Phi_- = \frac{\Phi}{2} = \pm \frac12\, (1 - \alpha) = \pm \frac{\delta}{4\pi} \,.
\ee

We see that a single 4D supersymmetry survives when the branes are identical --- {\em i.e.} have equal tensions\footnote{Non-rugby-ball solutions with differing tensions also have nontrivial dilaton profiles \cite{GGP, OtherConical, laterconical}, and so are excluded by the condition $\partial_\ssM \phi = 0$.} and fluxes --- and with localized fluxes related to their tensions by eq.~\pref{E:SUSYcond}. In particular, when $\Phi_b = 0$ then any nonzero brane tension --- $\alpha \ne 1$ --- breaks supersymmetry.

Finally, we remark on the remarkable equivalence of the flux-quantization condition, eq.~\pref{E:SSqntznresult}, and the supersymmetry condition, eq.~\pref{E:SUSYcond}, on $\Phi$. This states that the value to which $\Phi$ is dynamically driven along the classical flat direction by flux quantization is precisely the one supersymmetric point on this flat direction. In particular, when $T_b$ and $\cA_b$ are $\phi$-independent (which ensures compatibility with vanishing gradients for $\phi$) this flat direction stabilizes at the supersymmetric position {\em for any choice} (consistent with the rugby-ball condition $L_+ = L_-$) for the constant coefficients $T_b$ and $\cA_b$.

\subsubsection*{Control of approximations}

We close this section with a brief summary of the domain of validity of the previous discussions, which has two important components: weak coupling and slowly-varying fields.

First, since we work within the semi-classical approximation, slowly varying fields are required to trust the effective 6D supergravity approximation for whatever theory (presumably a string theory) provides its ultraviolet completion. In practice, without knowing the details of this UV completion, we demand fields vary slowly relative to the length scale set by $\kappa$. This is the analogue of the $\alpha'$ expansion in string theory, and in the Jordan frame it requires $\hat r^2 \gg \kappa$ where $\hat r$ is the size of the extra dimensions as measured with the Jordan-frame metric, $\hat g_{\ssM \ssN}$. In terms of the Einstein-frame radius, $r$, used elsewhere in the text, this condition instead is $t := r^2 e^\phi/\kappa \gg 1$. If the classical rugby-ball solutions are to fall within this regime, eq.~\pref{E:phircond} shows that we must require
\be
 \kappa \gg 4 \gR^2 \,.
\ee

Second, since (as remarked earlier) each bulk loop in the 6D supergravity of interest comes accompanied by a factor of $e^{2\phi}$, the semiclassical approximation additionally requires weak coupling: $e^\phi \ll 1$. (This is the analogue of the condition of small string coupling for string compactifications.) This implies a semiclassical understanding of the flat direction labeled by $r$ (or $\phi$) is possible within the regime
\be
 \frac{\kappa}{r^2} \ll e^\phi \ll 1 \,.
\ee

Next, once brane sources are included we must also demand them not to curve excessively the background geometry, and for branes with tension $T$ this requires %
\be
 \kappa^2 T_b \ll 1 \,.
\ee
For rugby-ball geometries this ensures the defect angle satisfies $\delta \ll 2\pi$.

Finally, the semiclassical approximation also restricts the properties of particles that can circulate within loops, even if these do not appear among the background fields. Most notably their masses cannot be too large if quantum effects associated with gravity are to remain under control \cite{GREFTrev}. For particles of mass $m^2 = M^2 e^\phi$ this requires
\be
 \gR^2 \, m^2 = \gR^2 \, M^2 e^\phi \ll \kappa m^2 = \kappa M^2 e^\phi \ll 1 \,.
\ee

\section{Mode sums and renormalization}
\label{sec:ModeSums}

This section summarizes the results of the companion paper \cite{Companion}, so readers familiar with \cite{Companion} should feel free to skip to \S4. Our goal is to compute the UV-sensitive part of the 1PI quantum action, $\Gamma = S + \Sigma$, due to bulk loops. Our starting point is the following expression
\eq{
    i\Sigma = -i \int \exd^4 x \, \Vone
     = -\, \frac{1}{2} \, (-)^\ssF \, \hbox{Tr}\;
    \hbox{Log} \,\left( \frac{ -\Box_6 + X + m^2}{\mu^2} \right) \,,
    \label{eqn: sigma} }
\eeq
for the one-loop action arising from a loop of low-spin 6D fields moving in the background rugby-ball geometry. The calculation is quite general, assuming only that the field has statistics $(-)^\ssF = \pm$ with upper (lower) sign applying for bosons (fermions), and its kinetic operator (or, for fermions, its square) can be written in the form $- \Box + X + m^2$. We also assume the six-dimensional d'Alembertian splits into the sum of four- and two-dimensional pieces: $\Box_6 = \Box_4 + \Box_2$; $X$ is some local quantity (perhaps a curvature or background flux); and $m$ is a 6D mass. This is sufficiently general to include the spin-zero, -half and -one particles of interest in later sections.

\subsection*{One-loop mode sums}

Specializing to rugby-ball backgrounds and Wick rotating to Euclidean signature, we have
\eqa
    \Vone &=& \frac12 \, (-)^\ssF \, \mu^{4-d} \sum_{jn} \int \frac{\exd^d k_\ssE}{(2\pi)^d} \, \ln \left( \frac{k_\ssE^2 + m^2 + m_{jn}^2}{\mu^2}  \right) \nn\\
    &=& (-)^{\ssF+1} \frac{\mu^{4-d}}{2(4 \pi r^2)^{d/2}}  \int_0^\infty \frac{\exd t}{t^{1 + d/2}} \, e^{- t (m r)^2} \, S(t) \,,
\eea
where $m_{jn}^2 = \lambda_{jn}/r^2$ denote the eigenvalues of $-\Box_2 + X$ in the compactified space and $d = 4 - 2 \, \varepsilon$ with regularization parameter, $\varepsilon$, taken to zero after all divergences in this limit are renormalized. The function $S(t)$ is defined by
\bea
 S(t) &:=& (-)^\ssF \sum_{jn} \exp \left[ - t \lambda_{jn}  \right] \nn\\
 &=&  \frac{s_{-1}}{t} +
 \frac{s_{-1/2}}{\sqrt t} + s_0 + s_{1/2} \sqrt{t} + s_1 \, t + s_{3/2} t^{3/2} + s_2 \, t^2 + \cO(t^{5/2}) \,,
\eea
and its small-$t$ limit is of interest because this controls the UV divergences appearing in $\Vone$:
\be \label{eq:Vinfty}
 \Vone = \frac{\cC}{(4\pi r^2)^2} \left[ \frac1{4-d} + \ln\left(\frac\mu{m}\right)\right] + \hat\cV_f
 = \frac{\cC}{(4\pi r^2)^2} \left[ \frac1{4-d} \right] + \cV_f \,,
\ee
where $\hat\cV_f$ is finite and $\mu$-independent when $d \to 4$ and $\cV_f := \hat\cV_f + \cC \ln(\mu/m)/(4\pi r^2)^2$. The constant $\cC$ is given in terms of the $s_i$ by
\be \label{eq:Cform}
 \cC := \frac{s_{-1}}{6} (m r)^6 - \frac{s_0}{2}(m r)^4 + s_1 (mr)^2 - s_2 \,.
\ee
The coefficients $s_i$ are functions of the rugby ball's defect angle, $\delta = 2\pi(1 - \alpha)$, and the background flux quantum, $N$, and are calculated explicitly in \cite{Companion} for loops of 6D spin-zero, -half and -one bulk particles.

These ultraviolet divergences also track the dominant dependence on $m$ in the limit that $m \gg 1/r$, since both UV divergences and large masses involve the short-wavelength part of a loop that can be captured as the renormalization of some local effective interaction.

\subsection*{Renormalization}

What is perhaps unusual about the renormalizations required to absorb the UV divergences (and large-$m$ limit) of $\Vone$ is that they are not done using the couplings of effective interactions in the 4D theory. Because the wavelengths of interest are much shorter than the extra-dimensional size, divergences are instead absorbed into counter-terms in both the 6D bulk and 4D brane actions. Ref.~\cite{Companion} shows how to use the dependence of the $s_i$'s on $\alpha$, $N$ and $r$ to disentangle which bulk and brane interactions absorb the divergences found in eq.~\pref{eq:Vinfty}, which for completeness we now briefly summarize.

\medskip\noindent{\em Bulk counterterms}

\medskip\noindent
The relevant bulk counterterms are identified by writing the most general derivative expansion of both the bulk lagrangian that is nonzero when evaluated at the rugby ball background:
\bea \label{eq:bulkLct}
 \frac{\cL_\ssB}{\sqrt{-g}} &=&  - U - \frac{1}{2 \kappa^2} \, R - \frac{\cH}{4} \, \left(1 + \frac{\kappa \zeta_{\ssA\ssR}}{2} \, R \right) \, F_{\ssM \ssN} F^{\ssM \ssN} - \frac{\zeta_{\ssR^2}}\kappa \ol R^2 -\zeta_{\ssR^3} \ol R^3  + \cdots \,,
\eea
where $U  = (2\gR^2/\kappa^4) \, \cU \, e^\phi + \delta U$ is the bulk potential, $\cH = e^{-\phi}/\gB^2 + \delta \cH$ is the background gauge coupling, and $\ol R^2$ (or $\ol R^3$) are as defined in \cite{Companion}: a linear combination of the most general quadratic (cubic) gravitational terms, which together evaluate to $\ol R^2 = R^2_{\rm sph} = 4/r^4$ (or $\ol R^3 = R^3_{\rm sph} = -8/r^6$) on the rugby-ball background.

In principle, all of the coefficients in eq.~\pref{eq:bulkLct} can depend on $\phi$, but because $e^{2\phi}$ acts as the loop-counting parameter this dependence is dictated as a series in $e^{2\phi}$ whose order is dictated by the number of loops being computed. Keeping in mind that \pref{eq:bulkLct} is written in the Einstein frame, and the powers of $e^{\phi}$ already present in the classical Einstein-frame action, eq.~\pref{E:Baction}, this leads us to expect that at one loop
\be
 \delta U \propto e^{3\phi} \,, \quad
 \delta \kappa^{-2} \propto e^{2\phi} \,, \quad
 \delta \cH , \,\zeta_{\ssR^2} \propto e^\phi \,,
\ee
while the leading term in $\zeta_{\ssA\ssR}$ and $\zeta_{\ssR^3}$ is $e^\phi$-independent, and so on.

In practice, in the Einstein frame this $\phi$ dependence arises through the mass of the particle circulating in the loop, since a particle with a $\phi$-independent Jordan-frame mass $M$ has Einstein-frame mass $m = M e^{\phi/2}$. So the one-loop $\phi$-dependence required by loop counting in the Einstein frame agrees with the $m$ dependence required by dimensional analysis. For instance, in dimensional regularization one-loop corrections to $U$ are dimensionally of order $\delta U \propto m^6 = M^6 e^{3\phi}$, and this agrees with the power of $e^\phi$ required by loop counting. Similarly, $\delta \kappa^{-2} \propto m^4 = M^4 e^{2\phi}$ and $\delta \cH \propto m^2 = M^2 e^\phi$, and so on.

A crucial feature of bulk counterterms is that none of the parameters like $\delta U$, $\delta \cH$ {\em etc.} can depend on brane properties like $\alpha$ or $\Phi_b$ \cite{Companion}. This is most easily seen if they are computed using Gilkey-de Witt heat-kernel techniques  \cite{GilkeydeWitt, GdWrev} -- since this calculation is explicitly boundary-condition independent (for bulk counterterms). Physically, it is because these counterterms capture the effects of very short-wavelength modes, which don't extend far enough through the extra dimensions to `know' about conditions imposed at the boundaries. Ref.~\cite{Companion} provides a calculation of what heat kernel techniques give for generic bulk counterterms when specialized to a rugby ball geometry, and the specialization to 6D supergravity is summarized in Appendix \ref{app:gilkeydewitt}.

This means that the renormalized lagrangian evaluated on a rugby-ball background takes the form
\bea \label{VbulkAtEom}
 \cV_\ssB = - \int \exd^2 x \, \cL_\ssB &=& \left( 4 \pi \alpha \, r^2 \right) \left\{ U - \frac{1}{\kappa^2 r^2} + \frac{f^2 \cH}{2} \left[ 1 - \frac{\kappa \zeta_{\ssA\ssR} }{r^2} \right]  + \frac{4 \zeta_{\ssR^2}}{\kappa \,r^4}  - \frac{8\zeta_{\ssR^3}}{r^6} + \cdots \right\} \nn\\
 &=& \left( 4 \pi \alpha \, r^2 \right) \left\{ U - \frac{1}{\kappa^2 r^2} + \frac{\cN^2 \cH}{8 \, r^4} \left[ 1 - \frac{\kappa \zeta_{\ssA\ssR} }{r^2} \right]  + \frac{4 \zeta_{\ssR^2}}{\kappa \,r^4} - \frac{8\zeta_{\ssR^3}}{r^6} +\cdots \right\}\!\!,\qquad\,\,
\eea
where the only dependence on $\alpha$ arises from the overall volume integration. With this in mind it is useful to split up the quantities $s_i$ in the following way:
\be \label{E:ssplit}
 s_i(\alpha,\cN,\Phi_b) = \alpha \, s_i^{\rm sph}(\cN) + \delta s_i(\alpha, \cN,\Phi_b) \,,
\ee
where $s_i^{\rm sph}$ is the $\alpha$-independent contribution renormalized by bulk counterterms, and the pre-factor of $\alpha$ corresponds to the rugby-ball volume, $4\pi \alpha \, r^2$, appearing in eq.~\pref{VbulkAtEom}. Because $s_i^{\rm sph}$ doesn't depend on $\alpha$, it can be evaluated using a Casimir energy calculation in the absence of branes --- that is, on the sphere (or, equivalently, by evaluating the rugby-ball result at $\alpha = 1$).

Given $s_i^{\rm sph}$, the contributions to $U$, $\kappa^{-2}$, $\cH$, $\zeta_{\ssA\ssR}$ and $\zeta_{\ssR^2}$ can be read off by identifying the coefficients of $r^2$, $r^0$, $\cN^2/r^2$, $\cN^2/r^4$ and $1/r^2$, respectively. Because the $\mu$ dependence of the renormalized quantities must cancel the explicit $\mu$-dependence of $\Vone$, they therefore satisfy \cite{Companion}
\bea
 \mu \, \frac{\partial U}{\partial \mu} = -\frac{m^6}{6 (4\pi)^3} \; s_{-1}^{\rm sph,\, 0} \,,&&\quad
 \mu \, \frac{\partial}{\partial \mu} \left( \frac{1}{\kappa^2} \right) = -\frac{ m^4}{2 (4\pi)^3} \; s_{0}^{\rm sph,\,0} \,,\\
 \mu\,\frac{\partial }{\partial \mu}\left(\frac{\zeta_{\ssR^2}}\kappa\right) = -\frac{m^2}{4(4\pi)^3} \; s_1^{\rm sph,\,0} \,,&&\quad
\quad \mu \, \frac{\partial \cH}{\partial \mu} = -\frac{8\,m^2}{(4\pi)^3 \cN^2} \; s_{1}^{\rm sph,\,2} \,,\\
  \mu\,\frac{\partial \zeta_{\ssR^3}}{\partial\mu} = -\frac{1}{8(4\pi)^3} \;s_2^{\rm sph,\,0} \,,&&\quad
 \quad \mu \, \frac{\partial}{\partial \mu} \left(\kappa \zeta_{\ssA\ssR} \cH \right) = -\frac{8}{(4\pi)^3 \cN^2} \; s_{2}^{\rm sph,\,2} \,.
\eea
Here the quantities $s_i^{{\rm sph,}k}$ denote those terms in $s_i^{\rm sph}$ that are proportional to $k$ powers of $\cN$.

\medskip\noindent{\em Brane counterterms}

\medskip\noindent
The divergences contained in $\delta s_i$ from eq.~\pref{E:ssplit} are absorbed in a similar way by counterterms in the brane action, whose generic derivative expansion is:
\bea \label{E:Lbren}
 \frac{\cL_b}{\sqrt{-\gamma}} &=& - T_b + \frac{\cA_b}{2\gB^2} \, \epsilon^{mn} F_{mn} - \frac{\zeta_{\ssR b}}\kappa \, R - \frac{\kappa\zeta_{\ssA b}}{4\gB^2} F_{\ssM \ssN} F^{\ssM \ssN} + \frac{\kappa \zeta_{\tilde{\ssA}\ssR\, b}}{2\gB^2} \, R \epsilon^{mn} F_{mn} -\zeta_{\ssR^2 b} \, \ol R^2 \nn\\
 &&\qquad - \frac{\kappa^2 \, \zeta_{\ssA \ssR \,b}}{8 \gB^2} \, R \, F_{\ssM \ssN} F^{\ssM \ssN} + \cdots \,.
\eea
Evaluating these at the background rugby-ball solution gives a contribution
\bea
 \cV_b &=& T_b - \frac{\cA_b f}{\gB^2} - \frac{2 \zeta_{\ssR \, b}}{\kappa \,r^2} + \frac{\kappa \zeta_{\ssA b} f^2}{2 \gB^2} + \frac{2 \, \kappa \zeta_{\tilde{\ssA}\ssR \, b} f}{\gB^2 r^2} + \frac{4 \zeta_{\ssR^2 b}}{r^4} - \frac{\kappa^2\zeta_{\ssA \ssR \, b} f^2}{2 \, \gB^2 r^2} + \cdots \nn\\
 &=& T_b - \frac{\cA_b \cN}{2 \, \gB^2 \,r^2} - \frac{2 \zeta_{\ssR \, b}}{\kappa \,r^2} + \frac{\kappa \zeta_{\ssA b} \cN^2}{8 \, \gB^2 r^4} + \frac{ \kappa \zeta_{\tilde{\ssA}\ssR \, b} \cN}{\gB^2 r^4} + \frac{4 \zeta_{\ssR^2 b}}{r^4} - \frac{\kappa^2 \zeta_{\ssA \ssR \, b} \cN^2}{8 \,  \gB^2 r^6} + \cdots \,,
\eea
to the 1PI 4D effective potential, with the complete result summed over all branes: $\cV_{\rm branes} = \sum_b \cV_b$. Again, each contribution can be disentangled by separating terms with different powers of $r$ and $\cN$ in $\delta s_i$.

The running of the brane couplings that results is
\bea \label{E:genbraneren}
 \mu \, \frac{\partial \, T_b}{\partial \mu} = \frac{m^4}{2(4\pi)^2} \; \delta s^0_{0}\,,\quad&& \mu\,\frac{\partial}{\partial \mu} \left( \frac{\cA_b}{\gB^2}\right) = \frac{2\, m^2}{(4\pi)^2 \cN} \; \delta s^{1}_1 \,, \nn\\
 \mu\,\frac{\partial}{\partial \mu} \left( \frac{\zeta_{\ssR b}}{\kappa}\right) = \frac{m^2}{2(4\pi)^2} \; \delta s^0_1 \,,\quad&& \mu\,\frac{\partial}{\partial \mu} \left( \frac{\kappa \zeta_{\tilde{\ssA} \ssR \, b}}{\gB^2}\right) = \frac{1}{(4\pi)^2 \cN} \; \delta s^{1}_{2} \,, \\
\mu\,\frac{\partial \zeta_{\ssR^2 b}}{\partial \mu} = \frac1{4(4\pi)^2}\; \delta s_2^0 \,,\quad&& \mu\,\frac{\partial}{\partial \mu} \left( \frac{\kappa \zeta_{\ssA b}}{\gB^2}\right) = \frac{8}{(4\pi)^2 \cN^2} \; \delta s^{2}_{2} \,,\nn
\eea
where, as before, $\delta s_i^{ k}$ denotes that part of $\delta s_i$ proportional to $\cN^k$.

It remains to compute the $s_i$ explicitly for various light bulk fields by performing the Kaluza-Klein mode sum. This was done in \cite{Companion} for spin-zero, -half and -one fields, with results which are summarized for completeness in Appendix \ref{App:CompanionResults}. In the next section we assemble the $s_i$'s using the field content of various 6D supersymmetric multiplets, to determine how these multiplets renormalize both bulk and brane counterterms.

\section{Supermultiplets}
\label{sec:SUSYmult}

We now use the results for the $s_i$'s for low-spin bulk fields, listed in Appendix \ref{App:CompanionResults}, and combine them into the field content of various 6D matter supermultiplets. We consider in particular two massless multiplets - the hypermultiplet and gauge multiplet, for which only $s_2$ is relevant to renormalizations. We then combine these results to examine the renormalization due to a massive 6D multiplet.

\subsection*{Hypermultiplet scalars}

Before combining into supermultiplets, we must first specialize the result for generic scalar fields given in Appendix \ref{App:CompanionResults} so that they can apply to the hyperscalars that appear in supersymmetric hypermultiplets.

The part of the action, eq.~\pref{E:Baction}, relevant for small hyperscalar fluctuations is
\be
    S_{\rm hyp} = - \int \exd^6x \, \sqrt{-g} \; \left[ \frac12 \, \cG_{\ssI\ssJ}(\Phi) \,
    D_\ssM \Phi^\ssI D^\ssM \Phi^\ssJ + \frac{2 \gR^2}{\kappa^4} \; e^\phi \, \cU(\Phi) \right] \,,
\ee
where
\be
  \cU= 1 + \frac{\kappa^2}{2} \, \cG_{\ssI\ssJ} \Phi^\ssI \Phi^\ssJ + \dots \,,
\ee
near $\Phi^\ssI = 0$, and as before, $\gR$ is the $U(1)_\ssR$ gauge coupling constant. Notice, in particular, that this expansion of $\cU(\Phi)$ near zero introduces a small universal 6D mass term for $\Phi^\ssI$ given by
\be \label{eq:hypermass}
 \delta m^2 = \frac{2 \gR^2}{\kappa^2} \; e^\phi = \frac{1}{2 r^2} \,,
\ee
where the last equality uses eq.~\pref{E:phircond} relating the background value of $e^\phi$ to $1/r^2$. Regarding this mass, $\delta m^2$, as a shift in the hyperscalar KK spectrum, $m_{jn}^2 = \lambda_{jn}^{\rm hs} / r^2$, and using the expression for the scalar spectrum, $\lambda_{jn}^{\rm s}$, given for minimally coupled scalars in \cite{Companion}, we find that\footnote{The quantities $N$ and $\Phi_b$ used in this paper differ from those in \cite{Companion} by a factor of $q$. (Therein, they are called $N_1$ and $\Phi_{1b}$.) Unless stated otherwise, we do not to track this $q$-dependence, since the $R$-charges of interest are simply $q\in\{\pm1,0\}$.}
\bea \label{hsspecrel}
 \lambda_{jn}^{\rm hs}(\omega,N,\Phi_b) &=& \lambda_{jn}^{\rm s}(\omega,N,\Phi_b) +\frac12 \\
 &=& \left(j +\frac{\omega}{2} \left| n - \Phi_- \right| + \frac{\omega}{2} \left| n -  N +\Phi_+ \right| + \frac12\right)^2 + \frac{(1-\cN^2)}{4} \,.\nn
\eea
This assumes a background flux with quantum $N$ (where $N = \pm 1$ for the supersymmetric case of $U(1)_\ssR$ flux), as well as the previously-mentioned definitions: $\Phi_b := \cA_b \,e^\phi /(2\pi)$ with $b=\pm$ denoting the north and south branes. Finally, $\cN := \omega(N-\Phi)$ with $\Phi := \sum_b \Phi_b$. The quantity $\cA_b$ enters the spectrum through the boundary condition $\cA_b := A_\varphi(\cos\theta = b)$.

When computing the small-$t$ limit of the mode sum
\be
S_{\rm hs}(\omega,N,\Phi_b,t) = \sum_{j=0}^\infty \sum_{n=-\infty}^\infty e^{-t\lambda^{\rm hs}_{jn}} \,,
\ee
we can use relation \pref{hsspecrel} to relate hyperscalar and scalar sums by $S_{\rm hs} = e^{-t/2} S_{\rm s}$, so using the expressions for $s^{\rm s}_i$ from Appendix \ref{App:CompanionResults} gives the following small-$t$ coefficients for the hyperscalar:
\begin{eqnarray}
 s^\mathrm{hs}_{-1}(\omega, N,\Phi_b) &=& s^\mathrm{s}_{-1} = \frac{1}{\omega} \,,\\
 s^\mathrm{hs}_0(\omega,N,\Phi_b) &=& s^\mathrm{s}_0 - \frac{s^\mathrm{s}_{-1}}2 = \frac{1}{\omega} \left[ -\frac13 + \frac{\omega^2}6 (1-3F) \right] \,,\\
 s^\mathrm{hs}_1(\omega,N,\Phi_b) &=& s^\mathrm{s}_1 -\frac{s^\mathrm{s}_0}2 + \frac{s^\mathrm{s}_{-1}}8  = \frac{1}{\omega} \Bigg[ \frac{17}{360} - \frac{\cN^2}{24} -\frac{\omega^2}{36} (1-3F) \nn\\
 &&\qquad\qquad\qquad\qquad\quad - \frac{\omega^3\cN}{12} \sum_b \Phi_b \, G(|\Phi_b|) + \frac{\omega^4}{180}\left(1-15 F^{(2)} \right) \Bigg] \,,\\
 s^\mathrm{hs}_2(\omega,N,\Phi_b) &=& s^\mathrm{s}_2 - \frac{s^\mathrm{s}_1}2 + \frac{s^\mathrm{s}_0}8 - \frac{s^\mathrm{s}_{-1}}{48} \nn\\
  &=& \frac{1}{\omega} \Bigg[ -\frac1{210} + \frac{\cN^2}{180} + \left(\frac1{240} - \frac{\cN^2}{144} \right) (1-3F)\omega^2 - \frac{\omega^4\cN^2}{360}\left( 1-15 F^{(2)}\right)  \nn\\
 &&\qquad  - \frac{\omega^5\cN}{120} \sum_b \Phi_b \, G(|\Phi_b|) (1+3 F_b) +\bigg(\frac{1}{1260} - \frac{F^{(2)}}{120} - \frac{F^{(3)}}{60} \bigg)\omega^6 \Bigg] \,.
\end{eqnarray}
In the above, we adopt the following shorthand:
\be \label{Fbdefs}
F_b:=|\Phi_b|\left(1-|\Phi_b|\right)\,,\quad F^{(n)}:= \sum_b F_b^n \,,\quad F^{(1)}:=F \,, \quad G(x):=(1-x)(1-2x) \,.
\ee
In the limit $\omega \to 1$, $\Phi_b \to 0$ these become $s_{-1}^{\rm sph,\,0} = 1$, $s_0^{\rm sph,\,0} = - 1/6$,
\be
 s_1^{\rm sph,\,0} = \frac{1}{40} \,, \quad
 s_1^{\rm sph,\,2} = - \frac{\cN^2}{24} \,, \quad
 s_2^{\rm sph,\,0} = \frac{1}{5040}  \quad
 \hbox{and} \quad
 s_2^{\rm sph,\,2} =  - \frac{\cN^2}{240}  \,.
\ee
These agree with the corresponding Gilkey-de Witt calculation of Appendix \ref{app:gilkeydewitt}. The corresponding contributions to the running of the bulk couplings are
\bea
 \mu \, \frac{\partial U}{\partial \mu} = -\frac{m^6}{6 (4 \pi)^3} \,,\quad && \mu\,\frac{\partial}{\partial \mu} \left( \frac{1}{\kappa^2} \right) = \frac{m^4}{12 (4\pi)^3} \,,\\
 \mu\,\frac{\partial}{\partial\mu} \left(\frac{\zeta_{\ssR^2}}{\kappa}\right) = -\frac{m^2}{160(4\pi)^3} \,,\quad && \mu\,\frac{\partial \zeta_{\ssR^3}}{\partial\mu} =- \frac1{40320(4\pi)^3} \,,\\
  \mu\,\frac{\partial \cH}{\partial\mu}  =
  \mu\,\frac{\partial }{\partial\mu}\left(\frac{e^{-\phi}}{\gB^2}\right) = \frac{q^2_{\rm hs} m^2}{3 (4\pi)^3}  \,,\quad &&
  \mu \, \frac{\partial}{\partial \mu} \Bigl( \kappa \cH \zeta_{\ssA\ssR} \Bigr) =  \frac{q^2_{\rm hs}}{30 (4\pi)^3} \,,
\eea
where we quote the result for the general case where the hyperscalar couples to the background field with strength $q_{\rm hs} \gB$.

For the brane counterterms, we similarly find $\delta s_{-1} = \delta s_1^{\rm 2} = 0$,
\bea
\delta s_0 &=&  \frac1{\omega}\left( \frac{\delta\omega}6 + \frac{\delta\omega^2}{12} -\frac{\omega^2 F_b}2 \right) \simeq \frac{\delta\omega}6 - \frac{|\Phi_b|}2 \,,\\
\delta s_1^0 &=& \frac1\omega\left( -\frac{\delta\omega}{60} + \frac{\delta\omega^2}{360} + \frac{\delta\omega^3}{90} + \frac{\delta\omega^4}{360} +\frac{\omega^2 F_b}{12} -\frac{\omega^4 F_b^2}{12} \right) \simeq - \frac{\delta\omega}{60} + \frac{|\Phi_b|}{12} \,,\\
\delta s_1^{\rm 1} &=& - \frac{\omega^2\cN}{12} \Phi_b \, G(|\Phi_b|) \simeq - \frac{\cN \Phi_b}{12} \,,\\
\delta s_2^0 &=& \frac1\omega\left( \frac{11\,\delta\omega}{1680} + \frac{9 \,\delta\omega^2}{1120} + \frac{\delta\omega^3}{126} + \frac{\delta\omega^4}{168} + \frac{\delta\omega^5}{420} + \frac{\delta\omega^6}{2520} - \frac{\omega^2 F_b}{80} - \frac{\omega^6 F^2_b}{120} - \frac{\omega^6 F^3_b}{60} \right)  \nn\\
&\simeq& \frac{11\,\delta\omega}{1680} - \frac{|\Phi_b|}{80} \,,\\
\delta s_2^{\rm 1} &=& -\frac{\omega^4 \cN}{120} \Phi_b \, G(|\Phi_b|) (1+3 F_b) \simeq -\frac{\cN\Phi_b}{120}  \,,\\
\delta s_2^{\rm 2} &=& -\frac{\cN^2}\omega \left( \frac{\delta\omega}{80} + \frac{17\,\delta\omega^2}{1440} +\frac{\delta\omega^3}{180} + \frac{ \delta\omega^4}{720} - \frac{\omega^2 F_b}{48} - \frac{\omega^4 F^2_b}{24} \right) \simeq -\cN^2 \left( \frac{\delta\omega}{80} - \frac{|\Phi_b|}{48}\right) \,,\qquad
\eea
and so (again generalizing to coupling $q_{\rm hs} \,\gB$)
\bea
 \mu \, \frac{\partial T_b}{\partial \mu} &=& \frac{m^4}{2 (4\pi)^2 \omega} \left( \frac{\delta\omega}6 + \frac{\delta\omega^2}{12} - \frac{\omega^2 F_b}2 \right) \simeq \frac{m^4}{4 (4\pi)^2} \left(\frac{\delta\omega}3-|\Phi_b|\right) \,,\\
   \mu\,\frac{\pd}{\pd\mu}\left(\frac{\zeta_{\tilde \ssA b}}{\gB^2}\right) &=&  -\frac{q_{\rm hs} \Phi_b\,\omega^2 m^2}{6(4\pi)^2}  G(|\Phi_b|) \simeq -\frac{q_{\rm hs}^2 m^2\,\cA_b}{3(4\pi)^3} \,,\\
  \mu \, \frac{\partial }{\partial \mu} \left(\frac{\zeta_{\ssR b}}{\kappa}\right) &=& -\frac{m^2}{2(4\pi)^2\omega} \left( \frac{\delta\omega}{60} - \frac{\delta\omega^2}{360} - \frac{\delta\omega^3}{90} - \frac{\delta\omega^4}{360} -\frac{\omega^2 F_b}{12} +\frac{\omega^4 F_b^2}{12} \right)  \nn\\
  &\simeq& -\frac{m^2}{2(4\pi)^2}\left( \frac{\delta\omega}{60} - \frac{|\Phi_b|}{12} \right) \,,\\
    \mu\,\frac{\pd}{\pd\mu}\left(\frac{\kappa \zeta_{\tilde\ssA\ssR b}}{\gB^2}\right) &=& -\frac{q_{\rm hs}\Phi_b \,\omega^4}{120(4\pi)^2} \, G(|\Phi_b|) (1+3 F_b)  \simeq -\frac{q_{\rm hs}^2 \,\cA_b }{60(4\pi)^3} \,,\\
  \mu\,\frac{\partial \,\zeta_{\ssR^2 b}}{\partial \mu} &=& -\frac{1}{4(4\pi)^2\omega} \Bigg(  \frac{11\,\delta\omega}{1680} + \frac{9 \,\delta\omega^2}{1120} + \frac{\delta\omega^3}{126} + \frac{\delta\omega^4}{168} + \frac{\delta\omega^5}{420} + \frac{\delta\omega^6}{2520} - \frac{\omega^2 F_b}{80}   \nn\\
  &&\qquad\qquad\qquad - \frac{\omega^6 F^2_b}{120} - \frac{\omega^6 F^3_b}{60} \Bigg) \simeq -\frac1{4(4\pi)^2} \left( \frac{11\,\delta\omega}{1680} - \frac{|\Phi_b|}{80} \right) \,, \\
  \mu \, \frac{\partial }{\partial \mu} \left(\frac{\kappa\zeta_{\ssA b}}{\gB^2}\right) &=& -\frac{8\,q_{\rm hs}^2}{(4\pi)^2\omega}\left( \frac{\delta\omega}{80} + \frac{17\,\delta\omega^2}{1440} +\frac{\delta\omega^3}{180} + \frac{ \delta\omega^4}{720} - \frac{\omega^2 F_b}{48} - \frac{\omega^4 F^2_b}{24} \right)  \nn\\
  &\simeq& -\frac{q_{\rm hs}^2}{(4\pi)^2} \left( \frac{\delta\omega}{10} - \frac{|\Phi_b|}{6}\right)\,.
\eea
In these expressions $\delta\omega:=\omega-1$, and the approximate equalities give the leading terms when $|\delta \omega| \ll 1$ and $|\Phi_b| \ll 1$.

\subsection{Hypermultiplet}
\label{sec:Hypermultiplet}

We are now in a position to sum the particle content of a 6D hypermultiplet, which consists of four massless hyperscalars together with a 6D Weyl fermion:
\be
s_i^{\mathrm{hyp}} := 4 \, s_i^{\mathrm{hs}} + s_i^{\rm f} \,,
\ee
with $s_i^{\rm hs}$ computed above and $s_i^{\rm f}$ given in Appendix \ref{App:CompanionResults}. Although we need only really be interested in $s_2^{\rm hyp}$ for massless fields, we nonetheless keep track of $s_{-1}^{\rm hyp}$, $s_{0}^{\rm hyp}$ and $s_1^{\rm hyp}$ as well, since these are needed when assembling a massive multiplet.

The above combination yields $s_{-1} = s_0^{\rm sph,\,0}=0$ (where we drop the `hyp' superscript), along with
\be
 s_1^{\rm sph,\,0} = \frac16  \,,\quad s_1^{\rm sph,\,2} = - (q^2_{\rm hs} + 2 \, q^2_{\rm f}) \frac{\cN^2}6  \,,\quad s_2^{\rm sph,\,0} = \frac1{60}  \,,\quad s_2^{\rm sph,\,2} = - (q_{\rm hs}^2 + 4 q_{\rm f}^2) \frac{\cN^2}{60}  \,,
\ee
where $\cN :=2 r^2 f = \pm \gB/\gR$ characterizes the bulk flux (with the second equality using the field equation, eq.~\pref{E:phircond}) while $q_{\rm hs}$ and $q_{\rm f}$ are the charges (in units of $\gB$) of the scalar and fermion, respectively, under the $U(1)$ gauged by the background flux. The corresponding bulk renormalizations are
\bea
 \mu\,\frac{\partial \zeta_{\ssR^3}}{\partial\mu} &=& -\frac1{480(4\pi)^3} \\
 \mu\,\frac{\partial}{\partial\mu} \Bigl( \kappa \cH \zeta_{\ssA\ssR} \Bigr) &=& \mu\,\frac{\partial}{\partial\mu} \left( \frac{\kappa e^{-\phi} \zeta_{\ssA\ssR}}{\gB^2}\right) =- \frac8{(4\pi)^3} \left(  \frac{s_2^{\rm sph,\,2}}{\cN^2} \right)=  \frac{2(q_{\rm hs}^2+4 q_{\rm f}^2)}{15(4\pi)^3} \,.
\eea
From here, we can use eq.~\pref{VbulkAtEom} compute the hypermultiplet contribution to the running of $\cV_\ssB$:
\bea
 \left(\mu\,\frac{\partial \cV_\ssB}{\partial\mu}\right) &=& \left( 4 \pi \alpha \, r^2 \right) \left[ - \frac{f^2}{2r^2}\left(\mu\,\frac{\partial}{\partial\mu}\left(\frac{\kappa \zeta_{\ssA\ssR}}{\gB^2}\right)\right)- \frac{8}{r^6} \left(\mu\,\frac{\partial \zeta_{\ssR^3}}{\partial\mu}\right)\right] \nn\\
 &=& \left( 1-q_{\rm hs}^2\cN^2 - 4 \, q_{\rm f}^2\cN^2 \right) \frac{\alpha}{60 (4\pi r^2)^2} \,.
\eea

Three choices for $q_{\rm hs}$ and $q_{\rm f}$ are of particular interest:
\begin{itemize}
\item No couplings to background fluxes: $q_{\rm hs} = q_{\rm f} = 0$;
\item Couplings to a background $U(1)$ that commutes with supersymmetry, in which case $q_{\rm hs} = q_{\rm f} = q$ and $\cN = \pm \gB/\gR$;
\item Couplings to a background flux that preserves supersymmetry, for which the background flux gauges the $U(1)_\ssR$ symmetry, and so $q_{\rm hs} = \pm 1$, $q_{\rm f} = 0$ and $\cN = \pm 1$.
\end{itemize}

In the supersymmetric case a cancellation occurs, generalizing to the rugby ball a result known to apply more generally to Ricci-flat geometries \cite{UVsensitivity}. In this case $s_1^{\rm sph,\,0} = - s_1^{\rm sph,\,2} = \frac16$ and $s_2^{\rm sph,\,0} = -  s_2^{\rm sph,\,2} = \frac1{60}$ and so
\be
s_{-1}^{\rm sph} = s_{0}^{\rm sph} = s_{1}^{\rm sph} = s_{2}^{\rm sph} = 0
\ee
and $\mu (\partial \cV_\ssB/\partial \mu) = 0$ vanishes once summed over the field content of a hypermultiplet. Notice that although the effective couplings for both $R F_{\ssM \ssN} F^{\ssM \ssN}$ and $\ol R^3$ do renormalize, their contributions cancel in $\cV_\ssB$.

Specializing to the supersymmetric charge assignments the brane-renormalized divergences are $\delta s_i = 4 \,\delta s_i^{\rm hs} + \delta s_i^{\rm f}$, and so
\bea
 \delta s_0 &=&  \frac1{\omega}\left( \delta\omega + \frac{\delta\omega^2}{2} -2\omega^2 F_b \right) \simeq \delta\omega - 2|\Phi_b| \,,\\
 \delta s_1^0 &=& \frac1\omega\left(  \frac{\delta\omega^2}{12} + \frac{\delta\omega^3}{12} + \frac{\delta\omega^4}{48} +\frac{\omega^2 F_b}{3} -\frac{\omega^4 F_b^2}{3} \right) \simeq   \frac{|\Phi_b|}{3} \,,\\
 \delta s_1^{\rm 1} &=& - \frac{\omega^2\cN}{3} \Phi_b \, G(|\Phi_b|) \simeq - \frac{\cN \Phi_b}{3} \,,\\
 \delta s_2^0 &=& \frac1\omega\left( \frac{\delta\omega}{20} + \frac{13 \,\delta\omega^2}{180} + \frac{13\,\delta\omega^3}{180} + \frac{71\,\delta\omega^4}{1440} + \frac{3\,\delta\omega^5}{160} + \frac{\delta\omega^6}{320} - \frac{\omega^2 F_b}{20} - \frac{\omega^6 F^2_b}{30} - \frac{\omega^6 F^3_b}{15} \right)  \nn\\
&\simeq& \frac{\delta\omega}{20} - \frac{|\Phi_b|}{20} \,,\\
 \delta s_2^{\rm 1} &=& -\frac{\omega^4 \cN}{30} \Phi_b \, G(|\Phi_b|) (1+3 F_b) \simeq -\frac{\cN\Phi_b}{30}  \,,\\
 \delta s_2^{\rm 2} &=& -\frac{\cN^2}\omega \left( \frac{\delta\omega}{20} + \frac{17\,\delta\omega^2}{360} +\frac{\delta\omega^3}{45} + \frac{ \delta\omega^4}{180} - \frac{\omega^2 F_b}{12} - \frac{\omega^4 F^2_b}{6} \right) \simeq -\cN^2 \left( \frac{\delta\omega}{20} - \frac{|\Phi_b|}{12}\right) \,,\qquad
\eea
where $\cN = \pm \gB/\gR = \pm 1$.

Notice that all of the brane contributions also vanish, $\delta s_i =0$ (for all $\omega$), in the special case that the brane fluxes are equal and the total brane flux satisfies the supersymmetric and flux-quantization conditions, eqs.~\pref{E:SSqntznresult} and \pref{E:SUSYcond}:
\be
 \Phi_+ = \Phi_- = \frac{\Phi_{\rm susy}}{2} = \pm \frac12 \,(1-\omega^{-1}) \,.
\ee
as well as $\cN = \hbox{sgn}(\Phi_\pm)$. Similarly $\mu (\partial \cV_{\rm branes}/\partial \mu) = 0$. Once again, although $\delta s_i = 0$ this is {\em not} true for $\delta s_i^{\rm 1}$ and $\delta s_i^{\rm 2}$ separately.

In general this cancellation fails when $\Phi_\pm$ are not equal, since this choice breaks supersymmetry. In order to track how $\cV_{\rm branes}$ deviates from zero once supersymmetry breaks, we write
\be
 \Phi_b = \frac12\Phi_{\rm susy} \pm \frac{b\eta}{\omega}
\ee
and allow $\eta$ to parameterize the difference through $\eta = \pm \frac12 \, \omega \Delta\Phi$ where $\Delta \Phi := \Phi_+ - \Phi_-$. To ensure that $\eta \geq 0$, we label branes such that $\Phi_+ \geq \Phi_-$ ($\Phi_+ \leq \Phi_-$) when $\cN=+1$ ($\cN=-1$). Also, notice that the condition $|\Phi_b| \leq 1$ implies that $\eta \leq 1 +\delta\omega/2$.

At the north ($b=+1$) brane, we find that the brane contributions to the coefficients $\delta s_i$ are given in terms of $\eta$ by
\bea
 \delta s_0 &=&  -\frac{2F_\eta}\omega  \,,\\
 \delta s_1^0 &=& \frac1\omega\bigg[ \frac{\delta\omega}6 +  \frac{\delta\omega^2}{12} + \frac{F_\eta}2 - \frac{F_\eta^2}3 -\frac{\omega^2 F_\eta}6 \bigg] \,, \\
 \delta s_1^{\rm 1} &=& \frac1\omega\bigg(-\frac{\delta\omega}6 -  \frac{\delta\omega^2}{12} -\frac{F_{\eta}}3  \bigg)(1-2\eta)  \,,\\
 \delta s_2^0 &=& \frac1\omega\bigg[ \frac{\delta\omega}{40} + \frac{37\,\delta\omega^2}{720} + \frac{7\,\delta\omega^3}{180} + \frac{7\,\delta\omega^4}{720} -\frac{F_\eta}{16} + \frac{F_\eta^2}{20} - \frac{F_\eta^3}{15} \nn\\
&& \qquad\qquad\qquad\qquad\qquad\qquad\qquad
 + \bigg( \frac{F_\eta}{24} - \frac{F_\eta^2}{12} \bigg) \omega^2 - \frac{7\,\omega^4 F_\eta}{240} \bigg] \,, \\
 \delta s_2^{\rm 1} &=& \frac1\omega \bigg( -\frac{\delta\omega}{60}- \frac{\delta\omega^2}{15} - \frac{7\,\delta\omega^3}{120} - \frac{7\,\delta\omega^4}{480} +\frac{F_{\eta}}{20} -\frac{F_{\eta}^2}{10} - \frac{\omega^2F_{\eta}}{12} \bigg)(1-2\eta) \,, \\
 \delta s_2^{\rm 2} &=& \frac1\omega \bigg[ -\frac{\delta\omega}{120} + \frac{11\,\delta\omega^2}{720}+ \frac{7\,\delta\omega^3}{360} + \frac{7\,\delta\omega^4}{1440} + \frac{F_\eta^2}6 + \frac{\omega^2 F_\eta}{12} \bigg]  \,,
\eea
where
\be
 F_\eta := \eta (1-\eta) \,.
\ee
The result for the south brane can be obtained from this using $\eta \to -\eta$ if $\eta \leq \delta\omega/2$. If instead $\delta\omega/2 \leq \eta \leq 1 + \delta\omega/2$ on the south brane, we find that a more convenient quantity is
\be
 \hat \eta := \eta-\delta\omega
\ee
(along with $F_{\hat \eta} := \hat \eta (1-\hat\eta)$), in which case
\bea
 \delta s_0 &=&  -\frac{2F_{\hat\eta}}\omega  \,,\\
 \delta s_1^0 &=& \frac1\omega\bigg[ \frac{\delta\omega}6 +  \frac{\delta\omega^2}{12} + \frac{F_{\hat \eta}}2 - \frac{F_{\hat \eta}^2}3 -\frac{\omega^2 F_{\hat \eta}}6 \bigg] \,, \\
 \delta s_1^{\rm 1} &=& \frac1\omega\bigg(\frac{\delta\omega}6 +  \frac{\delta\omega^2}{12} +\frac{F_{\hat\eta}}3  \bigg)(1-2\hat\eta)  \,,\\
 \delta s_2^0 &=& \frac1\omega\bigg[ \frac{\delta\omega}{40} + \frac{37\,\delta\omega^2}{720} + \frac{7\,\delta\omega^3}{180} + \frac{7\,\delta\omega^4}{720} -\frac{F_{\hat \eta}}{16} + \frac{F_{\hat \eta}^2}{20} - \frac{F_{\hat \eta}^3}{15} \nn\\
 && \qquad\qquad\qquad\qquad\qquad\qquad\qquad
 + \bigg( \frac{F_{\hat \eta}}{24} - \frac{F_{\hat \eta}^2}{12} \bigg) \omega^2 - \frac{7\,\omega^4 F_{\hat \eta}}{240} \bigg] \,, \\
 \delta s_2^{\rm 1} &=& \frac1\omega \bigg( \frac{\delta\omega}{60} + \frac{\delta\omega^2}{15} + \frac{7\,\delta\omega^3}{120} + \frac{7\,\delta\omega^4}{480} -\frac{F_{\hat\eta}}{20} +\frac{F_{\hat\eta}^2}{10} + \frac{\omega^2F_{\hat\eta}}{12} \bigg)(1-2\hat\eta) \,, \\
 \delta s_2^{\rm 2} &=& \frac1\omega \bigg[ -\frac{\delta\omega}{120} + \frac{11\,\delta\omega^2}{720}+ \frac{7\,\delta\omega^3}{360} + \frac{7\,\delta\omega^4}{1440} + \frac{F_{\hat \eta}^2}6 + \frac{\omega^2 F_{\hat \eta}}{12} \bigg]  \,.
\eea
These expressions are identical to those obtained for the north brane, given the replacements $\eta \to \hat \eta$, $s_i^{\rm 1} \to -s_i^{\rm 1}$. This follows from
\be
 \Phi_- = \pm \left(\frac{\delta\omega}{2\omega} - \frac\eta\omega\right) = \pm \left(-\frac{\delta\omega}{2\omega} - \frac{(\eta-\delta\omega)}\omega\right) = \mp \left( \frac{\delta\omega}{2\omega} + \frac{\hat\eta}\omega\right)
\ee
and because $s_i^{\rm 1} \propto \Phi_b$ rather than depending only on $|\Phi_b|$, like the others.

%
%

The running of the couplings on the north --- the $b=+1$ --- brane (which we choose to be the brane whose flux is larger in magnitude) are
\bea
 \mu \frac{\pd \zeta_{\ssR^2 +}}{\pd \mu} &=& \frac1{4(4\pi)^2\omega} \bigg[ \frac{\delta\omega}{40} + \frac{37\,\delta\omega^2}{720} + \frac{7\,\delta\omega^3}{180} + \frac{7\,\delta\omega^4}{720} -\frac{F_\eta}{16} + \frac{F_\eta^2}{20} - \frac{F_\eta^3}{15} \nn\\
 &&\qquad\qquad\qquad+ \bigg( \frac{F_\eta}{24} - \frac{F_\eta^2}{12}
 \bigg) \omega^2 - \frac{7\,\omega^4 F_\eta}{240} \bigg] \,,\\
 \mu \frac\pd{\pd\mu}\left( \frac{\kappa \zeta_{\tilde \ssA\ssR +}}{\gB^2}\right)  &=& \frac{1}{(4\pi)^2\omega} \bigg( -\frac{\delta\omega}{60}- \frac{\delta\omega^2}{15} - \frac{7\,\delta\omega^3}{120} - \frac{7\,\delta\omega^4}{480} +\frac{F_{\eta}}{20} -\frac{F_{\eta}^2}{10} - \frac{\omega^2F_{\eta}}{12} \bigg)(1-2\eta) \,,\nn \\
&&\\
 \mu \frac{\pd}{\pd\mu} \left( \frac{\kappa \zeta_{\ssA +}}{\gB^2} \right) &=&  \frac{8}{(4\pi)^2\omega} \bigg[ -\frac{\delta\omega}{120} + \frac{11\,\delta\omega^2}{720}+ \frac{7\,\delta\omega^3}{360} + \frac{7\,\delta\omega^4}{1440} + \frac{F_\eta^2}6 + \frac{\omega^2 F_\eta}{12} \bigg] \,.
\eea
As before, the corresponding expressions for the south brane when $\eta \leq \delta\omega/2$ ($\eta \geq \delta\omega/2$) can be found by taking $\eta \to -\eta$ ($\eta \to \hat \eta=(\eta-\delta\omega)$ and $\zeta_{\tilde \ssA\ssR +}\to-\zeta_{\tilde \ssA\ssR -}$) in the above.

For both branes --- {\em i.e.} for both $b=\pm$ --- and $\eta \leq \delta\omega/2$, the hypermultiplet contribution to the total renormalized lagrangian becomes
\bea
 \mu\,\frac{\partial \cV_b}{\partial\mu} &=& \frac4{r^4} \left[ \mu \frac{\pd \zeta_{\ssR^2 b}}{\pd \mu} \right] + \frac1{r^4} \left[ \mu \frac\pd{\pd\mu}\left( \frac{\kappa \zeta_{\tilde \ssA\ssR b}}{\gB^2}\right) \right] + \frac1{8 r^4} \left[ \mu \frac{\pd}{\pd\mu} \left( \frac{\kappa \zeta_{\ssA b}}{\gB^2} \right) \right] \nn\\
 &=& \frac{\eta^2}{(4\pi r^2)^2\omega} \bigg[ \frac{7}{240} - \frac{\eta^2}{12} + \frac{\eta^4}{15} + \left( \frac{1}{24} - \frac{\eta^2}{12} \right) \omega^2 + \frac{7\,\omega^4}{240}  \bigg] \label{V1bsmalleta} \,.
\eea
This expression is positive-definite on our domain of validity, $0 \leq \eta \leq 1 +\delta\omega/2$, as can be checked by showing: ($i$) $\mu\,\pd \cV_b /\pd\mu \geq 0$ at $\eta=1$ for any $\omega \geq 1$; and ($ii$) the smallest root of the bracketed factor, $\eta_0(\omega) \geq 1 +\delta\omega/2$ for any $\omega \geq 1$. Also, when $\omega=1$, this expression vanishes as $\eta \to 0$ or 1, as expected.

When $\delta\omega/2 \leq \eta \leq 1 +\delta\omega/2$ on the south brane, we instead find that the renormalizations are
\bea
 \mu \frac{\pd \zeta_{\ssR^2 -}}{\pd \mu} &=& \frac1{4(4\pi)^2\omega} \bigg[ \frac{\delta\omega}{40} + \frac{37\,\delta\omega^2}{720} + \frac{7\,\delta\omega^3}{180} + \frac{7\,\delta\omega^4}{720} -\frac{F_{\hat \eta}}{16} + \frac{F_{\hat \eta}^2}{20} - \frac{F_{\hat \eta}^3}{15} \nn\\
 &&\qquad + \bigg( \frac{F_{\hat \eta}}{24} - \frac{F_{\hat \eta}^2}{12} \bigg) \omega^2 - \frac{7\,\omega^4 F_{\hat \eta}}{240} \bigg] \,,\\
 \mu \frac\pd{\pd\mu}\left( \frac{\kappa \zeta_{\tilde \ssA\ssR -}}{\gB^2}\right)  &=& \frac{1}{(4\pi)^2\omega} \bigg( \frac{\delta\omega}{60} + \frac{\delta\omega^2}{15} + \frac{7\,\delta\omega^3}{120} + \frac{7\,\delta\omega^4}{480} -\frac{F_{\hat\eta}}{20} +\frac{F_{\hat\eta}^2}{10} + \frac{\omega^2F_{\hat\eta}}{12} \bigg)(1-2\hat\eta) \,, \nn\\
&&\\
 \mu \frac{\pd}{\pd\mu} \left( \frac{\kappa \zeta_{\ssA -}}{\gB^2} \right) &=& \frac{8}{(4\pi)^2\omega} \bigg[ -\frac{\delta\omega}{120} + \frac{11\,\delta\omega^2}{720}+ \frac{7\,\delta\omega^3}{360} + \frac{7\,\delta\omega^4}{1440} + \frac{F_{\hat \eta}^2}6 + \frac{\omega^2 F_{\hat \eta}}{12} \bigg] \,.
\eea
Therefore, when $\eta \geq \delta\omega/2$ on the south brane, the beta function for the hypermultiplet contribution to the renormalized brane lagrangian is
\bea
 \mu\,\frac{\partial \cV_-}{\partial\mu} &=& \frac4{r^4} \left[ \mu \frac{\pd \zeta_{\ssR^2 -}}{\pd \mu} \right] + \frac1{r^4} \left[ \mu \frac\pd{\pd\mu}\left( \frac{\kappa \zeta_{\tilde \ssA\ssR -}}{\gB^2}\right) \right] + \frac1{8 r^4} \left[ \mu \frac{\pd}{\pd\mu} \left( \frac{\kappa \zeta_{\ssA -}}{\gB^2} \right) \right] \nn\\
 &=& \frac{(\eta-\omega)^2}{(4\pi r^2)^2\omega} \bigg[ \frac{7}{240} - \frac{(\eta-\omega)^2}{12} + \frac{(\eta-\omega)^4}{15} + \left( \frac{1}{24} - \frac{(\eta-\omega)^2}{12} \right) \omega^2 + \frac{7\,\omega^4}{240}  \bigg] \,.\quad\label{V1bbiggereta}
\eea
This expression is identical to the one found for the north brane, but with the replacement $\eta \to \eta-\omega$, so its positive-definiteness can be shown in the same way as before. (Recall that it is inconsistent to consider the limit $\eta\to0$ with fixed $\omega$ in the above expression, because it is only valid when $\eta\geq \delta\omega/2$.)

Finally, we sum the total contribution from both branes to compute how $\cV_{\rm branes} = \sum_b \cV_b$ runs. When $\eta \leq \delta\omega/2$, this is simply twice the result found in eq.~\pref{V1bsmalleta}. However, when $\eta \geq\delta\omega/2$, we must sum eq.~\pref{V1bsmalleta} with our last result, eq.~\pref{V1bbiggereta}. As expected, if we sum an even function of $\eta$ with the same function evaluated at $\eta-\omega$, we get an even function about $\eta = \omega/2$ (which happens to be the midpoint of our domain of validity, $\delta\omega/2\leq\eta\leq 1+\delta\omega/2$):
\bea
 \mu \frac{\pd \cV_{\rm branes}}{\pd\mu} &=& \frac1{(4\pi r^2)^2\omega} \bigg[ \frac{7\,\omega^2}{480} + \frac{\omega^4}{96} + \frac{\omega^6}{160} + \left(\frac7{120} -\frac{\omega^2}6 - \frac{\omega^4}{15} \right) \left(\eta-\frac\omega2\right)^2 \nn\\
 &&\qquad\qquad\qquad\qquad\qquad- \left(\frac16 - \frac{\omega^2}3\right) \left(\eta-\frac\omega2\right)^4 + \frac2{15} \left(\eta-\frac\omega2\right)^6 \bigg] \,.
\eea
This expression is positive-definite, because it is the sum of two separately positive-definite expressions.

\subsection{Massless gauge multiplet}

We next compile similar results for a massless gauge multiplet, by summing the contribution of a gauge field and a 6D Weyl fermion:
\be
 s_2^{\mathrm{gm}} := s_2^{\mathrm{gf}} + s_2^{\rm f} \,.
\ee
Dropping the `gm' superscript this combination yields $s_{-1}^{\rm sph} = s_0^{\rm sph} =0$, and
\be
 s_1^{\rm sph,\,0} = \frac13  \,, \quad s_1^{\rm sph,\,2} = -\frac{q_{\rm f}^2 \cN^2}3  \,,\quad s_2^{\rm sph,\,0} = \frac1{15}  \,,\quad s_2^{\rm sph,\,2} = - \frac{q_{\rm f}^2 \cN^2}{15}  \,,
\ee
where, as before, $\cN = \pm \gB/\gR$ and where $q_{\rm f} \, \gB$ is the charge of the fermion under the $U(1)$ gauged by the background flux. Since we assume the background flux gauges an abelian, $U(1)$, factor of the gauge group, we also choose $q_{\rm gf} = 0$ for the gauge boson.

The corresponding bulk beta functions are
\bea
 \mu\,\frac{\partial \zeta_{\ssR^3}}{\partial\mu} &=& -\frac1{120(4\pi)^3} \\
 \mu\,\frac{\partial}{\partial\mu} \Bigl( \kappa \cH \zeta_{\ssA\ssR} \Bigr)
 &=& \mu\,\frac{\partial}{\partial\mu}\left(\frac{\kappa e^{-\phi} \zeta_{\ssA\ssR}}{\gB^2}\right)
 = \frac{ 8 \, q_{\rm f}^2 }{15(4\pi)^3} \,,
\eea
which, when combined using eq.~\pref{VbulkAtEom}, give the gauge multiplet contribution to the running of $\cV_\ssB$:
\bea
\mu\,\frac{\partial \cV_\ssB}{\partial\mu} &=& \left( 4 \pi\alpha  r^2 \right) \left[ - \frac{f^2}{2r^2}\left(\mu\,\frac{\partial}{\partial\mu}\Bigl( \kappa \cH \zeta_{\ssA\ssR} \Bigr)\right)- \frac{8}{r^6} \left(\mu\,\frac{\partial \zeta_{\ssR^3}}{\partial\mu}\right)\right] \nn\\
&=& \frac{\alpha(1-q_{\rm f}^2 \cN^2)}{15 (4\pi r^2)^2} \,.
\eea
This again vanishes given the supersymmetric choices $q_{\rm f}^2 = 1 $ and $\cN = \pm 1$ (and $q_{\rm gf} = 0$), although this comes as a cancellation between the running of $\zeta_{\ssA \ssR}$ and $\zeta_{\ssR^3}$, which both separately renormalize.

We similarly compute the brane contributions, $\delta s_i = \delta s_i^{\rm gf} + \delta s_i^{\rm f}$, using the expressions in \cite{Companion}. However, as discussed therein, there are two cases to consider because of the shift in fluxes required to calculate the fermionic mode sum. In \cite{Companion}, we use the quantity $\Phi_b^{\rm f\sigma} := \Phi_b - \sigma \Phi_0^{\rm f}$ --- where $\Phi_0^{\rm f}:= (1-\omega^{-1})/2$ and $\sigma$ is the spinor's helicity  --- to calculate the fermionic mode sum, so this mode sum is dependent on whether $|\Phi_b| \leq \Phi_0^{\rm f}$ or $|\Phi_b| \geq \Phi_0^{\rm f}$. As it happens, we encounter the exact same distinction when considering unbalanced fluxes, when $|\Phi_b| >  \Phi_0^{\rm f}$ ($|\Phi_b| <  \Phi_0^{\rm f}$) on the north (south) brane, respectively. As for the hypermultiplet we specialize to the supersymmetric background flux, and write the brane-localized fluxes relative to the common supersymmetric value through the substitutions
\be
 \cN = \pm 1 \,,\quad \Phi_b = \frac{\Phi_{\rm susy}}2 \pm \frac{b\eta}\omega \,,\quad \Phi_{\rm susy} = \pm (1-\omega^{-1}) \quad (\eta\geq 0) \,.
\ee
At the north brane (where $b=+1$), we find --- using the notation \cite{Companion}, $\tilde \Phi_b = \omega(\Phi_b - \rho_b \Phi_0^{\rm f})$, $\rho_b = \Phi_b/|\Phi_b|$, $\tilde F_b = |\tilde \Phi_b| (1-|\tilde \Phi_b|)$, etc., as well as the shorthand $F_\eta = \eta(1-\eta)$,
\bea
 \delta s_0 &=& (\delta s_0)_{\rm gf} + (\delta s_0)_{\rm f}  = \frac1\omega \left(-\frac{\delta\omega}3 +\frac{\delta\omega^2}3\right) + \frac1\omega \left( \frac{\delta\omega}3 - \frac{\delta\omega^2}3 + 2 \tilde F_b \right) = \frac{2 F_\eta}{\omega} \,.
\eea
Similarly, by making the identifications $\rho_b = \pm$, $|\tilde \Phi_b| = \eta$, we find
\bea
\delta s_1^0 &=& \frac1\omega\left( \frac{\delta\omega}3 + \frac{\delta\omega^2}6 + \frac{F_\eta^2}3 - \frac{\omega^2 F_\eta}3 \right) \,,\\
\delta s_1^{\rm 1} &=& \frac1\omega\left( -\frac{\delta\omega}3 - \frac{\delta\omega^2}6 +\frac{F_{\eta}}3 \right)(1-2\eta) \,,\\
\delta s_2^0 &=& \frac1\omega\bigg( \frac{\delta\omega}{10} + \frac{17\,\delta\omega^2}{180} + \frac{2\,\delta\omega^3}{45} + \frac{\delta\omega^4}{90} + \frac{F_\eta^2}{30} + \frac{F_\eta^3}{15} - \frac{\omega^2 F_\eta^2}6 -\frac{\omega^4 F_\eta}{30}  \bigg) \,,\\
\delta s_2^{\rm 1} &=& \frac1\omega\bigg( -\frac{\delta\omega}{15} -\frac{\delta\omega^2}{10} - \frac{\delta\omega^3}{15} - \frac{\delta\omega^4}{60} +\frac{F_{\eta}}{30} +\frac{F_{\eta}^2}{10} - \frac{\omega^2F_{\eta}}6   \bigg)(1-2\eta) \,,\\
\delta s_2^{\rm 2} &=& \frac1\omega \left( -\frac{\delta\omega}{30} + \frac{\delta\omega^2}{180} + \frac{\delta\omega^3}{45} + \frac{\delta\omega^4}{180} - \frac{F_\eta^2}6 + \frac{\omega^2F_\eta}6 \right)
\eea
using the same approach. When summed, these coefficients give
\be
\delta s_0 = \delta s_1 = \delta s_2 = 0
\ee
for all $\omega$ in the supersymmetric limit of equal fluxes (for which $\eta=0$).

For nonzero $\eta$ supersymmetry is broken. When $\eta \geq \delta \omega$ at the south brane, we would instead make the identifications $\rho_b = \mp$, $|\tilde \Phi_b| = \eta - \delta\omega$, and find results similar to those above, but with $\eta \to \hat\eta:=\eta-\delta\omega$, $s_i^{\rm 1} \to - s_i^{\rm 1}$. However, when $\eta \leq \delta\omega$, we must instead use the fermionic result valid when $|\Phi_b| \leq \Phi_0^{\rm f}$. Substituting, we find that the coefficients at the south brane are, e.g.,
\bea
 \delta s_0 &=& (\delta s_0)_{\rm gf} + (\delta s_0)_{\rm f}  = \frac1\omega \left(-\frac{\delta\omega}3 +\frac{\delta\omega^2}3\right) + \frac1\omega \left( \frac{\delta\omega}3 + \frac{\delta\omega^2}6 + 2 \omega^2 \Phi_b^2 \right) = \frac1\omega \left( F_{-\eta} +2\omega\eta \right)   \nn \\
&&
\eea
using
\be
 F_{-\eta} = -\eta (1+\eta) \,,
\ee
and, similarly,
\bea
\delta s_1^0 &=& \frac1\omega\left( \frac{\delta\omega}3 + \frac{\delta\omega^2}6 + \frac{F_{-\eta}^2}3 +\frac{\omega F_{-\eta}}3 (1+2\eta) -\frac{\omega^2 F_{-\eta}}3 \right) \,,\\
\delta s_1^{\rm 1} &=& \frac1\omega\left( -\frac{\delta\omega}3- \frac{\delta\omega^2}6 + \frac{F_{-\eta}}3  \right)(1+2\eta) \,,\\
\delta s_2^0 &=& \frac1\omega\bigg[ \frac{\delta\omega}{10} + \frac{17\,\delta\omega^2}{180} + \frac{2\,\delta\omega^3}{45} + \frac{\delta\omega^4}{90} + \frac{F_{-\eta}^2}{30} + \frac{F_{-\eta}^3}{15} \nn\\
&&\qquad\quad  + \bigg( \frac{F_{-\eta}}{30} + \frac{F_{-\eta}^2}{10} \bigg)\omega(1+2\eta) - \frac{\omega^2 F_{-\eta}^2}6 -\frac{\omega^4 F_{-\eta}}{30}  \bigg] \,,\\
\delta s_2^{\rm 1} &=& \frac1\omega\bigg[\bigg( -\frac{\delta\omega}{15} -\frac{\delta\omega^2}{10} - \frac{\delta\omega^3}{15} - \frac{\delta\omega^4}{60} +\frac{F_{-\eta}}{30} +\frac{F_{-\eta}^2}{10} - \frac{\omega^2F_{-\eta}}6   \bigg)(1+2\eta) -\frac{\omega F_{-\eta}}2 \bigg] \,,\qquad\\
\delta s_2^{\rm 2} &=& \frac1\omega \left[ -\frac{\delta\omega}{30} + \frac{\delta\omega^2}{180} + \frac{\delta\omega^3}{45} + \frac{\delta\omega^4}{180} - \frac{F_{-\eta}^2}6 -\frac{\omega F_{-\eta}}6 (1+2\eta) + \frac{\omega^2F_{-\eta}}6 \right] \,.
\eea
The condition $|\Phi_b^{\rm f\sigma}|\leq 1$ means that $\eta\leq 1$, so all of these expressions lose their validity beyond this upper limit.

The running of the renormalized brane couplings from a gauge supermultiplet then become, on the north brane:
\bea
 \mu \frac{\pd \zeta_{\ssR^2 +}}{\pd \mu} &=& \frac1{4(4\pi)^2\omega} \bigg( \frac{\delta\omega}{10} + \frac{17\,\delta\omega^2}{180} + \frac{2\,\delta\omega^3}{45} + \frac{\delta\omega^4}{90} + \frac{F_\eta^2}{30} + \frac{F_\eta^3}{15} - \frac{\omega^2 F_\eta^2}6 -\frac{\omega^4 F_\eta}{30}  \bigg) \,,\nn\\
&&\\
 \mu \frac\pd{\pd\mu}\left( \frac{\kappa \zeta_{\tilde \ssA\ssR +}}{\gB^2}\right)  &=&\frac{1}{(4\pi)^2\omega} \bigg( -\frac{\delta\omega}{15} -\frac{\delta\omega^2}{10} - \frac{\delta\omega^3}{15} - \frac{\delta\omega^4}{60} +\frac{F_{\eta}}{30} +\frac{F_{\eta}^2}{10} - \frac{\omega^2F_{\eta}}6   \bigg)(1-2\eta) ,\qquad \\
 \mu \frac{\pd}{\pd\mu} \left( \frac{\kappa \zeta_{\ssA +}}{\gB^2} \right) &=& \frac{8}{(4\pi)^2\omega} \left( -\frac{\delta\omega}{30} + \frac{\delta\omega^2}{180} + \frac{\delta\omega^3}{45} + \frac{\delta\omega^4}{180} - \frac{F_\eta^2}6 + \frac{\omega^2F_\eta}6 \right)  \,.
\eea
Therefore, the beta function for the gauge multiplet contribution to the running of $\cV_b$ is
\bea \label{VnorthGM}
 \mu\,\frac{\partial \cV_+}{\partial\mu} &=& \frac4{r^4} \left[ \mu \frac{\pd \zeta_{\ssR^2 +}}{\pd \mu} \right] + \frac1{r^4} \left[ \mu \frac\pd{\pd\mu}\left( \frac{\kappa \zeta_{\tilde \ssA\ssR +}}{\gB^2}\right) \right] + \frac1{8 r^4} \left[ \mu \frac{\pd}{\pd\mu} \left( \frac{\kappa \zeta_{\ssA +}}{\gB^2} \right) \right] \nn\\
&=& \frac{\eta^2}{(4\pi r^2)^2\omega} \bigg[ -\frac2{15} + \frac{\eta^2}6 - \frac{\eta^4}{15} + \left(\frac16 - \frac{\eta^2}6 \right) \omega^2 +\frac{\omega^4}{30} \bigg]  \,.
\eea

Several features of eq.~\pref{VnorthGM} bear emphasis. First, it vanishes for all $\omega$ as $\eta \to 0$, as appropriate to the supersymmetric limit, and when $\omega=1$ it also vanishes as $\eta \to 0$ or 1, as also expected for the Salam-Sezgin sphere in the absence of branes. Second, for nonzero $\eta$ it is positive definite throughout its domain of validity, $0 \leq \eta \leq 1$, so long as $\omega > 1$ ({\em i.e.} the branes have positive tension). This definiteness of sign can be checked by showing that: ($i$) $\mu\,\pd \cV_+ /\pd\mu \geq 0$ at, e.g., $\eta=1$ for any $\omega \geq 1$; and ($ii$) the smallest root of the bracketed factor, $\eta_0(\omega) \geq 1$ for any $\omega \geq 1$.

For the south brane, the result can be obtained from the above in the case $\eta \geq \delta\omega$, simply by taking $\eta \to \eta-\delta\omega$ and $\zeta_{\tilde \ssA\ssR +} \to -\zeta_{\tilde \ssA\ssR -}$. When $\eta \leq \delta\omega$ we instead find for the south brane
\bea
 \mu \frac{\pd \zeta_{\ssR^2 -}}{\pd \mu} &=& \frac1{4(4\pi)^2\omega} \bigg[ \frac{\delta\omega}{10} + \frac{17\,\delta\omega^2}{180} + \frac{2\,\delta\omega^3}{45} + \frac{\delta\omega^4}{90} + \frac{F_{-\eta}^2}{30} + \frac{F_{-\eta}^3}{15} \nn\\
 &&\qquad\qquad\qquad  + \bigg( \frac{F_{-\eta}}{30} + \frac{F_{-\eta}^2}{10} \bigg)\omega(1+2\eta) - \frac{\omega^2 F_{-\eta}^2}6 -\frac{\omega^4 F_{-\eta}}{30}  \bigg] \,,\\
 \mu \frac\pd{\pd\mu}\left( \frac{\kappa \zeta_{\tilde \ssA\ssR -}}{\gB^2}\right)  &=& \frac{1}{(4\pi)^2\omega} \bigg[\bigg( -\frac{\delta\omega}{15} -\frac{\delta\omega^2}{10} - \frac{\delta\omega^3}{15} - \frac{\delta\omega^4}{60} +\frac{F_{-\eta}}{30} +\frac{F_{-\eta}^2}{10} - \frac{\omega^2F_{-\eta}}6   \bigg)(1+2\eta) \nn\\
 &&\qquad\qquad\qquad -\frac{\omega F_{-\eta}}2 \bigg] \,, \\
 \mu \frac{\pd}{\pd\mu} \left( \frac{\kappa \zeta_{\ssA -}}{\gB^2} \right) &=& \frac{8}{(4\pi)^2\omega} \left[ -\frac{\delta\omega}{30} + \frac{\delta\omega^2}{180} + \frac{\delta\omega^3}{45} + \frac{\delta\omega^4}{180} - \frac{F_{-\eta}^2}6 -\frac{\omega F_{-\eta}}6 (1+2\eta) + \frac{\omega^2F_{-\eta}}6 \right]  \,.\nn\\
&&
\eea
Therefore, the gauge multiplet contribution to the running of $\cV_b$ for the south brane in this case is
\bea
 \mu\,\frac{\partial \cV_-}{\partial\mu} &=& \frac4{r^4} \left[ \mu \frac{\pd \zeta_{\ssR^2 -}}{\pd \mu} \right] + \frac1{r^4} \left[ \mu \frac\pd{\pd\mu}\left( \frac{\kappa \zeta_{\tilde \ssA\ssR -}}{\gB^2}\right) \right] + \frac1{8 r^4} \left[ \mu \frac{\pd}{\pd\mu} \left( \frac{\kappa \zeta_{\ssA -}}{\gB^2} \right) \right] \\
 &=& \frac{1}{(4\pi r^2)^2\omega} \bigg[ -\frac{2\,\eta^2}{15} + \frac{\eta^4}6 - \frac{\eta^6}{15} +\left( \frac{2\,\eta}{15} - \frac{\eta^3}3 + \frac{\eta^5}5 \right) \omega  + \left(\frac{\eta^2}6 - \frac{\eta^4}6 \right) \omega^2 +\frac{\omega^4\eta^2}{30} \bigg]  \,.\nn
\eea
When $\omega=1$, this expression vanishes as $\eta \to 0$ or 1, again as expected. It can also be shown to be positive definite, in the same way as is done for eq.~\pref{VnorthGM}.

Finally, we sum the contributions at each brane to find the total renormalization of $\cV_{\rm branes} = \sum_b \cV_b$ for a massless gauge multiplet. When $\eta \leq \delta\omega$, this is
\bea
 \mu\,\frac{\partial \cV_{\rm branes}}{\partial\mu} &=&  \frac{1}{(4\pi r^2)^2\omega} \bigg[ -\frac{4\,\eta^2}{15} +\frac{\eta^4}3 - \frac{2\,\eta^6}{15} + \left( \frac{2\,\eta}{15} - \frac{\eta^3}3 + \frac{\eta^5}5 \right) \omega \nn\\
 &&\qquad\qquad\qquad + \left(\frac{\eta^2}3 - \frac{\eta^4}3\right) \omega^2 + \frac{\omega^4\eta^2}{15}   \bigg]
\eea
which vanishes, as expected, when $\eta \to 0$. For $\delta \omega \leq \eta \leq 1$, we instead have
\bea
 \mu\,\frac{\partial \cV_{\rm branes}}{\partial\mu} &=&  \frac{1}{(4\pi r^2)^2\omega} \bigg[ -\frac{\omega^2}{15} + \frac{5\,\omega^4}{48} - \frac{\omega^6}{160} - \left( \frac4{15} - \frac{5\,\omega^2}6  + \frac{67\,\omega^4}{120} \right) \left(\eta-\frac\omega2\right)^2 \nn\\
 &&\qquad\qquad\qquad+ \left( \frac13 - \frac{5\,\omega^2}6 \right) \left(\eta-\frac\omega2\right)^4 - \frac2{15} \left(\eta-\frac\omega2\right)^6 \bigg]  \,,
\eea
a function that is even about $\eta=\omega/2$ (the center of its domain of validity). These expressions agree, as they should, when $\eta=\delta\omega$.

\subsection{Massive matter multiplet}

The previous sections have the drawback that they involve only massless supermultiplets, in the sense that supersymmetry forbids their 6D masses being parametrically large compared with the KK scale, $1/r$. As a consequence only dimensionless couplings get renormalized when these fields are integrated out (using dimensional regularization and minimal subtraction).

To get beyond this, in this section we compute the $m$-dependence of the Casimir coefficient for a massive 6D supermultiplet. Recall that the field content for this multiplet is a massive gauge field (mgf), two Weyl fermions (2f) and 3 real hyperscalars (3hs). Keeping in mind that the Higgs mechanism makes a massive vector equivalent to a massless vector plus a hyperscalar, the particle content of a massive supermultiplet is equivalent to the combined field content of a gauge and a hypermultiplet.\footnote{At first sight this is hard to reconcile with the particle $U(1)_\ssR$ assignments. However, because hyperscalars carry nonzero $U(1)_\ssR$ charge, we also expect the standard $U(1)_\ssR$ symmetry to be spontaneously broken if the gauge field acquires its mass due to a nonzero hyperscalar vev. Particle states would then be labeled by the unbroken linear combination of the $R$ symmetry and the naive generator for the heavy gauge field.}

We use this observation to compute the coefficients $s_i$ for the massive multiplet in terms of those found above for gauge- and hypermultiplets, by taking
\be \label{eq:massive=g+h}
 s_i^{\mathrm{mm}} = s_i^{\rm hm} + s_i^{\rm gm} \,.
\ee
Our previous work in tracking all Gilkey-de Witt coefficients --- even though we then only needed $s_2$ for massless fields --- pays off here, since they all contribute for a massive multiplet.

Given this prescription, a massive matter multiplet gives the following nonzero bulk coefficients, $s_i^{\rm sph,\,k}$,
\be
 s_1^{\rm sph,\,0} = \frac12  \,, \quad s_1^{\rm sph,\,2} = -\frac{1}2  \,,\quad s_2^{\rm sph,\,0} = \frac1{12}  \,,\quad s_2^{\rm sph,\,2} = - \frac{1}{12}  \,,
\ee
and so contributes the following renormalizations to the bulk couplings:
\be
 \mu\, \frac{\partial \, U}{\partial\mu}  =  \mu\, \frac{\partial}{\partial\mu} \left(\frac{1}{\kappa^2}\right) = 0 \,,
\ee
\be
  \mu\,\frac{\partial}{\partial\mu}\left(\frac{\zeta_{\ssR^2}}{\kappa} \right) = -\frac{m^2}{8(4\pi)^3} \,, \quad
  \mu\,\frac{\partial \cH}{\partial\mu}  = \mu\,\frac{\partial}{\partial\mu}\left(\frac{e^{-\phi}}{\gB^2}\right) = \frac{4 m^2}{(4\pi)^3} \,,
\ee
and
\be
 \mu\,\frac{\partial \zeta_{\ssR^3}}{\partial\mu}  = -\frac1{96(4\pi)^3} \,, \quad
 \mu\,\frac{\partial}{\partial\mu}\left(\frac{\kappa \zeta_{\ssA\ssR}}{\gB^2}\right) = \frac{2}{3(4\pi)^3} \,.
\ee

Notice in particular that the nonzero renormalization of $\cH$ and generation of curvature-squared terms is consistent with the known loop corrections to the gauge kinetic functions \cite{HigherTerms} required by 6D anomaly cancellation. However unbroken supersymmetry implies these contributions precisely cancel in the renormalizations of the total bulk lagrangian evaluated at the rugby ball background:
\be
 \mu\, \frac{\partial \cV_\ssB}{\partial\mu} = 0 \,.
\ee
This is equally true when the branes break supersymmetry ({\em i.e.} $\eta \ne 0$), since bulk renormalizations do not know about brane boundary conditions.

The situation is more complicated for the brane renormalizations, however, for which $\mu \, (\partial \cV_b/\partial \mu)$ should not vanish for unequal brane-localized fluxes. We start by quoting the $\delta s_i$'s for a massive multiplet, and then computing the corresponding beta functions. On the north brane, we have
\bea
 \delta s_0 &=&  0  \,,\\
 \delta s_1^0 &=& \frac1\omega\left( \frac{\delta\omega}2 + \frac{\delta\omega^2}4 + \frac{F_\eta}2 - \frac{\omega^2 F_\eta}2 \right) \,,\\
 \delta s_1^{\rm 1} &=& \frac1\omega\left( -\frac{\delta\omega}2 - \frac{\delta\omega^2}4  \right) (1-2\eta) \,,\\
 \delta s_2^0 &=& \frac1\omega\bigg[ \frac{\delta\omega}{8} + \frac{7\,\delta\omega^2}{48} + \frac{\delta\omega^3}{12} + \frac{\delta\omega^4}{48} - \frac{F_\eta}{16} + \frac{F_\eta^2}{12} + \bigg( \frac{F_\eta}{24} -\frac{F_\eta^2}4 \bigg) \omega^2 - \frac{\omega^4 F_\eta}{16} \bigg] \,,\\
 \delta s_2^{\rm 1} &=& \frac1\omega\bigg( -\frac{\delta\omega}{12} -\frac{\delta\omega^2}{6} - \frac{\delta\omega^3}{8} - \frac{\delta\omega^4}{32} +\frac{F_\eta}{12} -\frac{\omega^2 F_\eta}4  \bigg) (1-2\eta) ,\qquad\\
 \delta s_2^{\rm 2} &=& \frac1\omega \left( -\frac{\delta\omega}{24} + \frac{\delta\omega^2}{48} + \frac{\delta\omega^3}{24} + \frac{\delta\omega^4}{96}  + \frac{\omega^2F_\eta}4 \right) \,,
\eea
which give the following renormalizations: $\mu\,\pd T_+/(\pd\mu) =0$,
\bea
 \mu \frac\pd{\pd\mu}\left( \frac{\zeta_{\ssR +}}\kappa \right) &=& \frac{m^2}{2(4\pi)^2\omega} \left( \frac{\delta\omega}2 + \frac{\delta\omega^2}4 + \frac{F_\eta}2 - \frac{\omega^2 F_\eta}2 \right) \\
 \mu \frac\pd{\pd\mu}\left( \frac{\zeta_{\tilde \ssA +}}{\gB^2} \right) &=& \frac{2\, m^2}{(4\pi)^2\omega} \left( -\frac{\delta\omega}2 - \frac{\delta\omega^2}4  \right) (1-2\eta)  \\
 \mu \frac{\pd \zeta_{\ssR^2 +}}{\pd \mu} &=& \frac1{4(4\pi)^2\omega} \bigg[ \frac{\delta\omega}{8} + \frac{7\,\delta\omega^2}{48} + \frac{\delta\omega^3}{12} + \frac{\delta\omega^4}{48} - \frac{F_\eta}{16} + \frac{F_\eta^2}{12} \nn\\
&&\qquad\qquad\quad+ \bigg( \frac{F_\eta}{24} -\frac{F_\eta^2}4 \bigg) \omega^2 - \frac{\omega^4 F_\eta}{16} \bigg] \,,\\
 \mu \frac\pd{\pd\mu}\left( \frac{\kappa \zeta_{\tilde \ssA\ssR +}}{\gB^2}\right)  &=& \frac{1}{(4\pi)^2\omega} \bigg( -\frac{\delta\omega}{12} -\frac{\delta\omega^2}{6} - \frac{\delta\omega^3}{8} - \frac{\delta\omega^4}{32} +\frac{F_\eta}{12} -\frac{\omega^2 F_\eta}4  \bigg) (1-2\eta) \,, \\
 \mu \frac{\pd}{\pd\mu} \left( \frac{\kappa \zeta_{\ssA +}}{\gB^2} \right) &=&  \frac{8}{(4\pi)^2\omega} \left( -\frac{\delta\omega}{24} + \frac{\delta\omega^2}{48} + \frac{\delta\omega^3}{24} + \frac{\delta\omega^4}{96}  + \frac{\omega^2F_\eta}4 \right)  \,.
\eea
Therefore, the total contribution of a massive matter multiplet to the running of $\cV_+$ is
\bea
 \mu\,\frac{\partial \cV_+}{\partial\mu} &=& -\frac1{2\, r^2} \left[\mu \frac\pd{\pd\mu} \left( \frac{\zeta_{\tilde \ssA +}}{\gB^2} \right) \right] - \frac2{r^2} \left[\mu \frac\pd{\pd\mu} \left( \frac{\zeta_{\ssR +}}\kappa \right) \right] + \frac4{r^4} \left[ \mu \frac{\pd \zeta_{\ssR^2 +}}{\pd \mu} \right] \nn\\
 && \qquad\qquad\qquad + \frac1{r^4} \left[ \mu \frac\pd{\pd\mu}\left( \frac{\kappa \zeta_{\tilde \ssA\ssR +}}{\gB^2}\right) \right]  + \frac1{8 r^4} \left[ \mu \frac{\pd}{\pd\mu} \left( \frac{\kappa \zeta_{\ssA +}}{\gB^2} \right) \right] \nn\\
 &=& \frac{\eta^2}{(4\pi r^2)^2\omega} \bigg[ -\frac{5}{48} + \frac{5\,\omega^2}{24} +\frac{\omega^4}{16} + \left( \frac{1}{12} -\frac{\omega^2}4 \right)\eta^2 - \frac{(\omega^2-1)}2 (mr)^2 \bigg]  \label{Vplusmm}\,.
\eea
Notice the appearance here of terms proportional to $m^2$, although the entire quantity vanishes in the supersymmetric limit $\eta \to 0$.

Following the prescription $\delta s_i^{\rm mm} = \delta s_i^{\rm hm} + \delta s_i^{\rm gm}$ on the south brane, we obtain different results when either: 1) $\eta \leq \delta\omega/2$; 2) $\delta\omega/2 \leq \eta \leq \delta\omega$; or 3) $\delta\omega \leq \eta \leq 1$. For the sake of brevity, we will only consider case 1 in what follows (since only in this case can the limit $\eta \to 0$ be taken, with fixed $\omega\neq 1$), and refer the avid reader to Appendix \ref{app:massiveresults} for the complete results that include cases 2 and 3 as well.

When $\eta \leq \delta\omega/2$, we find that
\bea
 \delta s_0 &=&  2\eta  \,,\\
 \delta s_1^0 &=& \frac1\omega\left[ \frac{\delta\omega}2 +\frac{\delta\omega^2}4 +\frac{F_{-\eta}}2 + \frac{\omega F_{-\eta}}3 (1+2\eta) -\frac{\omega^2 F_{-\eta}}2 \right] \,,\\
 \delta s_1^{\rm 1} &=& \frac1\omega\left[ \left( -\frac{\delta\omega}2 -\frac{\delta\omega^2}4 \right)(1+2\eta) - \omega F_{-\eta} \right] \,,\\
 \delta s_2^0 &=& \frac1\omega\bigg[ \frac{\delta\omega}8 + \frac{7\,\delta\omega^2}{48} + \frac{\delta\omega^3}{12} + \frac{\delta\omega^4}{48} -\frac{F_{-\eta}}{16} + \frac{F_{-\eta}^2}{12} + \bigg( \frac{F_{-\eta}}{30} +\frac{F_{-\eta}^2}{10}\bigg)\omega(1+2\eta)  \nn\\
&&\qquad + \bigg( \frac{F_{-\eta}}{24} -\frac{F_{-\eta}^2}4 \bigg) \omega^2 -\frac{\omega^4 F_{-\eta}}{16} \bigg] \,,\\
 \delta s_2^{\rm 1} &=& \frac1\omega\bigg[ \left(-\frac{\delta\omega}{12} -\frac{\delta\omega^2}6 - \frac{\delta\omega^3}8 - \frac{\delta\omega^4}{32} + \frac{F_{-\eta}}{12} -\frac{\omega^2 F_{-\eta}}4 \right) (1+2\eta) -\frac{\omega F_{-\eta}^2}2 \bigg] \,,\\
 \delta s_2^{\rm 2} &=& \frac1\omega\left[ -\frac{\delta\omega}{24} + \frac{\delta\omega^2}{48} + \frac{\delta\omega^3}{24} + \frac{\delta\omega^4}{96} -\frac{\omega F_{-\eta}}6 (1+2\eta) + \frac{\omega^2 F_{-\eta}}4 \right] \,,
\eea
which give the following renormalizations:
\bea
 \mu \frac{\pd T_-}{\pd\mu} &=& \frac{m^4\,\eta}{(4\pi)^2} \\
 \mu \frac\pd{\pd\mu}\left( \frac{\zeta_{\ssR -}}\kappa \right) &=& \frac{m^2}{2(4\pi)^2\omega} \left[ \frac{\delta\omega}2 +\frac{\delta\omega^2}4 +\frac{F_{-\eta}}2 + \frac{\omega F_{-\eta}}3 (1+2\eta) -\frac{\omega^2 F_{-\eta}}2 \right] \\
 \mu \frac\pd{\pd\mu}\left( \frac{\zeta_{\tilde \ssA -}}{\gB^2} \right) &=& \frac{2\, m^2}{(4\pi)^2\omega} \left[ \left( -\frac{\delta\omega}2 -\frac{\delta\omega^2}4 \right)(1+2\eta) - \omega F_{-\eta} \right]  \\
 \mu \frac{\pd \zeta_{\ssR^2 -}}{\pd \mu} &=& \frac1{4(4\pi)^2\omega} \bigg[ \frac{\delta\omega}8 + \frac{7\,\delta\omega^2}{48} + \frac{\delta\omega^3}{12} + \frac{\delta\omega^4}{48} -\frac{F_{-\eta}}{16} + \frac{F_{-\eta}^2}{12} \nn\\
&&\quad + \bigg( \frac{F_{-\eta}}{30} +\frac{F_{-\eta}^2}{10}\bigg)\omega(1+2\eta) + \bigg( \frac{F_{-\eta}}{24} -\frac{F_{-\eta}^2}4 \bigg) \omega^2 -\frac{\omega^4 F_{-\eta}}{16} \bigg] \,,\\
 \mu \frac\pd{\pd\mu}\left( \frac{\kappa \zeta_{\tilde \ssA\ssR -}}{\gB^2}\right)  &=& \frac{1}{(4\pi)^2\omega} \bigg[ \left(-\frac{\delta\omega}{12} -\frac{\delta\omega^2}6 - \frac{\delta\omega^3}8 - \frac{\delta\omega^4}{32} + \frac{F_{-\eta}}{12} -\frac{\omega^2 F_{-\eta}}4 \right) (1+2\eta) -\frac{\omega F_{-\eta}^2}2 \bigg] \,, \nn\\
 &&\\
 \mu \frac{\pd}{\pd\mu} \left( \frac{\kappa \zeta_{\ssA -}}{\gB^2} \right) &=&  \frac{8}{(4\pi)^2\omega} \left[ -\frac{\delta\omega}{24} + \frac{\delta\omega^2}{48} + \frac{\delta\omega^3}{24} + \frac{\delta\omega^4}{96} -\frac{\omega F_{-\eta}}6 (1+2\eta) + \frac{\omega^2 F_{-\eta}}4 \right]  \,.
\eea
Therefore, the total contribution of a massive matter multiplet to the running of $\cV_-$ (when $\eta \leq \delta\omega/2$; see Appendix \ref{app:massiveresults} for the result when $\eta\geq\delta\omega/2$) is
\bea
 \mu\,\frac{\partial \cV_-}{\partial\mu} &=& \mu\,\frac{\pd T_-}{\pd\mu} -\frac1{2\, r^2} \left[\mu \frac\pd{\pd\mu} \left( \frac{\zeta_{\tilde \ssA -}}{\gB^2} \right) \right] - \frac2{r^2} \left[\mu \frac\pd{\pd\mu} \left( \frac{\zeta_{\ssR -}}\kappa \right) \right] + \frac4{r^4} \left[ \mu \frac{\pd \zeta_{\ssR^2 -}}{\pd \mu} \right] \nn\\
 && \qquad\qquad\qquad + \frac1{r^4} \left[ \mu \frac\pd{\pd\mu}\left( \frac{\kappa \zeta_{\tilde \ssA\ssR -}}{\gB^2}\right) \right]  + \frac1{8 r^4} \left[ \mu \frac{\pd}{\pd\mu} \left( \frac{\kappa \zeta_{\ssA -}}{\gB^2} \right) \right] \nn\\
 &=& \frac{1}{(4\pi r^2)^2\omega} \Bigg[\left(-\frac5{48} +\frac{5\,\omega^2}{24} + \frac{\omega^4}{16} \right)\eta^2 + \left(\frac1{12} -\frac{\omega^2}4 \right)\eta^4 -\frac{(\omega^2-1)}2\eta^2(mr)^2 \nn \\
&&\qquad +\omega\eta \left( \frac2{15}-\frac{\eta^2}3 +\frac{\eta^4}5 -\left( \frac23 -\frac{2\,\eta^2}3 \right)(mr)^2 + (mr)^4 \right) \Bigg]  \label{Vminusmm}\,.
\eea
The first line in eq.~\pref{Vminusmm} is identical to the result in eq.~\pref{Vplusmm} for the running at the north brane, and the additional piece in the second line of eq.~\pref{Vminusmm} is odd in $\eta$.

Lastly, we can assemble the total contribution of both branes to the running of $\cV_{\rm branes} = \sum_b \cV_b$:
\bea
\mu\,\frac{\partial \cV_{\rm branes}}{\partial\mu} &=& \frac{1}{(4\pi r^2)^2\omega} \Bigg[ \left(-\frac5{24} +\frac{5\,\omega^2}{12} + \frac{\omega^4}{8} \right)\eta^2 + \left(\frac1{6} -\frac{\omega^2}2 \right)\eta^4 -(\omega^2-1)\eta^2(mr)^2 \nn\\
&&\qquad  +\omega\eta\left( \frac2{15}-\frac{\eta^2}3 +\frac{\eta^4}5 -\left( \frac23 -\frac{2\,\eta^2}3 \right)(mr)^2 + (mr)^4 \right) \Bigg] \label{Vbranesmm}
\eea
Regarding the positive-definiteness of this result, we can rest assured that any $m$-independent contribution is positive definite, since it is simply a sum of two positive-definite contributions from the massless hyper- and gauge-multiplets. As it turns out, the $m$-dependent terms are also positive definite. (Checking this must be done with some care, but it can be shown that the non-trivial zero, $\eta_0(\omega)$, of the coefficient of $(mr)^2$ in eq.~\pref{Vbranesmm} satisfies both $\eta_0(1)=1$ and $\exd \eta_0 /\exd\omega >0$ for all $\omega \geq 1$.). However, for $\eta$ larger than $\delta\omega/2$, the results in Appendix \ref{app:massiveresults} indicate that there is always some choice of $m$ for which the beta function is negative.

An exception to this is the special case where $\omega=1$, in which case $\eta \geq \delta\omega$ $(=0)$ for any $\eta$. As seen in Appendix \ref{app:massiveresults}, this regime is also positive definite, since the contribution of a massive matter multiplet to the renormalization of the 1PI effective potential in this case is simply the sum of the contributions of a massless hypermultiplet and a massless gauge multiplet.

\section{The 4D vacuum energy}
\label{sec:4Dvaceng}

The previous sections give the divergent part of $\Vone$ obtained by integrating out low-spin bulk fields and show how these divergences are absorbed by renormalization of various bulk and brane interactions. This section computes the implication of these renormalizations for the effective 4D cosmological constant, $\Lambda$, and on-brane curvature, as seen by a low-energy 4D observer.

As argued more generally in \cite{Companion}, for codimension-2 branes this is {\em not} simply given by the sum\footnote{This result would be appropriate in the `probe' approximation, but this approximation often fails for codimension-2 objects.} $\cV := \cV_\ssB + \cV_{\rm branes} + \cV_f$, where $\cV_f$ is the finite part of $\Vone$. Instead, the changes to the branes captured by $\cV_{\rm branes}$ must be combined with the contributions of bulk back-reaction -- along the lines of refs.~\cite{localizedflux} -- which in general need {\em not} be suppressed relative to the direct effects of $\Vone$, $\cV_\ssB$ or $\cV_{\rm branes}$ itself \cite{Towards, SLEDrefs, TNCC}. Indeed, this back-reaction is what allows flat solutions to exist at all at the classical level, despite the large classical positive tensions carried by each brane.


The complete back-reacted response to $\cV$ is not yet as well understood as is the response to a localized brane source. For this reason it is worth focussing exclusively on the large logarithm, $\ln( M/ m)$ in $\Lambda$ that our renormalization-group mechanism tracks. That is, although the $\mu$-dependence in $\cV_\ssB$ and $\cV_{\rm branes}$ always cancels the explicit $\ln(\mu/m)$ appearing in $\cV_f$, there is a ($\mu$-independent) large logarithm of order $\ln(M/m)$ that survives once it has done so, where $M$ is a typical UV scale, of generic order $M_6$. Because part of the log always comes from the brane and bulk renormalizations, its coefficient can be tracked purely using the RG calculations as given above. And the logarithm can be the dominant part of the answer when $M$ is much greater than $m$.

\subsection{Classical bulk back-reaction}

We first recap the general results of ref.~\cite{localizedflux}, to establish a common notation and to emphasize those features that are special to the supersymmetric case. Ref.~\cite{localizedflux} starts with a rugby-ball solution and asks how its properties respond to small changes in the brane action, $\delta S_b = - \int \exd^4x \sqrt{-\gamma} \; \delta L_b$, where $\gamma_{ab}$ is the metric induced on the brane from the 6D Einstein-frame metric in the bulk and
\be
 \delta L_b = \delta T_b - \frac12 \;  \delta \left( \frac{\cA_b}{\gB^2} \right) \; \epsilon^{mn} F_{mn} \,.
\ee
In particular, it asks how the effective 4D cosmological constant is affected by such a change, given that it vanishes for the unperturbed system.

The back-reaction caused by $\delta S_b$ is evaluated by tracking how it affects the bulk boundary conditions \cite{uvcaps}, and then solving the linearized 6D field equations to compute the change to the predicted value for the curvature, $R_{\mu\nu}$, along the brane directions. In particular, it is {\em not} assumed that the perturbed geometry has a rugby ball form. The effective 4D cosmological constant is then defined as the quantity that would give the same curvature in the low-energy 4D theory. This is a special case of a `matching' calculation between the effective theory and its UV completion \cite{EFTrev, GREFTrev}.

The result found in \cite{localizedflux} is easy to state when the bulk lagrangian density has the Einstein-Maxwell-scalar form of interest here:
\be \label{E:Lambdaeffeq}
 \Lambda = \sum_b \left[ \delta L_b - \frac{\cN}{2\, r^2}  \;  \delta \left( \frac{\cA_b}{\gB^2} \right) \right]_{\phi_*}
 = \sum_b \left[ \delta T_b - \frac{\cN}{r^2} \;  \delta \left( \frac{\cA_b}{\gB^2} \right) \right]_{\phi_*} \,,
\ee
where $\frac12 \, \epsilon^{mn} F_{mn} = f = \cN/(2 r^2)$ is the background rugby-ball bulk flux and the subscript $\phi_*$ indicates that $\delta T_b$ and $\delta (\cA_b/\gB^2)$ are to be evaluated at the classical background configuration for any bulk scalar(s).

At first sight the only difference between this and the naive `probe-brane' expectation, $\Lambda_{\rm probe} = \sum_b \delta L_b$, seems to be small: the additional contribution of the $\delta \cA_b$ term in the first equality of eq.~\pref{E:Lambdaeffeq}. (Physically, this additional contribution to $\Lambda$ arises from the energy cost imposed by adjustments to the bulk flux caused by flux quantization when the bulk volume changes in response to the altered defect angle due to $\delta L_b$.) Furthermore, the suppression by $1/r^2$ of the $\delta \cA_b$ terms relative to the $\delta T_b$ terms make it tempting to conclude that the new terms are always negligible.

There are two reasons why this intuition breaks down for the 6D supergravity of interest in this paper. First, the supergravity has a classical flat direction in the absence of the branes which is stabilized by the brane-bulk couplings, implying that the brane action is itself important in determining the value for $\phi_*$. For 6D supergravity, the position $\phi_*$ where this stabilization occurs is related to $\delta L_b$ by \cite{localizedflux}
\be \label{E:Lambdaphi*}
  \sum_b \left[ \delta L_b - \frac{\cN}{2\, r^2}  \;  \delta \left( \frac{\cA_b}{\gB^2} \right) + \frac12 \, L_b' \right]_{\phi_*} = 0 \,,
\ee
where $L_b' := \partial L_b/\partial \phi$ where $\phi$ is the 6D dilaton. When the value of $\phi_*$ is determined by the interplay of the brane action and the flux-quantization condition the second reason for believing the $\delta \cA_b$ term to be unsuppressed becomes operative: flux quantization ensures $\phi_*$ takes a value that makes $\cA_b$ and $T_b$ the same order of magnitude, as found above in eqs.~\pref{E:deltavsTgN} and \pref{E:deltavsalphaSUSY}. The same is not true for higher terms in the derivative expansion of $S_b$ because these do not enter into the flux-quantization condition.

Using eq.~\pref{E:Lambdaphi*} in eq.~\pref{E:Lambdaeffeq} gives the classical supergravity result of \cite{localizedflux}:
\be \label{E:LambdaSUSY}
 \Lambda = - \sum_b \left(\frac12 \, L_b' \right)_{\phi_*} \,.
\ee
Notice if $L_b \propto e^{n\phi}$ this takes the simple form
\be \label{E:LambdaSUSY1loop}
 \Lambda = - \sum_b \left(\frac12 \, L_b' \right)_{\phi_*} = - \frac{n}{2} \left. \sum_b L_b \right|_{\phi_*} \,,
\ee
and so vanishes in particular when $L_b$ is independent of $\phi$ (as is the case for the zeroeth-order brane action for classical rugby ball solutions), and $\Lambda = - \sum_b L_b$ if $L_b \propto e^{2\phi}$ (as is true for one-loop corrections to the tension in these solutions).

\subsection{Application to supersymmetric renormalizations}

There are two complications to be checked before applying the results of \cite{localizedflux} to the renormalized one-loop action of this paper: ($i$) both bulk and brane actions are renormalized; and ($ii$) both bulk and brane actions include corrections that are higher order in the derivative expansion. We next deal with the relevance of each of these in turn, specializing to the supersymmetric case where the bulk Maxwell field is chosen to lie in the $U(1)_\ssR$ direction, with unit flux $\cN = N = \pm 1$.

\subsubsection*{Bulk counterterms}

Renormalizations of bulk terms in a generic action can modify the classical field equations, and so in general also their linearization around the bulk rugby-ball solutions. In particular they can cause the on-brane curvature to become nonzero. This is simplest to see in situations where a flat on-brane metric is achieved by tuning the bulk cosmological constant, since for a generic theory this tuning need not be preserved under renormalization. Ref.~\cite{Companion} gives expressions for these corrections for a slightly broader class of theories than is considered in \cite{localizedflux}.

With this in mind there is much good news for the renormalizations of the 6D supergravity of interest here. First, the vanishing of $s_{-1}^{\rm sph}$ and $s_0^{\rm sph}$ (once summed over a supermultiplet) automatically ensures that neither $U$ nor $1/\kappa^2$ get renormalized at all at one loop by the low-spin massive matter supermultiplet considered here.

The bulk Maxwell action does get renormalized by a massive matter multiplet, however, as do the curvature-squared and higher-derivative terms,
\be \label{E:deltaLB}
 \delta \cL_\ssB = - \sqrt{-g} \; \left[ M^2 \, e^\phi \left( - F_{\ssM \ssN} F^{\ssM \ssN} + \frac{\ol R^2}{8} \right) + \frac{1}{12} \left( - R \, F_{\ssM \ssN} F^{\ssM \ssN} + \frac{\ol R^3}{8} \right) \right] \frac{\fL}{(4\pi)^3} \,,
\ee
%
%
where we use $m^2 = M^2 e^\phi$ and define $\fL := \ln(M/m)$. We assume $M/m$ to be independent of $\phi$ when differentiating, as is plausibly the case if the UV scale is a string theory (since then all Einstein-frame masses come with the same factor of $e^\phi$). As noted earlier, this vanishes once evaluated at a supersymmetric rugby ball --- for which $F_{\ssM \ssN} F^{\ssM \ssN} = 2 f^2 = 1/(2 \, r^4)$ and $\ol R = R = -2/r^2$. We now ask how these terms change the solutions to the background equations of motion.

Differentiating eq.~\pref{E:deltaLB} with respect to $\phi$ and evaluating at the rugby ball background gives a vanishing result, and shows that $\delta \cL_\ssB$ does not affect the background dilaton solution. Next, differentiating with respect to $A_\ssM$ gives
\ba \label{E:ddeltaLBdA}
 \delta \int \exd^2 x \, \Bigl( \delta \cL_\ssB \Bigr) &=& + \int \exd^2 x \, \sqrt{-g} \; \left[ M^2 \, e^\phi + \frac{R}{12} \right] \frac{4 \fL}{(4\pi)^3} \; F^{\ssM \ssN} \Bigl( \partial_\ssM \delta A_\ssN \Bigr) \nn\\
 &=& - \left[ 1 - \frac{3\kappa^2 M^2}{2 \gR^2} \right] \frac{\fL}{3(4\pi)^3r^4} \int \exd^2 x \, \sqrt{-g} \; \epsilon^{mn} \Bigl( \partial_m \delta A_n \Bigr) \,,
\ea
which also vanishes at the rugby ball, for which the integrand is a total derivative.\footnote{The surface terms associated with these total derivatives are not negligible, and contribute brane-localized terms once singular behaviour near the branes is excised by surrounding them with small Gaussian pillboxes. There they combine with the brane action and lead to the near-brane boundary conditions, such as the analogues of \pref{E:hsbranebc} for $A_\ssM$ and $g_{\ssM \ssN}$.} A similar argument applies when the metric is varied. In this case the variation of terms like $\sqrt{-g} \; g^{mn} g^{pq}$ vanish once evaluated at the rugby ball background, leaving variation of the 2D curvature as the only nontrivial quantity. Because in two dimensions one always has $R_{mnpq} = \frac12 \, R (g_{mp} \, g_{nq} - g_{mq} \, g_{np})$ we may write $\ol R = R$ for this variation and so find
\ba \label{E:ddeltaLBdg}
 \delta \int \exd^2x \, \Bigl( \delta \cL_\ssB \Bigr) &=& - \int \exd^2x \sqrt{-g} \; \left[ M^2 \, e^\phi \, R - \frac{1}{3} \, F_{mn} F^{mn} + \frac{R^2}{8} \right] \frac{ \fL}{4(4\pi)^3} \;  \delta R \nn\\
 &=& - \left[1 - \frac{3\kappa^2 M^2}{2 \gR^2} \right] \frac{ \fL}{3(4\pi)^3 r^4} \int \exd^2x \sqrt{-g}  \; \delta R \,,
\ea
which again involves the integral of a total derivative.

\subsubsection*{Higher-derivative brane counterterms}

Since the background is unchanged by the renormalizations of the bulk action, the results of ref.~\cite{localizedflux} are almost directly applicable. The only remaining caveat is that renormalizations don't just renormalize the first two terms of the action, eq.~\pref{E:Lbren}, but also generate the higher-derivative brane-bulk couplings. These can be neglected because (unlike for the brane-localized flux term, $\cA_b$) their effects really are suppressed by powers of $\kappa/r^2$ or $\gB^2/r^2$. They are suppressed in this way because (unlike the $\cA_b$ term) none of them are amplified by the flux-quantization condition.

\subsection{Loop-corrected 4D cosmological constant}

We may now use eqs.~\pref{E:LambdaSUSY} and \pref{E:LambdaSUSY1loop} to write an expression for $\Lambda$ at one loop in 6D supergravity compactified on a rugby ball stabilized by $U(1)_\ssR$ flux, directly in terms of the renormalized quantities $\cV_{b}$ and $\cV_f$. We do so under the assumption (necessary, but not sufficient, for unbroken supersymmetry) that the classical brane action is independent of $\phi$: $\partial \cL_b/\partial \phi = 0$.

In this case the brane action acquires a dilaton dependence through its renormalization by bulk loops, with dilaton-dependence of the 1-loop corrected action arising through the powers of $m$, whose appearance is dicated on dimensional grounds: $\delta T_b \propto m^4 = M^4 e^{2\phi}$, $\delta (\cA/\gB^2) \propto m^2 = M^2 e^\phi$ and similarly for the higher-derivative interactions. Consequently, loop-corrected brane action takes the form of eq.~\pref{E:Lbren} with coefficients
\be \label{E:Tbseriesetc}
 T_b(\phi) = T_b^{(0)} + c_{\ssT b}^{(1)} \, M^4 \, e^{2\phi} \, \fL + \cdots
 \,,
\ee
\be
 \frac{\cA_b(\phi)}{2\gB^2} = \frac{\cA_b^{(0)}}{2\gB^2} +c_{\tilde \ssA b}^{(1)} \, M^2 \, e^{\phi} \, \fL + \cdots \,, \quad
 \frac{\zeta_{\ssR b}(\phi)}{\kappa} = \frac{\zeta_{\ssR b}^{(0)}}{\kappa} +c_{\ssR b}^{(1)} \, M^2 \, e^{\phi} \, \fL + \cdots \,,
\ee
\be
 \frac{\kappa \zeta_{\ssA_b}(\phi)}{4\gB^2} = \frac{\kappa \zeta_{\ssA_b}^{(0)}}{4\gB^2} + c_{\ssA b}^{(1)} \, \fL + \cdots \,, \quad
 \frac{\kappa \zeta_{\tilde \ssA \ssR b}(\phi)}{2 \gB^2} = \frac{\kappa \zeta_{\tilde \ssA \ssR b}^{(0)}}{2 \gB^2} + c_{\tilde \ssA \ssR b}^{(1)} \, \fL + \cdots \,,
\ee
and
\be
 \zeta_{\ssR^2 b}(\phi) = \zeta_{\ssR^2 b}^{(0)} + c_{\ssR^2 b}^{(1)} \, \fL + \cdots \,,
\ee
{\em etc}., with coefficients that are directly given by the brane renormalization equations, eqs.~\pref{E:genbraneren}:
\ba \label{E:cbcoeffs}
 c_{\ssT b}^{(1)} &=& - \frac{\delta s_0^0}{2(4\pi)^2}
 \,, \quad
 c_{\tilde \ssA b}^{(1)} = - \frac{\delta s_1^1}{(4\pi)^2 \cN}
 \,, \quad
 c_{\ssR b}^{(1)} = - \frac{\delta s_1^0}{2(4\pi)^2}
 \nn\\
 c_{\ssA b}^{(1)} &=& - \frac{2 \, \delta s_2^2}{(4\pi)^2 \cN^2}
 \,, \quad
 c_{\tilde \ssA \ssR b}^{(1)} = - \frac{\delta s_2^1}{2(4\pi)^2 \cN} \,, \quad
  c_{\ssR^2 b}^{(1)} = - \frac{\delta s_2^0}{4(4\pi)^2} \,,
\ea
and so on, where the second equality specializes to the result computed earlier for a massive matter multiplet. All of the 1-loop terms are therefore suppressed by $e^{2\phi}$ relative to the choices that would have been invariant under the classical scaling symmetry, eq.~\pref{eq:scalingsym}.

Incorporating the bulk back-reaction finally leads to a formula for the effective 4D cosmological constant of the form
\be
 \Lambda = \sum_b \Lambda_b + \Lambda_f \,,
\ee
where the explicit $\mu$-dependence that $\Lambda_f$ inherits from $\cV_f$ is canceled by the implicit $\mu$-dependence of the renormalized couplings in $\Lambda_b$ (just as the $\mu$-dependence in $\cV$ canceled between $\cV_f$ on one hand and $\cV_b$ and $\cV_\ssB$ on the other). This $\mu$-independence is most usefully exploited by choosing $\mu$, so that all of the large-$M$ dependence resides in $\Lambda_b$ rather than in $\Lambda_f$.

Explicitly, $\Lambda_b$ is obtained from $\cV_b$ by evaluating eq.~\pref{E:LambdaSUSY} at the rugby-ball background, as well as $\cV_\ssB = 0$. Using eqs.~\pref{E:cbcoeffs} in eq.~\pref{E:LambdaSUSY1loop} we find the most UV sensitive part of $\Lambda$ is
%
%
%
\ba \label{E:Lambdabfinal}
 \Lambda_b &=& - c_{\ssT b}^{(1)} \, M^4 \, e^{2\phi} \, \fL - \frac12 \left(- \frac{2 c_{\ssR b}^{(1)}}{r^2} - \frac{\cN c_{\tilde \ssA b}^{(1)}}{r^2} \right) \, M^2 \, e^{\phi} \, \fL \nn\\
 &:=& \frac{C}{(4 \pi r^2)^2}  \,,
\ea
where
%
%
%
\be \label{E:Cfinal}
  C = \frac{\delta s_0^0}{2} \left( \frac{\kappa M}{2 \,\gR} \right)^4 \, \fL - \frac12 \Bigl( \delta s_1^0 + \delta s_1^1 \Bigr) \left( \frac{\kappa M}{2 \,\gR} \right)^2 \, \fL  +\cdots \,.
\ee
There are several noteworthy features about this result. First, it vanishes (for all $\omega$) in the supersymmetric limit where $\eta \to 0$. Second, it is of order $1 / (4 \pi r^2)^2$ even if $\kappa M^2 \simeq \cO(1)$. As noted in \cite{TNCC}, this size is a consequence of the flux stabilization which, through eq.~\pref{E:phircond}, ensures the loop-counting parameter is $e^{2\phi} \propto 1/r^4$. Finally, notice that massless multiplets, such as the gauge- and (hundreds of) hyper-multiplets required for anomaly cancellation,
do not contribute at all, since all of the terms proportional to $\delta s_2^k$ are $\phi$-independent.

In the case of a massive matter multiplet with $\eta\leq\delta\omega/2$ ({\em i.e.}~the case considered in the main text), $C$ is given by
\be
C = \eta \left( \frac{\kappa M}{2 \,\gR} \right)^4 \, \fL - \left(  \frac{\eta}3  +\frac{(\omega^2-1)}{2\omega} \eta^2 - \frac{\eta^3}3 \right)\left( \frac{\kappa M}{2 \,\gR} \right)^2 \, \fL  +\cdots
\ee
Although nothing conclusive can be said about the sign of $\Lambda_{\rm branes}:=\sum_b \Lambda_b$ for arbitrary values of $\omega$, $\eta$ and $M$ (as was done for the effective potential beta functions in \S4), taking $(\kappa M/2g_\ssR)^2 > 1/3$ guarantees positive $\Lambda_{\rm branes}$ for a range of $\eta$'s near $\eta=0$. Similarly, taking $(\kappa M/2g_\ssR)^2 < 1/3$ guarantees negative $\Lambda_{\rm branes}$ near $\eta=0$. Also, if we take the sphere limit ({\em i.e.}~the case in Appendix \ref{app:massiveresults} where $\eta \geq \delta\omega$ with $\delta\omega \to 0$), we find that $\delta s_0= \delta s_1^0 + \delta s_1^1 = 0$ for any $\eta$.

\section{Conclusions}
\label{sec:Conclusions}

We close with a brief summary and a sketch of the most dangerous bulk-brane higher loops.

\subsection*{Summary}

This paper uses the recent results of \cite{Companion} to compute the one-loop 1PI quantum action for 6D gauged, chiral supergravity \cite{NS}, evaluated at a rugby-ball solution \cite{Towards} to its classical field equations. By carefully tracking the near-brane boundary conditions as a function of the brane action \cite{uvcaps, localizedflux}, we are able to include the effects of bulk back-reaction to these loop corrections.

Our main focus is to identify the UV-sensitive part of the result, to see how it generates local effective interactions in the bulk and on the brane and how these interactions depend on the assumed properties of the branes and bulk. Because the rugby ball geometries are curved, they capture many UV-sensitive interactions that are not seen in the more familiar quantum calculations on tori \cite{CETori}.

We find that bulk supersymmetry strongly constrains the renormalization of bulk interactions. Renormalizations, such as corrections to the gauge coupling function and to higher-curvature terms, do occur, but with coefficients that are related to one another by supersymmetry. This is consistent with general expectations based on anomaly cancellation arguments in six dimensions \cite{HigherTerms}. The total bulk contribution to the effective vacuum energy vanishes, due to cancellations these relations permit between between gauge and curvature renormalizations.

The branes are {\em not} similarly assumed to be supersymmetric {\em a-priori}, and so we use the most general brane action expanded in a derivative expansion. The first two terms of this expansion can be physically interpreted as the brane tension (no derivatives) and the amount of background flux that is localized on the brane (one derivative). In general the presence of both these terms are required to allow low-energy perturbations to exist that are consistent with flux quantization \cite{localizedflux}, and their dependence on the bulk dilaton, $\phi$, can be used to stabilize the one modulus of the bulk geometry through a 6D analogue of the 5D Goldberger-Wise \cite{GW} mechanism.

Although branes generically break all supersymmetries we find that it is possible to couple them to the bulk in a way that preserves the unbroken supersymmetry of the bulk \cite{SS}, provided three conditions are satisfied.
\begin{itemize}
\item The branes do not couple to the bulk dilaton at all;
\item The total brane-localized flux and brane tension are related by eq.~\pref{E:SUSYcond}.
\item The defect angles are the same size at both branes (as would be automatic if the two branes were identical).
\end{itemize}
What is surprising about the second condition is that (at least in the case of identical branes) it is selected automatically as the bulk modulus adjusts to satisfy the flux-quantization conditions that drive the Goldberger-Wise mechanism in 6D, leading to a supersymmetric configuration for arbitrary dilaton-independent brane tensions. This need not remain possible once two-derivative terms and higher --- such as $\delta \cL_b = \sqrt{- \gamma} \; \cB_b \, R/\kappa$, for example --- are included in the brane action, so supersymmetry is expected to break once these are included.

Not too surprisingly, the entire one-loop 1PI quantum action evaluated at the rugby ball vanishes when the brane also preserves supersymmetry, at least when the brane action is only kept out to one-derivative order. This ensures that the entire one-loop vacuum energy vanishes in this case. Since higher-derivative terms need not be supersymmetric we expect the one-loop vacuum energy not to vanish generically once their influence on bulk modes is included.

\subsection*{Dangerous higher loops}

In the absence of brane-localized particles this would be the end of the story. However when brane particles are present larger effects can be possible. The simplest way to compute these is to estimate how bulk loops would renormalize the properties of brane-localized fields, and in particular what dependence on $\phi$ they introduce. Then include this $\phi$-dependence into a brane loop (which doesn't itself cost an additional factor of $e^{2\phi}$), using a heavy and non-supersymmetric brane particle. We do so here by assuming that all $\phi$-dependence introduced by bulk loops enters (in 6D Einstein frame) through the $\phi$-dependent bulk mass $m^2(\phi) = M^2 e^\phi$, with powers of $m^2$ appearing wherever they can on dimensional grounds.

To see how this goes, consider as brane lagrangian the cartoon Standard-Model form
\be
 \cL_b = -\sqrt{- \gamma} \; \left( \frac12 \,  \partial_\mu h \partial^\mu h + \fM^2_0 h^2  + \ol\psi \, (\Dsl + \lambda h ) \psi + \frac14 \, \cF_{\mu\nu} \cF^{\mu\nu} \right) \,,
\ee
where $h$, $\psi$ and $\cA_\mu$ are brane-localized scalar, spin-half and gauge fields, with $\cF = \exd \cA$. On dimensional grounds bulk renormalizations would be expected to renormalize this action by an amount
\be
 \delta\cL_b = - \frac{1}{(4\pi)^2} \sqrt{- \gamma} \; \left( \frac{c_1}2 \,  \partial_\mu h \partial^\mu h + c_2 m^2(\phi) h^2  + \ol\psi \, ( c_3 \Dsl + c_4 \lambda h ) \psi + \frac{c_5}4 \, \cF_{\mu\nu} \cF^{\mu\nu} + \cdots \right) \,,
\ee
in addition to the renormalizations of the tension, $T$, and other brane-field independent coefficients considered earlier. Here the $c_i$ are dimensionless coefficients that could be calculated using techniques similar to those used in earlier sections. As usual, a power of $m^2$ only appears for the scalar mass term, since the fermion and gauge masses are respectively protected by chiral and gauge symmetries.

Proceeding now to performing a loop of brane fields \cite{TNc2B}, we focus on the scalar loop. We expect this to give contributions to the brane tension of order
\be \label{E:Tbranebulk}
 \delta T \simeq k \, \frac{\fM^4}{(4 \pi)^2}
 \simeq \frac{k}{(4\pi)^2} \, \left[ \fM_0^4 \left( 1 - \frac{2 c_1}{(4\pi)^2} \right) + \frac{2 c_2 \fM_0^2 m^2}{(4\pi)^2} + \cdots \right] \,,
\ee
where $k$ is a calculable number and we use
\be
 \fM^2 \simeq \frac{\fM_0^2 + {c_2 m^2}/{(4\pi)^2}}{1 + c_1/(4\pi)^2} \simeq \fM_0^2 \left[ 1 - \frac{c_1}{(4\pi)^2} \right] + \frac{c_2 m^2}{(4\pi)^2} + \cdots \,.
\ee
What is important is the term in \pref{E:Tbranebulk} proportional to $\fM_0^2 m^2 = \fM_0^2 M^2 e^\phi$, since this is of order $M^2/(16 \pi^2 r)^2$ (as opposed to $1/r^4$) when $\fM \sim M \sim 1/\gR \sim M_6$ are much larger than $1/r$. Terms independent of $m^2$ are not dangerous since the arguments of \S5 ensure that they drop out of $\Lambda$ once back-reaction is included. Terms involving four powers (or more) of $m$ are also not dangerous because they are proportional to (at least) $e^{2\phi} \propto 1/r^4$.

Notice the dangerous term requires both of the following two ingredients:
\begin{itemize}
\item A brane-localized scalar, since the dangerous term only comes from scalar masses on the brane which are the only super-renormalizable interaction that is not protected by a symmetry and so can be shifted by $m^2(\phi)$ (an intriguing connection to the ordinary hierarchy problem.)
\item A massive bulk supermultiplet, since the dangerous terms arise from powers of $m^2$.
\end{itemize}
The easiest way to avoid the dangerous terms is simply to postulate the absence of any massive multiplets (or that they do not couple to the branes, if present), since these are not required at all by particle physics (unlike the required existence of massive Standard Model particles on the branes). This would remove the low-energy part of the cosmological constant problem (see, {\em e.g.} \cite{TNCC} for a discussion), which is usually the hardest part. But it leaves open that part of the problem that asks why the UV completion (presumably a string theory) does not give a large contribution.

The upshot is this: the generic size of UV effects in a scenario with nonsupersymmetric branes coupled to a supersymmetric bulk is of order $M^2 m^2/(4\pi)^4$, where $M$ is a large brane scale and $m$ is the KK scale. In the absence of massive brane states this leading term can vanish, leaving terms of order the KK scale alone. None of this is generic to an arbitrary extra-dimensional setup. Three additional ingredients appear to be required \cite{Towards, TNCC}:
\begin{itemize}
\item The classical scale invariance of the bulk supergravity, and the associated zero mode that survives classical flux stabilization of the extra-dimensional bulk;
\item The systematic inclusion of brane back-reaction of the branes on the bulk geometry;
\item The possibility of having codimension-2 branes with brane-localized flux, $\Phi_b$, that can break this scale-invariance and so lift the flat direction through the interplay of back-reaction and flux-quantization.
\end{itemize}

The crucial role played by back-reaction in this mechanism underlines its importance, particularly for the dynamics of low-codimension objects (for which the inter-brane forces do not fall off appreciably with distance). Although this is understood reasonably well for codimension-1 objects (through the Israel junction conditions \cite{IJC}), it is just beginning to be explored for higher codimension (and in particular codimension-2 \cite{uvcaps, localizedflux}). It is rare to find such a vast region of unexplored territory in particle physics, and its exploration is likely to contain other surprises as well.

\section*{Acknowledgements}

We would like to thank Riccardo Barbieri,
Oriol Pujol\`as, Fernando Quevedo, Seifallah Randjbar-Daemi and George Thompson for helpful discussions and Hyun-Min Lee for much help trying to disentangle the supergravity sector in early stages of this work. The matter supermultiplet heat-kernel calculation on spheres given in Appendix \ref{app:gilkeydewitt} was first computed in unpublished work with Doug Hoover. Various combinations of us are grateful for the the support of, and the pleasant environs provided by, the Abdus Salam International Center for Theoretical Physics, Perimeter Institute and McMaster University. CB's research was supported in part by funds from the Natural Sciences and Engineering Research Council (NSERC) of Canada. Research at the Perimeter Institute is supported in part by the Government of Canada through Industry Canada, and by the Province of Ontario through the Ministry of Research and Information (MRI). SLP is funded by the Deutsche Forschungsgemeinschaft (DFG) inside the ``Graduiertenkolleg GRK 1463''. The work of AS was 
supported by the EU ITN ``Unification in the LHC Era", contract PITN-GA-2009-237920 (UNILHC) and by MIUR under contract 2006022501.


\appendix

\section{Heat kernels and bulk renormalization}
\label{app:gilkeydewitt}

In this appendix we collect for convenience the explicit
expressions for the heat-kernel coefficients for low-spin matter supermultiplets in 6D theories compactified on a 2-sphere.

\subsection*{Gilkey-de Witt coefficients}

Heat-kernel methods provide very general results for the form of UV divergences in the presence of various background fields. Consider, for example, a collection of $N$ fields, assembled into a column vector, $\Psi$, and coupled to a background spacetime metric, $g_{\ssM\ssN}$, scalars, $\varphi^i$, and gauge fields, $A^a_\ssM$, with background-covariant derivative, $D_\ssM$, of the form
\be
    D_\ssM \Psi = \partial_\ssM \Psi + \omega_\ssM \, \Psi
    - i A^a_\ssM \, t_a \Psi \,.
\ee
Here $\omega_\ssM$ is the spin connection, and the gauge group is represented by the hermitian matrices $t_a$. The commutator of two such derivatives defines the matrix-valued curvature, $Y_{\ssM\ssN} \Psi = [D_\ssM, D_\ssN] \Psi$, which has the following form:
\be
    Y_{\ssM\ssN} = \cR_{\ssM\ssN} -i F^a_{\ssM\ssN} \, t_a \,.
\ee
Here $\cR_{\ssM\ssN}$ is the curvature built from the spin connection $\omega_\ssM$, which is related to the Riemann curvature of the background spacetime in a way which is made explicit in the heat-kernel Appendix of ref.~\cite{Companion}.

Suppose further that the one loop quantum action for such a field is given by
\beq     
    i\Sigma = - (-)^F \, \frac{1}{2} \, \Tr \log
    \Bigl( - \Box + X + m^2 \Bigr) \,,
\eeq
where $(-)^F = +$ for bosons and $-$ for fermions, and $\Box = g^{\ssM\ssN} D_\ssM D_\ssN$, $X$ is some local quantity built from the background fields and $m^2$ is the 6D mass matrix. In dimensional regularization the divergent part of this quantity can be written as \cite{GdWrev}
\beq \label{sigmainfty}
    \Sigma_\infty =
    \frac{1}{2(4 \pi)^3} \, (-)^F
    \sum_{k=0}^{3} \Gamma ( k - 3 + \varepsilon)
    \int d^6 x \sqrt{-g} \; \tr[ m^{6-2k}\, a_k ]
\eeq
where the divergences as $\varepsilon \to 0$ arise from the poles of Euler's gamma function, $\Gamma(z)$, at non-positive integers. Specializing these expressions to a rugby ball and comparing to eqs.~\pref{eq:Vinfty} and \pref{eq:Cform}, clearly $a_k$ is proportional to $s_{k-1}$ of the main text.

What is most useful about this expression is that in the absence of branes the coefficients, $a_k$, are explicitly known matrix-valued local quantities constructed from $X$ and $Y_{\ssM\ssN}$. In our conventions the first four Gilkey coefficients are \cite{GdWrev}
\begin{eqnarray}
  \label{eqn: gilkey}
  a_0 &=& I \nonumber \\
  a_1 &=& -\frac{1}{6}(R+6X)  \\
  a_2 &=& \frac{1}{360} \left( 2 \Riem2 - 2 \Ricci2 + 5 R^2 -12\, \Box R \right)
   \nonumber \\  &&+ \frac{1}{6} R X + \frac{1}{2} X^2 - \frac{1}{6} \Box X +
  \frac{1}{12} \y2 \nonumber\\
  a_3 &=& \frac{1}{7!} \left( - 18 \, \Box^2 R + 17 D_\ssM R D^\ssM R
  -2 D_\ssL R_{\ssM\ssN} D^\ssL R^{\ssM\ssN}
  -4 D_\ssL R_{\ssM\ssN} D^\ssN R^{\ssM\ssL} \phantom{\frac12} \right. \nonumber\\
  && + 9 D_\ssK R_{\ssM\ssN\ssL\ssP} D^\ssK
    R^{\ssM\ssN\ssL\ssP} +28 R \Box R - 8 R_{\ssM\ssN} \Box R^{\ssM\ssN}
    +24 {R^\ssM}_{\ssN} D^\ssL D^\ssN R_{\ssM\ssL} \nonumber\\
    &&+ 12 R_{\ssM\ssN\ssL\ssP} \Box R^{\ssM\ssN\ssL\ssP}
    - \frac{35}{9} \, R^3 + \frac{14}{3} \, R \,\Ricci2
    - \frac{14}{3} \, R \, \Riem2 \nonumber \\
    && + \frac{208}{9} \, {R^\ssM}_\ssN \, R_{\ssM\ssL} \, R^{\ssN\ssL}
    - \frac{64}{3} \, R^{\ssM\ssN} \, R^{\ssK\ssL} \, R_{\ssM\ssK\ssN\ssL}
    + \frac{16}{3} \, {R^\ssM}_\ssN \, R_{\ssM\ssK\ssL\ssP} \, R^{\ssN\ssK\ssL\ssP} \nonumber\\
    && \left. - \frac{44}{9} \, {R^{\ssA\ssB}}_{\ssM\ssN} \, R_{\ssA\ssB\ssK\ssL}
    \, R^{\ssM\ssN\ssK\ssL} - \frac{80}{9} \, {{{R^\ssA}_\ssB}^\ssM}_\ssN \,
    R_{\ssA\ssK\ssM\ssP} \, R^{\ssB\ssK\ssN\ssP} \right)  \nonumber \\
    &&+ \frac{1}{360} \left( 8 D_\ssM Y_{\ssN\ssK} \, D^\ssM Y^{\ssN\ssK}
    +2 D^\ssM Y_{\ssN\ssM} \, D_\ssK Y^{\ssN\ssK} + 12 Y^{\ssM\ssN} \Box Y_{\ssM\ssN}
    \phantom{\frac12} \right.  \\
    && - 12 {Y^\ssM}_\ssN \, {Y^\ssN}_\ssK \, {Y^\ssK}_\ssM - 6 R^{\ssM\ssN\ssK\ssL} \, Y_{\ssM\ssN}
    \, Y_{\ssK\ssL} +4 {R^\ssM}_\ssN \, Y_{\ssM\ssK} \, Y^{\ssN\ssK} \nonumber \\
    && - 5 R \, Y^{\ssM\ssN} \, Y_{\ssM\ssN} - 6 \Box^2 X + 60 X \Box X
    +30 D_\ssM X \, D^\ssM X - 60 X^3
    \nonumber \\
    && - 30 X \, Y^{\ssM\ssN} \, Y_{\ssM\ssN} + 10 R \, \Box X + 4 R^{\ssM\ssN}
    \, D_\ssM D_\ssN X +12 D^\ssM R \, D_\ssM X
    -30 X^2 \, R \nonumber \\
    && \left. \phantom{\frac12}
    + 12 X \, \Box R - 5 X \, R^2 + 2 X \, \Ricci2
    -2 X \, \Riem2 \right) \,, \nonumber
\end{eqnarray}
where $I$ is the unit matrix.

Specialized to the product of 4D Minkowski space with a 2-sphere, the coefficients $a_1$ through $a_3$ simplify to
\bea \label{gilkeysc}
   a_0 &=& I \nonumber \\
   a_1 &=& -\frac{1}{6} \, R - X \nonumber  \\
   a_2 &=& \frac{1}{60} \, R^2 + \frac{1}{6} \, R X +
   \frac{1}{2} \, X^2  +
   \frac{1}{12} \, Y_{mn} Y^{mn}  \\
   a_3 &=& - \, \frac{1}{630} \, R^3  - \frac{1}{30} \,
   {Y^m}_n \, {Y^n}_l \, {Y^l}_m - \frac{1}{40}
      R \, Y_{mn} Y^{mn}  - \frac{1}{12} \, X \,
      Y_{mn} Y^{mn} \nonumber \\
   && \qquad - \frac16 \, X^3 -\frac{1}{12} \, X^2 \, R
      - \frac{1}{60} \, X \, R^2  \,. \nonumber
\eea

\subsection*{Spins zero through one}

We now collect the results for $X$, $Y_{\ssM \ssN}$ and the ultraviolet-divergent parts of the one-loop action, for the particles arising in 6D matter gauge- and hyper-multiplets. We assume also the fields in the loop do not mix appreciably with the supergravity sector, so in particular any gauge fields considered cannot be those whose background flux stabilizes the extra dimensions.

\medskip\noindent{\em Scalars}

\medskip\noindent
Consider $\cN$ scalars, $\Phi^\ssI$, with action,
\be \label{scalaraction}
     S = - \int \exd^6 x \sqrt{-g} \; \left[ \frac12 \, g^{\ssM \ssN} \,
     G_{ij} \, D_\ssM \Phi^i D_\ssN \Phi^j
     + V  + \frac12 \, U \, R +
     \frac14 \, W \, F^a_{\ssM \ssN} F_a^{\ssM \ssN}
     \right] \,.
\ee
The functions $U$, $V$, $W$ and the target-space metric, $G_{ij}$, are imagined to be known functions of the $\Phi^i$. The background-covariant derivative appropriate to this case is:
\be
    D_\ssM \Phi^i = \partial_\ssM \Phi^i - i A^a_\ssM \, {(t_a)^i}_j \Phi^j
     \,,
\ee
where the matrices ${(t_a)^i}_j$ represent the gauge group on the scalars.

The kinetic operator controlling small fluctuations about a classical background is given by
\be
    {\Delta^i}_{j} = - {\delta^i}_{j} \, \Box  + {X^i}_{j}
     \,,
\ee
with ${X^i}_{j}$ given by
\be
    {X^i}_{j} = G^{ik} \Bigl[ V_{kj}(\varphi) + \frac12 \, R \, U_{kj}(\varphi) + \frac14 \, F^a_{\ssM\ssN} F_a^{\ssM\ssN} \, W_{kj}(\varphi) \Bigr] \,,
\ee
where subscripts on $U$, $V$ and $W$ denote differentiation with respect to the background field $\varphi^i$. Specializing to the simple geometry and Maxwell fields of the rugby ball, these simplify to ${X^i}_{j} = G^{ik}[ V_{kj} + \frac12 \, R \,U_{kj} + \frac{1}{2} \, f^2 \, W_{kj}]$ and $Y_{mn} = -i\tilde{g} f \, Q \, \epsilon_{mn}$, where $\tilde{g}$ is the gauge coupling and $\tilde{g}Q = t_a$ is the hermitian, antisymmetric charge matrix for the background gauge field. Notice that these imply $Y_{mn} Y^{mn} = -2 \tilde{g}^2 f^2 \, Q^2$ and ${Y^m}_n \, {Y^n}_l \, {Y^l}_m = 0$.

With these expressions the coefficients $a_0$ through $a_3$ satisfy
\be \label{gilkeyscalar0}
   \tr a_0 = \cN \,, \qquad
   \tr a_1 = -\frac{\cN}{6} \, R  - \tr X \,,
\ee
and
\ba \label{gilkeyscalar}
   \tr a_2 &=& \frac{\cN}{60} \, R^2  + \frac{1}{6} \, R \, \tr X + \frac{1}{2} \, \tr X^2  -
   \frac{1}{6} \, \tilde{g}^2 f^2 \, \tr Q^2  \\
   \tr a_3 &=& - \, \frac{\cN}{630} \, R^3  + \frac{1}{20}
      R \, \tilde{g}^2 f^2 \, \tr Q^2  + \frac{1}{6} \, \tilde{g}^2 f^2 \, \tr (X Q^2) \nn\\
      && \qquad\qquad\qquad\qquad
      - \frac16 \,\tr X^3 -\frac{1}{12} \, R \, \tr X^2
      - \frac{1}{60} \, R^2 \, \tr X  \,. \nonumber
\ea
These give explicit functions of $\varphi$ once the above expression for $X$ is used.

\medskip\noindent{\em Spin-half fermions}

\medskip\noindent
For $\cN$ 6D massless spin-half Weyl fermions, $\psi^a$ with $a = 1,...,\cN$, we take the following action
\be
     S = - \int \exd^6x \sqrt{-g} \;  \frac12 \,
     G_{ab}(\varphi) \, \overline\psi^{\;a} \nott{D} \psi^b \,,
\ee
where $\nott{D} = {e_\ssA}^{\ssM} \, \gamma^\ssA D_\ssM$ with
\be
 D_\ssM \psi^a = \partial_\ssM \psi^a - \frac14 \, \omega_\ssM^{\ssA\ssB} \gamma_{\ssA\ssB} \psi - i A_\ssM^a t_a \psi \,,
\ee
where $\gamma^\ssA$ are the 6D Dirac matrices and ${e_\ssA}^\ssM$ the inverse sechsbein, $\gamma_{\ssA\ssB} = \frac12 [ \gamma_\ssA, \gamma_\ssB]$, and $t_a$
denotes the gauge-group generator acting on the spinor fields. Since 6D Weyl spinors have 4 complex components their representation of the 6D Lorentz group has $d = 8$ real dimensions.

The differential operator which governs the one-loop contributions is in this case $\nott{D} = {e_\ssA}^\ssM \gamma^\ssA D_\ssM$ and so in order to use the general results of the previous section we write (assuming there are no gauge or Lorentz anomalies) $\log \det \nott{D} = \frac{1}{2} \log \det(-\nott{D}^2)$, which implies
\begin{eqnarray}
\label{eqn: sigmaspin1/2}
    i \Sigma_{1/2} &=& \frac12 \, \Tr \log \nott{D}  =
    \frac{1}{4} \Tr \log \left( -
    {\nott{D}}^2 \right) \nonumber \\
        &=& \frac{1}{4} \Tr \log \left(-\Box - \frac{1}{4} R + \frac{1}{4} \gamma^{\ssA \ssB} F^a_{\ssA \ssB} t_a \right) \,.
\end{eqnarray}
This allows us to adopt the previous results for the ultraviolet divergences, provided we divide the result by an overall factor of 2 (and so effectively $d=4$ instead of 8), and use
\be
    X =  -\frac{1}{4} \, R  + \frac{1}{4} \,
    \gamma^{\ssA \ssB} \, F^a_{\ssA \ssB} \, t_a  \,,
\ee
and
\be
    Y_{\ssM\ssN} = -\, \frac{i}{2} \, R_{\ssM\ssN\ssA\ssB} \gamma^{\ssA\ssB} -i {F^a}_{\ssM\ssN} t_a
    \,.
\ee
The Gilkey coefficients become\footnote{We adopt the convention of using $\Tr[...]$ to denote a trace which includes the Lorentz and/or spacetime indices, while reserving $\tr[...]$ for those which run only over the `flavor' indices which count the fields of a given spin.}
\ba
    \Tr_{1/2}[\y2] &=& - 4 \, \tr_{1/2}(t_a t_b) \, F^a_{\ssM\ssN} F^{b\ssM\ssN}
    -\frac{\cN}{2} \, \Riem2 \nonumber \\
    &=& - 8\, \tilde{g}^2 f^2 \, \tr_{1/2}(Q^2) - \frac{\cN}{2} \, R^2  \,,
\ea
where the second line specializes to rugby ball background fields.

Keeping explicit the sign due to statistics, and dropping terms which vanish when traced, this leads to the following expressions for the divergent contributions of $\cN$ 6D Weyl fermions:
\begin{eqnarray}
    \label{gilkeyspinor}
    &&(-)^F \, \Tr_{1/2}[a_0] = - 4\cN \,, \qquad
    (-)^F \, \Tr_{1/2}[a_1] = -\frac{\cN}{3} \, R  \nonumber \\
    &&(-)^F \, \Tr_{1/2}[a_2] =  \frac{\cN}{60} \, R^2 - \frac{4}{3} \,
    \tilde{g}^2 f^2 \, \tr_{1/2}(Q^2) \\
    &&(-)^F \, \Tr_{1/2}[a_3] = -\, \frac{\cN}{504} \, R^3 + \frac{2}{15}
    \, \tilde{g}^2 f^2 \, R\, \tr_{1/2}(Q^2)  \,.
    \nonumber
\end{eqnarray}

\medskip\noindent{\em Massless gauge bosons}

\medskip\noindent
For $\cN$ gauge bosons, ${\cA}^a_\ssM$, with field strength
$\cF^a_{\ssM\ssN}$ and $a = 1,...,\cN$, we use the usual Yang-Mills action
\be
     S = - \int \exd^6x \sqrt{-g} \;  \frac14 \, W(\varphi)
     \, \cF^a_{\ssM\ssN} \cF_a^{\ssM\ssN}  \,,
\ee
expanded to quadratic order about the background fields: $\cA^a_\ssM = A_\ssM^a + \delta A^a_\ssM$. For an appropriate choice of gauge the differential operator which governs the loop contributions becomes
\be
    {\Delta^{a\ssM}}_{b\ssN} = - {\delta^a}_b \, {\delta^\ssM}_\ssN \Box
    + {X^{a\ssM}}_{b\ssN} \,,
\ee
with
\be
    {X^{a\ssM}}_{b \ssN} = -
    {R^\ssM}_{ \ssN} {\delta^a}_{ b} + 2i
    {({t}_c)^a}_{b} {F^{c \ssM}}_{\ssN}  \,,
\ee
where $t_c$ here denotes a gauge generator in the adjoint
representation.

Ref.~\cite{Companion} sums the contributions of the vector fields and ghosts to get the contribution of $\cN$ physical 6D massless gauge bosons:
\begin{eqnarray} \label{gilkeyspin1}
 &&(-)^F \, \Tr_1[a_0] = 4 \cN \,, \qquad
 (-)^F \, \Tr_1[a_1] =  \frac{\cN}{3} \, R \\
 &&(-)^F \, \Tr_1[a_2] = \frac{\cN}{15} \, R^2
 + \frac{10}{3} \, \tilde{g}^2 f^2 \, \tr_1(Q^2) \nonumber \\
 && (-)^F \, \Tr_1[a_3] = - \frac{2\cN}{315} \, R^3 + \frac{7}{10}
 \, \tilde{g}^2 f^2 \, R \, \tr_1(Q^2) \,. \nonumber
\end{eqnarray}

\subsection*{Supermultiplets}

In this section, we show that the Gilkey coefficients cancel when summed over the field content of a gauge- or hypermultiplets, providing that the background flux does not break supersymmetry. Recall that unbroken supersymmetry requires the background gauge field to lie in the $R$-symmetry direction.

\medskip\noindent{\em Gauge Multiplets}

\medskip\noindent
A gauge multiplet involves one Weyl spinor and one gauge boson. Specializing these earlier results to rugby-ball background fields, we have for $\cN_g$ Weyl fermions
\begin{eqnarray}
    \label{gilkeyspinorSS}
    &&(-)^F \, \Tr_{1/2}[a_0] = - 4\cN_g \,, \qquad
    (-)^F \, \Tr_{1/2}[a_1] = \frac{2\cN_g}{3\,r^2} \nonumber \\
    &&(-)^F \, \Tr_{1/2}[a_2] =  \frac{\cN_g}{15 \, r^4} -
    \frac{N^2}{3 \, r^4} \,  \tr_{1/2}(Q^2) \\
    &&(-)^F \, \Tr_{1/2}[a_3] = \frac{\cN_g}{63 \, r^6}
    - \frac{N^2}{15 \, r^6}  \, \tr_{1/2}(Q^2)\,,
    \nonumber
\end{eqnarray}
where $N^2 = \gB^2/\gR^2$ is the background flux quantum number.

Similarly, $\cN_g$ massless spin-1 particles gives
\begin{eqnarray} \label{gilkeyspin1SS}
 &&(-)^F \, \Tr_1[a_0] = 4 \cN_g \,, \qquad
 (-)^F \, \Tr_1[a_1] =  - \,\frac{2\cN_g}{3 \, r^2}  \\
 &&(-)^F \, \Tr_1[a_2] = \frac{4 \cN_g}{15 \, r^4}
 + \frac{5 N^2}{6 \, r^4} \, \tr_1(Q^2) \nonumber \\
 && (-)^F \, \Tr_1[a_3] = \frac{16 \cN_g}{315 \, r^6} - \frac{7 N^2}{20 \, r^6}
  \, \tr_1(Q^2) \,. \nonumber
\end{eqnarray}
These sum to give the following result for $\cN_g$ massless 6D gauge supermultiplets:
\begin{eqnarray}
    \label{gilkey6Dgauge}
    &&\Tr_g[(-)^F \, a_0] = 0 \,, \qquad
    \Tr_g[(-)^F \, a_1] = 0 \nonumber \\
    &&\Tr_g[(-)^F \, a_2] =  \frac{\cN_g}{3 \, r^4} -
    \frac{N^2}{3 \, r^4} \,  \tr_{1/2}(Q^2) + \frac{5N^2}{6 \, r^4} \, \tr_1(Q^2)\\
    &&\Tr_g[(-)^F \, a_3] = \frac{\cN_g}{15 \, r^6}
    - \frac{N^2}{15 \, r^6}  \, \tr_{1/2}(Q^2) - \frac{7N^2}{20 \, r^6}
  \, \tr_1(Q^2)\,.
    \nonumber
\end{eqnarray}

For the supersymmetric compactification we must use unit background flux, $N^2 = 1$, and the charge assignments for $U(1)_R$: $\tr_1(Q^2) = 0$ and $\tr_{1/2}(Q^2) = \cN_g$. In this case the above formulae simplify to
\be
    \Tr_g[(-)^F \, a_0] = \Tr_g[(-)^F \, a_1]
    = \Tr_g[(-)^F \, a_2] = \Tr_g[(-)^F \, a_3] = 0 \,,
\ee
as claimed in the main text.

\medskip\noindent{\em Hypermultiplets}

\medskip\noindent
Hypermultiplets contain one Weyl fermion and two complex scalars, with the scalars carrying charge $\pm 1$ under the gauge group $U(1)_R$.

The fermionic contribution to the vacuum energy is as in eq.~\pref{gilkeyspinorSS}. For the scalars we may use the results of eq.~(\ref{gilkeyscalar}), specialized to the hypermultiplet lagrangian, for which
\be
 U = W = 0 \quad \hbox{and} \quad V = 2 \gR^2 \, e^\phi \, v(\Phi) \quad \hbox{where} \quad v(\Phi) = 1 + \frac12 \, G_{ij} \Phi^i \Phi^j + \cdots \,,
\ee
which imply
\be
 {X^i}_j = 2 \gR^2 \, \left. e^\phi \, G^{ik} v_{kj} \right|_{\Phi = 0} = \frac{1}{2 \, r^2} \, {\delta^i}_j \,.
\ee
It follows that for $\cN_h$ hyperscalars $\Tr_0 I = 4\cN_h$, $\Tr_0 X = 2 \cN_h/r^2$, $\Tr_0(X Q^2) = \Tr_0(Q^2)/(2\,r^2)$, $\Tr_0 (X^2) = \cN_h/r^4$ and $\Tr_0 (X^3) = \cN_h/(2\,r^6)$. Using these expressions we have the following spin-0 contribution for $\cN_h$ hyperscalars:
\begin{eqnarray}
    \label{gilkeyscalarhyper}
    &&\Tr_0[(-)^F \, a_0] = 4\cN_h \,, \qquad
    \Tr_0[(-)^F \, a_1] = - \frac{2 \cN_h}{3\,r^2} \nonumber \\
    &&\Tr_0[(-)^F \, a_2] = \frac{\cN_h}{10 \,r^4} -
   \frac{N^2}{24 r^4} \, \tr_0 (Q^2) \\
    &&\Tr_0[(-)^F \, a_3] = \frac{\cN_h}{1260 \,r^6} -
   \frac{N^2}{240 r^6} \, \tr_0  (Q^2)\,,
    \nonumber
\end{eqnarray}
where, as before, $N^2 = \gB^2/\gR^2$.

Summing this with eq.~\pref{gilkeyspinorSS} for $\cN_h$ Weyl fermions gives the result for $\cN_h$ hyper-multiplets
\begin{eqnarray}
    \label{gilkey6Dhypermults}
    &&\Tr_h[(-)^F \, a_0] = 0 \,, \qquad
    \Tr_h[(-)^F \, a_1] = 0 \nonumber \\
    &&\Tr_h[(-)^F \, a_2] =  \frac{\cN_h}{6 \, r^4} -
    \frac{N^2}{3 \, r^4} \,  \tr_{1/2}(Q^2) - \frac{N^2}{24 \, r^4} \, \tr_0(Q^2)\\
    &&\Tr_h[(-)^F \, a_3] = \frac{\cN_h}{60 \, r^6}
    - \frac{N^2}{15 \, r^6}  \, \tr_{1/2}(Q^2) - \frac{N^2}{240 \, r^6}
  \, \tr_0(Q^2)\,,
    \nonumber
\end{eqnarray}
which, with the supersymmetric choices $N^2 = 1$, $\tr_{1/2} (Q^2) = 0$ and $\tr_0 (Q^2) = 4 \cN_h$, gives the simple result
\begin{eqnarray}
    \label{gilkeyspinorhyper}
    \Tr_h[(-)^F \, a_0] = \Tr_h[(-)^F \, a_1]
    = \Tr_h[(-)^F \, a_2] = \Tr_h[(-)^F \, a_3] = 0 \,,
    \nonumber
\end{eqnarray}
used in the main text.

\section{Results for spins zero, half and one}
\label{App:CompanionResults}

This appendix briefly summarizes the results for $s_i$ for spins zero, half and one, as computed in ref.~\cite{Companion}.

\subsection*{Spin zero}

Consider first the simplest case of a single minimally coupled real scalar field, satisfying $(-\Box + m^2) \phi = 0$, that is coupled to the background gauge field with monopole number $N$ and brane--localized fluxes $\Phi_b$.

In this case, using the notation $\omega = 1/\alpha$ and
\be 
F_b:=|\Phi_b|\left(1-|\Phi_b|\right)\,,\quad F^{(n)}:= \sum_b F_b^n \,,\quad F^{(1)}:=F \,, \quad G(x):=(1-x)(1-2x) \,,
\ee
we find the following for the $s_i$ coefficients:
\bea \label{eq:simplescalars1}
s^{\rm s}_{-1} &=& \frac1\omega \,,\nn\\
s^{\rm s}_0(\omega,N,\Phi_b) &=& \frac1\omega \left[ \frac16 + \frac{\omega^2}6(1-3F) \right] \,, \nn\\
s^{\rm s}_1(\omega,N,\Phi_b) &=& \frac1\omega\Bigg[ \frac1{180} - \frac{\cN^2}{24} + \frac{\omega^2}{18}(1 - 3 F) -\frac{\omega^3\cN}{12} \sum_b \Phi_b \, G(|\Phi_b|) + \frac{\omega^4}{180} (1 -15F^{(2)}) \Bigg] ,\qquad \nn\\
s^{\rm s}_2(\omega,N,\Phi_b) &=& \frac1\omega \Bigg[ -\frac1{504} - \frac{11\,\cN^2}{720} + \left(\frac1{90} -\frac{\cN^2}{144} \right) (1-3F) \omega^2 -\frac{\omega^3\cN}{24} \sum_b \Phi_b \, G(|\Phi_b|)  \nn\\
&& \qquad  + \frac{\omega^4(1-\cN^2)}{360}(1 - 15F^{(2)}) - \frac{\omega^5\cN}{120}\sum_b \Phi_b \,G(|\Phi_b|)(1+3F_b)  \\
&&\qquad   + \Bigg(\frac1{1260} -\frac{F^{(2)}}{120} - \frac{F^{(3)}}{60} \Bigg)\omega^6 \Bigg] \label{eq:simplescalars2} \,.\nn
\eea
When $\omega = 1$ and $\Phi_b=0$, these become
\bea
 s_{-1}^{\rm sph} = 1 \,, \quad
 s_{0}^{\rm sph} &=& \frac13 \,, \quad
 s_{1}^{\rm sph,\,0} = \frac{1}{15} \,, \quad
 s_{1}^{\rm sph, \,2} = -\frac{\cN^2}{24} \,, \\
 \quad s_{2}^{\rm sph,\,0} &=& \frac{4}{315} \,,
  \quad \hbox{and} \quad s_{2}^{\rm sph,\, 2} = -\frac{\cN^2}{40} \nn
\eea
in agreement with the results in \cite{KandM, CandelasWeinberg}, as well as with the result as computed using the Gilkey-de Witt coefficients for a 6D scalar on a sphere using the general results found in \cite{GdWrev}. If the scalar couples to the background field with strength $q \gB$, its contribution to the running of the leading bulk counterterms therefore is
\bea
 \mu \, \frac{\partial U}{\partial \mu} = -\frac{m^6}{6 (4 \pi)^3} \,,\quad&&
 \mu \, \frac{\partial}{\partial \mu} \left( \frac{1}{\kappa^2} \right) = -\frac{m^4}{6 (4\pi)^3} \,,\nn\\
 \mu\,\frac{\partial }{\partial\mu}\left(\frac{\zeta_{\ssR^2}}\kappa\right) = -\frac{m^2}{60(4\pi)^3} \,,\quad && \mu \, \frac{\partial \zeta_{\ssR^3}}{\partial\mu} = -\frac1{630(4\pi)^3}  \,,\\
 \mu \, \frac{\partial}{\partial\mu}\left(\frac{1}{\gB^2}\right) = \frac{2\,q^2 m^2}{3(4\pi)^3} \,,\quad && \mu \, \frac{\partial}{\partial \mu} \left(\frac{\kappa \zeta_{\ssA\ssR}}{\gB^2}\right) = \frac{2\,q^2}{5(4\pi)^3} \,.\nn
\eea

The quantities relevant to brane renormalizations are $\delta s_{-1} = 0$,
\bea
 \delta s_0 &=&  \frac{\omega^2 - 1}{12\, \omega} -\frac{\omega F_b}2 = \frac1{\omega} \left( \frac{\delta\omega}6 + \frac{\delta\omega^2}{12} - \frac{\omega^2 F_b}2 \right) \simeq \frac{\delta\omega}6 - \frac{|\Phi_b|}2 \,,\nn\\
 \delta s^0_1 &=& \frac1{\omega}\left(\frac{\omega^2-1}{36} + \frac{\omega^4-1}{360}- \frac{\omega^2 F_b}6 -\frac{\omega^4 F_b^2}{12}\right)  \nn\\
 &=& \frac1\omega\left(\frac{\delta\omega}{15} + \frac{2\,\delta\omega^2}{45} + \frac{\delta\omega^3}{90} + \frac{\delta\omega^4}{360} - \frac{\omega^2 F_b}6 - \frac{\omega^4 F^2_b}{12} \right) \simeq \frac{\delta\omega}{15} - \frac{|\Phi_b|}6 \,,\nn\\
 \delta s^{\rm 1}_1 &=& -\frac{\omega^2 \cN}{12} \Phi_b \, G(|\Phi_b|) \,,\nn\\
 \delta s^{\rm 2}_1 &=& s^{\rm 2}_1 -\left(- \frac{\cN^2}{24\,\omega}\right) = 0 \,,\\
 \delta s^0_2 &=& \frac1\omega\Bigg[ \frac{\omega^2-1}{180} + \frac{\omega^4-1}{720} + \frac{\omega^6-1}{2520} -\omega^2\bigg(\frac{F_b}{30} +\frac{\omega^2F_b^2}{24} +\frac{\omega^4 F^2_b}{120} + \frac{\omega^4 F^3_b}{60}\bigg)\Bigg] \nn\\
 &=& \frac1\omega\Bigg[ \frac{2\,\delta\omega}{105} + \frac{5\,\delta\omega^2}{252} + \frac{17\,\delta\omega^3}{1260} + \frac{37\,\delta\omega^4}{5040} + \frac{\delta\omega^5}{420} + \frac{\delta\omega^6}{2520} \nn\\
 &&\qquad -\omega^2\bigg(\frac{F_b}{30} +\frac{\omega^2F_b^2}{24} +\frac{\omega^4 F^2_b}{120} + \frac{\omega^4 F^3_b}{60}\bigg)\Bigg] \,,\nn\\
 \delta s^{\rm 1}_2 &=& -\frac{\omega^2 \cN}{24}  \Phi_b \, G(|\Phi_b|) -\frac{\omega^4 \cN}{120}  \Phi_b \, G(|\Phi_b|) (1+3F_b) \,,\nn\\
 \delta s^{\rm 2}_2 &=&  -\frac{\cN^2}\omega \left( \frac{\omega^2-1}{288} + \frac{\omega^4-1}{720} - \frac{\omega^2 F_b}{48} -\frac{\omega^4 F_b^2}{24} \right) \nn\\
 &=& -\frac{\cN^2}\omega \left( \frac{\delta\omega}{80} + \frac{17\,\delta\omega^2}{1440} + \frac{\delta\omega^3}{180} + \frac{\delta\omega^4}{720} - \frac{\omega^2 F_b}{48} -\frac{\omega^4 F_b^2}{24} \right) .\nn
\eea
The corresponding contributions to the running of the brane counterterms are
\bea
 \mu \, \frac{\partial T_b}{\partial \mu} &=& \frac{m^4}{2 (4\pi)^2 \omega} \left( \frac{\delta\omega}6 + \frac{\delta\omega^2}{12} - \frac{\omega^2 F_b}2 \right) \simeq \frac{m^4}{4(4\pi)^2} \left(\frac{\delta\omega}{3}-|\Phi_b| \right) \,,\nn\\
   \mu\,\frac{\pd}{\pd\mu}\left(\frac{\cA_b}{\gB^2}\right) &=&  -\frac{q \Phi_b\,\omega^2 m^2}{6(4\pi)^2} \, G(|\Phi_b|) \simeq -\frac{q^2 m^2\, \cA_b}{3(4\pi)^3} \,,\nn\\
  \mu \, \frac{\partial }{\partial \mu} \left(\frac{\zeta_{\ssR b}}{\kappa}\right) &=& \frac{m^2}{2(4\pi)^2\omega} \left(\frac{\delta\omega}{15} + \frac{2\,\delta\omega^2}{45} + \frac{\delta\omega^3}{90} + \frac{\delta\omega^4}{360} - \frac{\omega^2 F_b}6 - \frac{\omega^4 F^2_b}{12} \right) \nn\\
  &\simeq& \frac{m^2}{2(4\pi)^2}\left(\frac{\delta\omega}{15} - \frac{|\Phi_b|}6\right) \,,\nn\\
    \mu\,\frac{\pd}{\pd\mu}\left(\frac{\kappa \zeta_{\tilde\ssA\ssR b}}{\gB^2}\right) &=& -\frac{q\Phi_b \,\omega^2}{24(4\pi)^2} \left(  G(|\Phi_b|) +\frac{\omega^2 }{5}   \, G(|\Phi_b|) (1+3F_b) \right) \simeq -\frac{q^2 \, \cA_b }{10(4\pi)^3} \,,\\
  \mu\,\frac{\partial \,\zeta_{\ssR^2 b}}{\partial \mu} &=& \frac{1}{4(4\pi)^2\omega} \Bigg[ \frac{2\,\delta\omega}{105} + \frac{5\,\delta\omega^2}{252} + \frac{17\,\delta\omega^3}{1260} + \frac{37\,\delta\omega^4}{5040} + \frac{\delta\omega^5}{420} + \frac{\delta\omega^6}{2520} \nn\\
&&\qquad\qquad\quad -\omega^2\bigg(\frac{F_b}{30} +\frac{\omega^2F_b^2}{24} +\frac{\omega^4 F^2_b}{120} + \frac{\omega^4 F^3_b}{60}\bigg)\Bigg]  \nn\\
  &\simeq& \frac{1}{4(4\pi)^2} \left(\frac{2\,\delta\omega}{105} - \frac{|\Phi_b|}{30}\right) \,,\nn\\
  \mu \, \frac{\partial }{\partial \mu} \left(\frac{\kappa\zeta_{\ssA b}}{\gB^2}\right) &=& -\frac{8\,q^2}{(4\pi)^2\omega}\left( \frac{\delta\omega}{80} + \frac{17\,\delta\omega^2}{1440} + \frac{\delta\omega^3}{180} + \frac{\delta\omega^4}{720} - \frac{\omega^2 F_b}{48} -\frac{\omega^4 F_b^2}{24} \right) \nn\\
  &\simeq& -\frac{q^2}{(4\pi)^2} \left(\frac{\delta\omega}{10} - \frac{|\Phi_b|}{6} \right)  \,.\nn
\eea

\subsection*{Spin half}

We next quote the results for a massive minimally coupled spin-half 6D Weyl field minimally coupled to the background. Using the notation
\be
 N_{\rm f\sigma} := N-\sigma \,,\quad \Phi_b^{\rm f\sigma} := \Phi_b - \sigma \Phi_0^{\rm f} \,, \quad \Phi_0^{\rm f} := \frac12 \left(1- \omega^{-1}\right)
 = \frac12 (1 - \alpha) = \frac{\delta}{4\pi} \,,
\ee
one finds different expressions depending on whether or not $|\Phi_b|$ is larger or smaller\footnote{Both results agree when $|\Phi_b| = \Phi_0^{\rm f}$.} than $\Phi_0^{\rm f}$. We quote only the case $|\Phi_b| < \Phi_0^{\rm f}$, and refer the reader to \cite{Companion} for the more general case.

For a 6D Weyl spinor the coefficients $s_i$ become: $s^{\rm f}_{-1} = -4/\omega$,
\bea
s^{\rm f}_0(\omega,N,\Phi_b) &=& \frac1\omega \left[ \frac13 + \left( \frac13 -2\sum_b\Phi_b^2 \right)\omega^2 \right] \,, \nn\\
s^{\rm f}_1(\omega,N,\Phi_b) &=& \frac1\omega\Bigg[ \frac7{360} -\frac{\omega\cN\Phi}2 - \frac{\cN^2}{3} + \left(\frac1{36} -\frac{1}6\sum_b\Phi_b^2  \right)\omega^2  \nn\\
&&\qquad -\frac{\omega^3\cN}{6}\sum_b \Phi_b (1-4\Phi_b^2) + \Bigg(\frac7{360} -\frac1{6}\sum_b \Phi_b^2 (1-2\Phi_b^2) \Bigg)\omega^4 \Bigg] \,, \\
s^{\rm f}_2(\omega,N,\Phi_b) &=& \frac1\omega \Bigg[ \frac{31}{10080} -\frac{\omega\cN\Phi}{16} - \frac{31\,\cN^2}{720}  + \left(\frac7{1440} -\frac{\cN^2}{72}\bigg(1-6\sum_b\Phi_b^2 \bigg) - \frac7{240} \sum_b\Phi_b^2 \right) \omega^2   \nn\\
&& -\frac{\omega^3\cN}{24}\sum_b \Phi_b (1-4\Phi^2_b)  + \Bigg(\frac{7}{1440} -\frac{7\,\cN^2}{720} -\frac{(1-2\cN^2)}{24}\sum_b \Phi_b^2(1-2\Phi_b^2) \Bigg) \omega^4 \nn\\
&&  -\omega^5\cN\left( \sum_b \Phi_b \bigg( \frac7{240} -\frac{\Phi_b^2}6 +\frac{\Phi_b^4}5\bigg) \right)  \nn\\
&&  + \Bigg(\frac{31}{10080} -\sum_b \Phi_b^2\left(\frac7{240} -\frac{\Phi_b^2}{12} + \frac{\Phi_b^4}{15} \right) \Bigg)\omega^6 \Bigg] \,.\nn
\eea

In the limit $\omega \to 1$, $\Phi_b \to 0$ one finds $s_{-1}^{\rm sph} = -4$,
\be
 s_0^{\rm sph} =  \frac23 \,, \quad
 s_1^{\rm sph,\,0} = \frac{1}{15} \,, \quad
 s_1^{\rm sph,\,2} = - \frac{\cN^2}3 \,, \quad
 s_2^{\rm sph,\,0} = \frac{1}{63}  \quad
 \hbox{and} \quad
 s_2^{\rm sph,\,2} =  - \frac{\cN^2}{15}  \,,
\ee
in agreement with Gilkey-de Witt methods and those found in \cite{KandM} for fermions on a sphere.

For a fermion with charge $q \gB$, the corresponding contributions to the running of the bulk couplings are
\bea
 \mu\, \frac{\partial U}{\partial\mu} = \frac{2m^6}{3(4\pi)^3} \,,\quad&& \mu\,\frac{\partial}{\partial\mu} \left(\frac{1}{\kappa^2}\right) = -\frac{m^4}{3(4\pi)^3} \,,\nn\\
 \mu\,\frac{\partial}{\partial\mu}\left(\frac{\zeta_{\ssR^2}}\kappa\right) = -\frac{m^2}{60(4\pi)^3} \,,\quad&& \mu\,\frac{\partial \zeta_{\ssR^3}}{\partial\mu} = -\frac1{504(4\pi)^3} \,,\\
 \mu\, \frac{\partial}{\partial\mu} \left(\frac{1}{\gB^2}\right) = \frac{8\,q^2m^2}{3(4\pi)^3} \,,\quad&& \mu\,\frac{\partial}{\partial\mu}\left(\frac{\kappa \zeta_{\ssA\ssR}}{\gB^2}\right) = \frac{8\,q^2}{15(4\pi)^3} \,. \nn
\eea

The quantities relevant for the brane action (when $|\Phi_b| \leq \Phi_0^{\rm f}$) are $\delta s_{-1} = 0$ and
\bea
 \delta s_0^0 &=& \frac1\omega\left(\frac{\omega^2-1}6-2\omega^2\Phi_b^2\right) = \frac{1}{\omega} \left(\frac{\delta\omega}3 + \frac{\delta\omega^2}6 -2\omega^2\Phi_b^2\right)  \,,\nn\\
 \delta s_1^0 &=& \frac1\omega\left( \frac{\omega^2-1}{72} + \frac{7(\omega^4-1)}{720} - \frac{\omega^2\Phi_b^2}6 - \frac{\omega^4 \Phi_b^2(1-2\Phi_b^2)}6\right)  \nn\\
 &=& \frac1\omega\left( \frac{\delta\omega}{15} + \frac{13\,\delta\omega^2}{180} + \frac{7\,\delta\omega^3}{180}+ \frac{7\,\delta\omega^4}{720} - \frac{\omega^2\Phi_b^2}6 - \frac{\omega^4 \Phi_b^2(1-2\Phi_b^2)}6\right)
 \nn\,,\\
 \delta s_1^{\rm 1} &=&-\frac{\cN\Phi_b}2 -\frac{\omega^2\cN\Phi_b}6 (1-4\Phi_b^2) \,,\qquad \delta s_1^{\rm 2} = 0 \,,\\
 \delta s_2^0 &=& \frac1\omega \Bigg[ \frac{7(\omega^2-1)}{2880} + \frac{7(\omega^4-1)}{2880} + \frac{31(\omega^6-1)}{20160} - \frac{7\,\omega^2\Phi_b^2}{240} - \frac{\omega^4}{24} \Phi_b^2 (1-2\Phi_b^2) \nn \nn\\
 &&\qquad-\omega^6 \Phi_b^2 \bigg(\frac7{240} -\frac{\Phi_b^2}{12} + \frac{\Phi_b^4}{15} \bigg)\Bigg] \nn\\
 &=& \frac1\omega \Bigg[ \frac{\delta\omega}{42} + \frac{101\,\delta\omega^2}{2520} + \frac{17\,\delta\omega^3}{420} + \frac{257\,\delta\omega^4}{10080} +\frac{31\delta\omega^5}{3360} + \frac{31\,\delta\omega^6}{20160}  \nn\\
 &&\qquad - \frac{7\,\omega^2\Phi_b^2}{240} - \frac{\omega^4}{24} \Phi_b^2 (1-2\Phi_b^2) -\omega^6 \Phi_b^2 \bigg(\frac7{240} -\frac{\Phi_b^2}{12} + \frac{\Phi_b^4}{15} \bigg)\Bigg]  \,,\nn\\
 \delta s_2^{\rm 1} &=& -\frac{\cN\Phi_b}{16} - \frac{\omega^2\cN\Phi_b}{24}\Big(1-4\Phi_b^2\Big) - \omega^4\cN\Phi_b \left( \frac7{240} -\frac{\Phi_b^2}6 +\frac{\Phi_b^4}5\right) \,,\nn\\
 \delta s_2^{\rm 2}  &=& -\frac{\cN^2}\omega\left[\frac{\omega^2-1}{144} +\frac{7(\omega^4-1)}{1440} - \frac{\omega^2\Phi_b^2}{12} -\frac{\omega^4\Phi_b^2}{12} \Big( 1-2\Phi_b^2 \Big) \right] \nn\\
 &=& -\frac{\cN^2}\omega \left[\frac{\delta\omega}{30} + \frac{13\,\delta\omega^2}{360}+\frac{7\,\delta\omega^3}{360} + \frac{7\,\delta\omega^4}{1440} - \frac{\omega^2\Phi_b^2}{12} -\frac{\omega^4\Phi_b^2}{12} \Big( 1-2\Phi_b^2 \Big)\right]  \,.\nn
\eea
The brane counterterms therefore renormalize as follows:
\bea
 \mu\,\frac{\partial T_b}{\partial\mu} &=& \frac{m^4}{2(4\pi)^2\omega} \left(\frac{\delta\omega}3 + \frac{\delta\omega^2}6 -2\omega^2\Phi_b^2\right) \simeq \frac{m^4}{(4\pi)^2}\left(\frac{\delta\omega}6 - \Phi_b^2\right) \,,\nn\\
  \mu \, \frac{\partial }{\partial \mu} \left(\frac{\cA_b}{\gB^2}\right) &=& -\frac{q\Phi_b \, m^2}{(4\pi)^2} \left(1 +\frac{\omega^2}3 (1-4\Phi_b^2)\right) \simeq  -\frac{8\,q^2m^2}{3(4\pi)^3} \cA_b \,,\nn\\
 \mu\,\frac\partial{\partial\mu} \left(\frac{\zeta_{\ssR b}}{\kappa}\right) &=& \frac{m^2}{2(4\pi)^2\omega} \left( \frac{\delta\omega}{15} + \frac{13\,\delta\omega^2}{180} + \frac{7\,\delta\omega^3}{180}+ \frac{7\,\delta\omega^4}{720} - \frac{\omega^2\Phi_b^2}6 - \frac{\omega^4 \Phi_b^2(1-2\Phi_b^2)}6\right) \nn\\
 &\simeq& \frac{m^2}{2(4\pi)^2} \left(\frac{\delta\omega}{15} - \frac{\Phi_b^2}3\right)  \,,\\
 \mu \, \frac{\partial }{\partial \mu} \left(\frac{\zeta_{\tilde\ssA\ssR b}}{\gB^2}\right) &=& -\frac{q\Phi_b }{(4\pi)^2} \left[ \frac{1}{16} + \frac{\omega^2}{24}\Big(1-4\Phi_b^2\Big) + \omega^4 \left( \frac7{240} -\frac{\Phi_b^2}6 +\frac{\Phi_b^4}5\right)  \right] \nn\\
 &\simeq& -\frac{4\,q^2}{15(4\pi)^3} \cA_b \,,\nn\\
 \mu \, \frac{\partial \zeta_{\ssR^2 b}}{\partial\mu} &=& \frac{1}{4(4\pi)^2\omega} \Bigg[ \frac{\delta\omega}{42} + \frac{101\,\delta\omega^2}{2520} + \frac{17\,\delta\omega^3}{420} + \frac{257\,\delta\omega^4}{10080} +\frac{31\delta\omega^5}{3360} + \frac{31\,\delta\omega^6}{20160}  \nn\\
 &&\qquad - \frac{7\,\omega^2\Phi_b^2}{240} - \frac{\omega^4}{24} \Phi_b^2 (1-2\Phi_b^2) -\omega^6 \Phi_b^2 \bigg(\frac7{240} -\frac{\Phi_b^2}{12} + \frac{\Phi_b^4}{15} \bigg)\Bigg] \nn \\
 &\simeq& \frac{1}{4(4\pi)^2}\left(\frac{\delta\omega}{42} - \frac{\Phi_b^2}{10}\right) \,,\nn\\
  \mu \, \frac{\partial }{\partial \mu} \left(\frac{\zeta_{\ssA b}}{\gB^2}\right) &=& -\frac{8\, q^2}{(4\pi)^2\omega} \left[\frac{\delta\omega}{30} + \frac{13\,\delta\omega^2}{360}+\frac{7\,\delta\omega^3}{360} + \frac{7\,\delta\omega^4}{1440} - \frac{\omega^2\Phi_b^2}{12} -\frac{\omega^4\Phi_b^2}{12} \Big( 1-2\Phi_b^2 \Big)\right]  \nn\\
&\simeq& -\frac{4\, q^2}{3(4\pi)^2}\left( \frac{\delta\omega}{5} - \Phi_b^2 \right)  \,.\nn
\eea

\subsection*{Spin one}

We next state the results for the Casimir coefficient for a gauge field, provided this gauge field is {\em not} the field whose flux stabilizes the background 2D geometry. We consider in turn the cases where the 6D gauge field is massless or massive (in the 6D sense).

\subsubsection*{Massless spin one}

We begin with the massless case. Defining
\be
 N_{\rm gf\xi} := N  \,,\quad \Phi_b^{\rm gf\xi} := \Phi_b - \xi \Phi_0^{\rm gf} \,, \quad \Phi_0^{\rm gf} = \omega^{-1} = \alpha = 1 - \frac{\delta}{2\pi}
\ee
one obtains the following contributions to the bulk divergences
\bea
s_{-1}^{\rm sph} = 4 \,,\quad s_0^{\rm sph} &=& -\frac23 \,,\quad s_1^{\rm sph,\,0} = \frac4{15} \,,\quad s_1^{\rm sph,\,2} = \frac{5\,\cN^2}6 \nn\,,\\
 s_2^{\rm sph,\,0} &=& \frac{16}{315} \,,\quad s_2^{\rm sph,\,2} = -\frac{7\,\cN^2}{20}
\eea
and after these are subtracted the brane renormalizations are obtained from
\bea
\delta s_0 &=& \frac1\omega \left( - (\omega-1) +\frac{(\omega^2-1)}3 -2\omega^2 F_b + \omega^2 |\Phi_b| \right) =\frac1\omega \left(-\frac{\delta\omega}3 + \frac{\delta\omega^2}3 -2\omega^2 F_b + \omega^2 |\Phi_b| \right)  \,,\nn\\
\delta s_1^0 &=& \frac1\omega \left( \frac{(\omega^2-1)}9 + \frac{\omega^4-1}{90} - \frac{2\,\omega^2F_b}3 +\frac{\omega^2 |\Phi_b|}3 - \frac{\omega^4F_b^2}3 -\frac{\omega^4 |\Phi_b|^3}3  \right)  \nn\\
 &=& \frac1\omega \left(\frac{4\,\delta\omega}{15} + \frac{8\,\delta\omega^2}{45} +\frac{2\,\delta\omega^3}{45} + \frac{\delta\omega^4}{90} - \frac{2\,\omega^2F_b}3 +\frac{\omega^2 |\Phi_b|}3 - \frac{\omega^4F_b^2}3 -\frac{\omega^4 |\Phi_b|^3}3 \right)  \,,\\
\delta s_1^{\rm 1} &=& -\frac{\omega^2\cN}3 \Phi_b G(|\Phi_b|) + \cN \Phi_b - \frac{\omega^2\cN}2 \Phi_b |\Phi_b|  \,,\nn\\
\delta s_2^0 &=& \frac1\omega\bigg( \frac{(\omega^2-1)}{45} + \frac{\omega^4-1}{180} + \frac{\omega^6-1}{630} -\frac{2\, \omega^2 F_b}{15} +\frac{\omega^2 |\Phi_b|}{15} -\frac{\omega^4F_b^2}{6}  - \frac{\omega^4|\Phi_b|^3}6 \nn\\
&&\qquad -\frac{\omega^6 F^2_b}{30} - \frac{\omega^6 F^3_b}{15}   + \frac{\omega^6|\Phi_b|^5}{10} \bigg) \nn\\
 &=& \frac1\omega\bigg(\frac{8\,\delta\omega}{105}+\frac{5\,\delta\omega^2}{63}+\frac{17\,\delta\omega^3}{315} + \frac{37\,\delta\omega^4}{1260}+\frac{\delta\omega^5}{105} + \frac{\delta\omega^6}{630} -\frac{2\, \omega^2 F_b}{15} +\frac{\omega^2 |\Phi_b|}{15} -\frac{\omega^4F_b^2}{6}  \nn\\
 &&\qquad - \frac{\omega^4|\Phi_b|^3}6 -\frac{\omega^6 F^2_b}{30} - \frac{\omega^6 F^3_b}{15}   + \frac{\omega^6|\Phi_b|^5}{10} \bigg)  \,,\nn\\
\delta s_2^{\rm 1} &=&  -\frac{\omega^2 \cN}{6}  \Phi_b \, G(|\Phi_b|) -\frac{\omega^4 \cN}{30}  \Phi_b \, G(|\Phi_b|) (1+3F_b) -\frac{\omega^2\cN\Phi_b}4 |\Phi_b| + \frac{\omega^4\cN\Phi_b}4 |\Phi_b|^3  \,,\nn\\
 \delta s_2^{\rm 2} &=& -\frac{\cN^2}\omega \left(\frac{\omega -1}{8} + \frac{\omega^2-1}{72} +
 \frac{\omega^4-1}{180} -\frac{\omega^2 F_b}{12} + \frac{\omega^2|\Phi_b|}{24} -\frac{\omega^4 F_b^2}6 -\frac{\omega^4|\Phi_b|^3}6 \right) \nn\\
 &=& -\frac{\cN^2}\omega\left(  \frac{7\, \delta\omega}{40} +\frac{17\,\delta\omega^2}{360}
 +\frac{\delta\omega^3}{45} +\frac{\delta\omega^4}{180} -\frac{\omega^2 F_b}{12} + \frac{\omega^2|\Phi_b|}{24} -\frac{\omega^4 F_b^2}6 -\frac{\omega^4|\Phi_b|^3}6 \right) \nn
\eea
along with $\delta s_{-1} = \delta s_1^{\rm 2} = 0$ (as usual).

Because the renormalizations coming from $s_k$ are proportional to $m^{4-2k}$, where $m$ is the 6D mass, for massless fields we need only follow the contributions of $s_2$, ensuring the only nonzero renormalizations are
\be
 \mu\,\frac{\partial \zeta_{\ssR^3}}{\partial\mu} = -\frac2{315(4\pi)^3} \quad \hbox{and} \quad \mu\,\frac{\partial}{\partial\mu}\left(\frac{\kappa \zeta_{\ssA\ssR}}{\gB^2}\right) = \frac{14\,q^2}{5(4\pi)^3} \,,
\ee
in the bulk, and
\bea
 \mu \, \frac{\partial \zeta_{\ssR^2 b}}{\partial\mu} &=& \frac{1}{4(4\pi)^2\omega} \bigg(\frac{8\,\delta\omega}{105}+\frac{5\,\delta\omega^2}{63}+\frac{17\,\delta\omega^3}{315} + \frac{37\,\delta\omega^4}{1260}+\frac{\delta\omega^5}{105} + \frac{\delta\omega^6}{630}    \nn\\
 &&\qquad -\frac{2\, \omega^2 F_b}{15} +\frac{\omega^2 |\Phi_b|}{15} -\frac{\omega^4F_b^2}{6} - \frac{\omega^4|\Phi_b|^3}6 -\frac{\omega^6 F^2_b}{30} - \frac{\omega^6 F^3_b}{15}   + \frac{\omega^6|\Phi_b|^5}{10} \bigg) \nn\\
 &\simeq& \frac1{(4\pi)^2} \left(\frac{2\,\delta\omega}{105} -\frac{|\Phi_b|}{60} \right) \,,\nn\\
 \mu \, \frac{\pd}{\pd\mu}\left(\frac{\kappa \zeta_{\tilde \ssA\ssR b}}{\gB^2}\right) &=& -\frac{q}{(4\pi)^2} \left( \frac{\omega^2}{6}  \Phi_b \, G(|\Phi_b|) +\frac{\omega^4}{30}  \Phi_b \, G(|\Phi_b|) (1+3F_b) +\frac{\omega^2\Phi_b}4 |\Phi_b| - \frac{\omega^4\Phi_b}4 |\Phi_b|^3 \right) \nn\\
 &\simeq& -\frac{q \,\Phi_b}{5(4\pi)^2} = -\frac{2\,q^2}{5(4\pi)^3} \cA_b  \,,\\
 \mu \, \frac{\partial }{\partial \mu} \left(\frac{\zeta_{\ssA b}}{\gB^2}\right) &=&
 -\frac{8\, q^2}{(4\pi)^2\omega} \bigg(  \frac{7\, \delta\omega}{40} +\frac{17\,\delta\omega^2}{360}
 +\frac{\delta\omega^3}{45} +\frac{\delta\omega^4}{180} -\frac{\omega^2 F_b}{12} + \frac{\omega^2|\Phi_b|}{24}  \nn\\
 &&\qquad -\frac{\omega^4 F_b^2}6 -\frac{\omega^4|\Phi_b|^3}6 \bigg) \simeq -\frac{q^2}{(4\pi)^2} \left(\frac{7\,\delta\omega}{5} -\frac{|\Phi_b|}{3}\right)  \,,\nn
\eea
on the brane.

\subsubsection*{Massive spin one}

In this case, in the sphere limit we have
\bea
 s_{-1}^{\rm sph,\, 0} = 5 \,, \quad
 s_{0}^{\rm sph,\, 0} &=& -\frac13 \,, \quad
 s_{1}^{\rm sph,\,0} = \frac{1}{3} \,, \quad s_{1}^{\rm sph,\,2} = \frac{19\, \cN^2}{24} \,, \nn \\
  s_{2}^{\rm sph,\,0} &=& \frac{4}{63} \,,
 \quad \mbox{and}\quad s_{2}^{\rm sph,\, 2} = - \frac{3\, \cN^2}{8}
\eea
and so we obtain the bulk renormalizations
\bea
 \mu \, \frac{\partial U}{\partial \mu} = -\frac{5\, m^6}{6 (4 \pi)^3} \,,\quad&&
 \mu \, \frac{\partial}{\partial \mu} \left( \frac{1}{\kappa^2} \right) = \frac{m^4}{6 (4\pi)^3} \,,\nn\\
 \mu\,\frac{\partial }{\partial\mu}\left(\frac{\zeta_{\ssR^2}}\kappa\right) = -\frac{m^2}{12(4\pi)^3} \,,\quad
 && \mu \, \frac{\partial \zeta_{\ssR^3}}{\partial\mu} = -\frac1{126(4\pi)^3}\, ,  \\
 \mu \, \frac{\partial}{\partial\mu}\left(\frac{1}{\gB^2}\right) = -\frac{19\,q^2m^2}{3(4\pi)^3} \,,\quad
 && \mu \, \frac{\partial }{\partial\mu}\left(\frac{\kappa  \zeta_{\ssA \ssR}}{\tilde{g}^2}\right)= \frac{3\, q^2}{(4\pi)^3} \,.\nn
 \eea

The running of the brane couplings is similarly obtained by computing the $\delta s_i$ coefficients:
 \bea
 \delta s_0 &=& \frac{1}{\omega}\left( -\frac{\delta\omega}6 + \frac{5\,\delta\omega^2}{12} - \frac{5\,\omega^2F_b}2 +\omega^2 |\Phi_b|  \right) \,,\nn\\
 \delta s^0_1 &=& \frac1\omega\left(\frac{\delta\omega}{3} + \frac{2\,\delta\omega^2}{9} + \frac{\delta\omega^3}{18} + \frac{\delta\omega^4}{72} - \frac{5\,\omega^2 F_b}6 +\frac{\omega^2 |\Phi_b|}3 - \frac{5\,\omega^4 F_b^2}{12} -\frac{\omega^4 |\Phi_b|^3}3 \right)\nn\\
 \delta s^{\rm 1}_1 &=& -\frac{5\,\omega^2\cN}{12} \Phi_b \, G(|\Phi_b|) + \cN \Phi_b - \frac{\omega^2\cN}2 \Phi_b |\Phi_b| \,,\nn\\
 %
 %
 \delta s^0_2 &=& \frac1\omega\bigg( \frac{2\,\delta\omega}{21} + \frac{25\,\delta\omega^2}{252} + \frac{17\,\delta\omega^3}{252}
+ \frac{37\,\delta\omega^4}{1008} + \frac{\delta\omega^5}{84} + \frac{\delta\omega^6}{504} -\frac{\omega^2 F_b}6 +\frac{\omega^2|\Phi_b|}{15} \nn\\
&&\qquad -\frac{5\,\omega^4 F_b^2}{24} - \frac{\omega^4|\Phi_b|^3}6 - \frac{\omega^6 F_b^2}{24} -\frac{\omega^6 F_b^3}{12} +\frac{\omega^6|\Phi_b|^5}{10} \bigg) \, , \\
 \delta s^{\rm 1}_2 &=& -\frac{5\,\omega^2 \cN}{24}  \Phi_b \, G(|\Phi_b|) -\frac{\omega^4 \cN}{24}  \Phi_b \, G(|\Phi_b|) (1+3F_b) -\frac{\omega^2\cN\Phi_b}4 |\Phi_b| + \frac{\omega^4\cN\Phi_b}4 |\Phi_b|^3  \,,\nn\\
\delta s^{\rm 2}_2 &=& -\frac{\cN^2}\omega \left(  \frac{3\,\delta\omega}{16} + \frac{17\,\delta\omega^2}{288} + \frac{\delta\omega^3}{36} + \frac{\delta\omega^4}{144} -\frac{5\,\omega^2 F_b}{48} + \frac{\omega^2|\Phi_b|}{24} -\frac{5\,\omega^4 F_b^2}{24} -\frac{\omega^4|\Phi_b|^3}6 \right) \,.\nn
\eea
These give
\bea
 \mu \, \frac{\partial T_b}{\partial \mu} &=& \frac{m^4}{2 (4\pi)^2 \omega} \left( -\frac{\delta\omega}6 + \frac{5\,\delta\omega^2}{12} - \frac{5\,\omega^2F_b}2 +\omega^2 |\Phi_b|  \right)  \,, \nn\\
 \mu\,\frac{\pd}{\pd\mu} \left(\frac{\cA_b}{\gB^2}\right) &=& \frac{2\, q m^2}{(4\pi)^2} \left( -\frac{5\,\omega^2}{12} \Phi_b\, G(|\Phi_b|) + \Phi_b - \frac{\omega^2}2 \Phi_b |\Phi_b| \right)\,, \nn\\
  \mu \, \frac{\partial }{\partial \mu} \left(\frac{\zeta_{\ssR b}}{\kappa}\right) &=&
  \frac{m^2}{2(4\pi)^2\omega} \bigg(\frac{\delta\omega}{3} + \frac{2\,\delta\omega^2}{9} + \frac{\delta\omega^3}{18} + \frac{\delta\omega^4}{72} - \frac{5\,\omega^2 F_b}6 +\frac{\omega^2 |\Phi_b|}3 \nn \nn\\
  &&\qquad - \frac{5\,\omega^4 F_b^2}{12} -\frac{\omega^4 |\Phi_b|^3}3 \bigg)   \,,\nn\\
  \mu\,\frac{\pd}{\pd\mu}\left(\frac{\kappa \zeta_{\tilde \ssA\ssR b}}{\gB^2}\right) &=& -\frac{q}{(4\pi)^2} \bigg(\frac{5\,\omega^2}{24}  \Phi_b \, G(|\Phi_b|) +\frac{\omega^4}{24}  \Phi_b \, G(|\Phi_b|) (1+3F_b) \nn\\
  &&\qquad +\frac{\omega^2\Phi_b}4 |\Phi_b| - \frac{\omega^4\Phi_b}4 |\Phi_b|^3\bigg) \,,\\
 \mu \, \frac{\partial \zeta_{\ssR^2 b}}{\partial\mu} &=& \frac{1}{4(4\pi)^2\omega} \bigg( \frac{2\,\delta\omega}{21} + \frac{25\,\delta\omega^2}{252} + \frac{17\,\delta\omega^3}{252}
+ \frac{37\,\delta\omega^4}{1008} + \frac{\delta\omega^5}{84} + \frac{\delta\omega^6}{504} -\frac{\omega^2 F_b}6 +\frac{\omega^2|\Phi_b|}{15} \nn\\
&&\qquad -\frac{5\,\omega^4 F_b^2}{24} - \frac{\omega^4|\Phi_b|^3}6 - \frac{\omega^6 F_b^2}{24} -\frac{\omega^6 F_b^3}{12} +\frac{\omega^6|\Phi_b|^5}{10} \bigg)  \,,\nn\\
 \mu \, \frac{\partial }{\partial \mu} \left(\frac{\zeta_{\ssA b}}{\gB^2}\right) &=&
 -\frac{8\, q^2}{(4\pi)^2\omega} \bigg(  \frac{3\,\delta\omega}{16} + \frac{17\,\delta\omega^2}{288} + \frac{\delta\omega^3}{36} + \frac{\delta\omega^4}{144} -\frac{5\,\omega^2 F_b}{48} + \frac{\omega^2|\Phi_b|}{24} \nn\\
 &&\qquad -\frac{5\,\omega^4 F_b^2}{24} -\frac{\omega^4|\Phi_b|^3}6 \bigg) \,.\nn
\eea

\section{Complete results for the massive multiplet}
\label{app:massiveresults}

In this appendix, we compile the complete result for the renormalization of the south ($b=-1$) brane when $\eta$ is not constrained to be $\eta\leq\delta\omega/2$, but instead $0\leq\eta\leq1$. At the end, we state the resulting beta function for $\cV_{\rm branes}:=\sum_b \cV_b$.

Given that $\delta s_i^{\rm mm} = \delta s_i^{\rm hm} + \delta s_i^{\rm gm}$, and using the results from the main text for the hyper- and gauge multiplets, we find that, on the south ($b=-1$) brane,
\bea
\delta s_0 &=&  \left\{
\begin{array}{l}
2\eta \,,\quad \eta \leq \delta\omega/2  \\
-2 \hat\eta \,,\quad \delta\omega/2 \leq \eta \leq \delta\omega  \\
0 \,,\quad \eta \geq \delta\omega
\end{array} \right. \\
\delta s_1^0 &=& \left\{
\begin{array}{l}
\frac1\omega\left[ \frac{\delta\omega}2 +\frac{\delta\omega^2}4 +\frac{F_{-\eta}}2 + \frac{\omega F_{-\eta}}3 (1+2\eta) -\frac{\omega^2 F_{-\eta}}2 \right] \,,\quad \eta \leq \delta\omega/2  \\
\frac1\omega\left[ \frac{\delta\omega}2 +\frac{\delta\omega^2}4 +\frac{F_{\hat\eta}}2 + \frac{\omega F_{\hat\eta}}3 (1-2\hat\eta) -\frac{\omega^2 F_{\hat\eta}}2 \right] \,,\quad \delta\omega/2 \leq \eta \leq \delta\omega  \\
\frac1\omega\left[ \frac{\delta\omega}2 +\frac{\delta\omega^2}4 +\frac{F_{\hat\eta}}2  -\frac{\omega^2 F_{\hat\eta}}2 \right] \,,\quad \eta \geq \delta\omega
\end{array} \right. \\
\delta s_1^{\rm 1} &=& \left\{
\begin{array}{l}
\frac1\omega\left[ \Big( -\frac{\delta\omega}2 -\frac{\delta\omega^2}4 \Big)(1+2\eta) - \omega F_{-\eta} \right] \,,\quad \eta \leq \delta\omega/2  \\
-\frac1\omega\left[ \Big( -\frac{\delta\omega}2 -\frac{\delta\omega^2}4 \Big)(1-2\hat\eta) - \omega F_{\hat\eta} \right] \,,\quad \delta\omega/2 \leq \eta \leq \delta\omega  \\
-\frac1\omega\left[ \Big( -\frac{\delta\omega}2 -\frac{\delta\omega^2}4 \Big)(1-2\hat\eta)  \right] \,,\quad \eta \geq \delta\omega
\end{array} \right. \\
\delta s_2^0 &=& \left\{
\begin{array}{l}
\frac1\omega\Big[ \frac{\delta\omega}8 + \frac{7\,\delta\omega^2}{48} + \frac{\delta\omega^3}{12} + \frac{\delta\omega^4}{48} -\frac{F_{-\eta}}{16} + \frac{F_{-\eta}^2}{12} + \Big( \frac{F_{-\eta}}{30} +\frac{F_{-\eta}^2}{10}\Big)\omega(1+2\eta)  \\
\qquad + \Big( \frac{F_{-\eta}}{24} -\frac{F_{-\eta}^2}4 \Big) \omega^2 -\frac{\omega^4 F_{-\eta}}{16} \Big] \,,\quad \eta \leq \delta\omega/2  \\
\frac1\omega\Big[ \frac{\delta\omega}8 + \frac{7\,\delta\omega^2}{48} + \frac{\delta\omega^3}{12} + \frac{\delta\omega^4}{48} -\frac{F_{\hat\eta}}{16} + \frac{F_{\hat\eta}^2}{12} + \Big( \frac{F_{\hat\eta}}{30} +\frac{F_{\hat\eta}^2}{10}\Big)\omega(1-2\hat\eta)  \\
\qquad + \Big( \frac{F_{\hat\eta}}{24} -\frac{F_{\hat\eta}^2}4 \Big) \omega^2 -\frac{\omega^4 F_{\hat\eta}}{16} \Big] \,,\quad \delta\omega/2 \leq \eta \leq \delta\omega  \\
\frac1\omega\Big[ \frac{\delta\omega}8 + \frac{7\,\delta\omega^2}{48} + \frac{\delta\omega^3}{12} + \frac{\delta\omega^4}{48} -\frac{F_{\hat\eta}}{16} + \frac{F_{\hat\eta}^2}{12}  + \Big( \frac{F_{\hat\eta}}{24} -\frac{F_{\hat\eta}^2}4 \Big) \omega^2 -\frac{\omega^4 F_{\hat\eta}}{16} \Big] \,,\quad \eta \geq \delta\omega
\end{array} \right. \\
\delta s_2^{\rm 1} &=& \left\{
\begin{array}{l}
\frac1\omega\Big[ \Big(-\frac{\delta\omega}{12} -\frac{\delta\omega^2}6 - \frac{\delta\omega^3}8 - \frac{\delta\omega^4}{32} + \frac{F_{-\eta}}{12} -\frac{\omega^2 F_{-\eta}}4 \Big) (1+2\eta) -\frac{\omega F_{-\eta}^2}2 \Big] \,,\quad \eta \leq \delta\omega/2  \\
-\frac1\omega\Big[ \Big(-\frac{\delta\omega}{12} -\frac{\delta\omega^2}6 - \frac{\delta\omega^3}8 - \frac{\delta\omega^4}{32} + \frac{F_{\hat\eta}}{12} -\frac{\omega^2 F_{\hat\eta}}4 \Big) (1-2\hat\eta) \\
\qquad-\frac{\omega F_{\hat\eta}^2}2 \Big] \,,\quad \delta\omega/2 \leq \eta \leq \delta\omega  \\
-\frac1\omega\Big[ \Big(-\frac{\delta\omega}{12} -\frac{\delta\omega^2}6 - \frac{\delta\omega^3}8 - \frac{\delta\omega^4}{32} + \frac{F_{\hat\eta}}{12} -\frac{\omega^2 F_{\hat\eta}}4 \Big) (1-2\hat\eta)  \Big] \,,\quad \eta \geq \delta\omega
\end{array} \right. \\
\delta s_2^{\rm 2} &=& \left\{
\begin{array}{l}
\frac1\omega\left[ -\frac{\delta\omega}{24} + \frac{\delta\omega^2}{48} + \frac{\delta\omega^3}{24} + \frac{\delta\omega^4}{96} -\frac{\omega F_{-\eta}}6 (1+2\eta) + \frac{\omega^2 F_{-\eta}}4 \right] \,,\quad \eta \leq \delta\omega/2  \\
\frac1\omega\left[ -\frac{\delta\omega}{24} + \frac{\delta\omega^2}{48} + \frac{\delta\omega^3}{24} + \frac{\delta\omega^4}{96} -\frac{\omega F_{\hat\eta}}6 (1-2\hat\eta) + \frac{\omega^2 F_{\hat\eta}}4 \right] \,,\quad \delta\omega/2 \leq \eta \leq \delta\omega  \\
\frac1\omega\left[ -\frac{\delta\omega}{24} + \frac{\delta\omega^2}{48} + \frac{\delta\omega^3}{24} + \frac{\delta\omega^4}{96}  + \frac{\omega^2 F_{\hat\eta}}4 \right] \,,\quad \eta \geq \delta\omega
\end{array} \right.
\eea
which give the following renormalizations (recall that $\hat \eta:= \eta-\delta\omega$):
\bea
\mu \frac{\pd T_-}{\pd\mu} &=& \frac{m^4}{2(4\pi)^2} \times \left\{
\begin{array}{l}
2\eta \,,\quad \eta \leq \delta\omega/2  \\
-2 \hat\eta \,,\quad \delta\omega/2 \leq \eta \leq \delta\omega  \\
0 \,,\quad \eta \geq \delta\omega
\end{array} \right. \\
\mu \frac\pd{\pd\mu}\left( \frac{\zeta_{\ssR -}}\kappa \right) &=&  \frac{m^2}{2(4\pi)^2\omega} \times \left\{
\begin{array}{l}
 \frac{\delta\omega}2 +\frac{\delta\omega^2}4 +\frac{F_{-\eta}}2 + \frac{\omega F_{-\eta}}3 (1+2\eta) -\frac{\omega^2 F_{-\eta}}2 \,,\quad \eta \leq \delta\omega/2  \\
 \frac{\delta\omega}2 +\frac{\delta\omega^2}4 +\frac{F_{\hat\eta}}2 + \frac{\omega F_{\hat\eta}}3 (1-2\hat\eta) -\frac{\omega^2 F_{\hat\eta}}2  \,,\quad \delta\omega/2 \leq \eta \leq \delta\omega \\
 \frac{\delta\omega}2 +\frac{\delta\omega^2}4 +\frac{F_{\hat\eta}}2  -\frac{\omega^2 F_{\hat\eta}}2 \,,\quad \eta \geq \delta\omega
\end{array} \right. \nn\\
&&\\
\mu \frac\pd{\pd\mu}\left( \frac{\zeta_{\tilde \ssA -}}{\gB^2} \right) &=& \frac{2\, m^2}{(4\pi)^2\omega} \times \left\{
\begin{array}{l}
 \Big( -\frac{\delta\omega}2 -\frac{\delta\omega^2}4 \Big)(1+2\eta) - \omega F_{-\eta}  \,,\quad \eta \leq \delta\omega/2  \\
- \Big( -\frac{\delta\omega}2 -\frac{\delta\omega^2}4 \Big)(1-2\hat\eta) + \omega F_{\hat\eta} \,,\quad \delta\omega/2 \leq \eta \leq \delta\omega  \\
- \Big( -\frac{\delta\omega}2 -\frac{\delta\omega^2}4 \Big)(1-2\hat\eta)  \,,\quad \eta \geq \delta\omega
\end{array} \right.  \\
\mu \frac{\pd \zeta_{\ssR^2 -}}{\pd \mu} &=& \frac1{4(4\pi)^2\omega} \times \left\{
\begin{array}{l}
\frac{\delta\omega}8 + \frac{7\,\delta\omega^2}{48} + \frac{\delta\omega^3}{12} + \frac{\delta\omega^4}{48} -\frac{F_{-\eta}}{16} + \frac{F_{-\eta}^2}{12}   \\
\qquad + \Big( \frac{F_{-\eta}}{30} +\frac{F_{-\eta}^2}{10}\Big)\omega(1+2\eta) + \Big( \frac{F_{-\eta}}{24} -\frac{F_{-\eta}^2}4 \Big) \omega^2 \\
\qquad-\frac{\omega^4 F_{-\eta}}{16} \,,\quad \eta \leq \delta\omega/2  \\
\frac{\delta\omega}8 + \frac{7\,\delta\omega^2}{48} + \frac{\delta\omega^3}{12} + \frac{\delta\omega^4}{48} -\frac{F_{\hat\eta}}{16} + \frac{F_{\hat\eta}^2}{12} + \Big( \frac{F_{\hat\eta}}{30} +\frac{F_{\hat\eta}^2}{10}\Big)\omega(1-2\hat\eta)  \\
\qquad + \Big( \frac{F_{\hat\eta}}{24} -\frac{F_{\hat\eta}^2}4 \Big) \omega^2 -\frac{\omega^4 F_{\hat\eta}}{16}  \,,\quad \delta\omega/2 \leq \eta \leq \delta\omega  \\
\frac{\delta\omega}8 + \frac{7\,\delta\omega^2}{48} + \frac{\delta\omega^3}{12} + \frac{\delta\omega^4}{48} -\frac{F_{\hat\eta}}{16} + \frac{F_{\hat\eta}^2}{12} \\
\qquad + \Big( \frac{F_{\hat\eta}}{24} -\frac{F_{\hat\eta}^2}4 \Big) \omega^2 -\frac{\omega^4 F_{\hat\eta}}{16} \,,\quad \eta \geq \delta\omega
\end{array} \right. \nn\\
&&\\
\mu \frac\pd{\pd\mu}\left( \frac{\kappa \zeta_{\tilde \ssA\ssR -}}{\gB^2}\right)  &=& \frac{1}{(4\pi)^2\omega} \times \left\{
\begin{array}{l}
\Big(-\frac{\delta\omega}{12} -\frac{\delta\omega^2}6 - \frac{\delta\omega^3}8 - \frac{\delta\omega^4}{32} + \frac{F_{-\eta}}{12} -\frac{\omega^2 F_{-\eta}}4 \Big) (1+2\eta) \\
\qquad -\frac{\omega F_{-\eta}^2}2  \,,\quad \eta \leq \delta\omega/2  \\
- \Big(-\frac{\delta\omega}{12} -\frac{\delta\omega^2}6 - \frac{\delta\omega^3}8 - \frac{\delta\omega^4}{32} + \frac{F_{\hat\eta}}{12} -\frac{\omega^2 F_{\hat\eta}}4 \Big) (1-2\hat\eta) \\
\qquad+\frac{\omega F_{\hat\eta}^2}2 \,,\quad \delta\omega/2 \leq \eta \leq \delta\omega  \\
- \Big(-\frac{\delta\omega}{12} -\frac{\delta\omega^2}6 - \frac{\delta\omega^3}8 - \frac{\delta\omega^4}{32} + \frac{F_{\hat\eta}}{12} -\frac{\omega^2 F_{\hat\eta}}4 \Big) (1-2\hat\eta) \,,\,\,\, \eta \geq \delta\omega
\end{array} \right. \\
\mu \frac{\pd}{\pd\mu} \left( \frac{\kappa \zeta_{\ssA -}}{\gB^2} \right) &=& \frac{8}{(4\pi)^2\omega} \times \left\{
\begin{array}{l}
-\frac{\delta\omega}{24} + \frac{\delta\omega^2}{48} + \frac{\delta\omega^3}{24} + \frac{\delta\omega^4}{96} -\frac{\omega F_{-\eta}}6 (1+2\eta) \\
\qquad + \frac{\omega^2 F_{-\eta}}4  \,,\quad \eta \leq \delta\omega/2  \\
-\frac{\delta\omega}{24} + \frac{\delta\omega^2}{48} + \frac{\delta\omega^3}{24} + \frac{\delta\omega^4}{96} -\frac{\omega F_{\hat\eta}}6 (1-2\hat\eta) \\
\qquad + \frac{\omega^2 F_{\hat\eta}}4 \,,\quad \delta\omega/2 \leq \eta \leq \delta\omega  \\
-\frac{\delta\omega}{24} + \frac{\delta\omega^2}{48} + \frac{\delta\omega^3}{24} + \frac{\delta\omega^4}{96}  + \frac{\omega^2 F_{\hat\eta}}4 \,,\quad \eta \geq \delta\omega
\end{array} \right.
\eea
Therefore, the total contribution of a massive matter multiplet to the running of $\cV_b$ on the south brane is
\bea
 \mu\,\frac{\partial \cV_-}{\partial\mu} &=& \mu\,\frac{\pd T_-}{\pd\mu} -\frac1{2\, r^2} \left[\mu \frac\pd{\pd\mu} \left( \frac{\zeta_{\tilde \ssA -}}{\gB^2} \right) \right] - \frac2{r^2} \left[\mu \frac\pd{\pd\mu} \left( \frac{\zeta_{\ssR -}}\kappa \right) \right] + \frac4{r^4} \left[ \mu \frac{\pd \zeta_{\ssR^2 -}}{\pd \mu} \right] \nn\\
&& + \frac1{r^4} \left[ \mu \frac\pd{\pd\mu}\left( \frac{\kappa \zeta_{\tilde \ssA\ssR -}}{\gB^2}\right) \right]   + \frac1{8 r^4} \left[ \mu \frac{\pd}{\pd\mu} \left( \frac{\kappa \zeta_{\ssA -}}{\gB^2} \right) \right] \nn\\
&=& \frac{1}{(4\pi r^2)^2\omega} \times \left\{
\begin{array}{l}
 \left(-\frac5{48} +\frac{5\,\omega^2}{24} + \frac{\omega^4}{16} \right)\eta^2 + \left(\frac1{12} -\frac{\omega^2}4 \right)\eta^4 \\
\qquad -\frac{(\omega^2-1)}2\eta^2(mr)^2 +\omega\eta\Big[ \frac2{15}-\frac{\eta^2}3 +\frac{\eta^4}5 \\
\qquad -\Big( \frac23 -\frac{2\,\eta^2}3 \Big)(mr)^2 + (mr)^4 \Big]  \,,\quad \eta \leq \delta\omega/2  \\
 \\
 \left(-\frac5{48} +\frac{5\,\omega^2}{24} + \frac{\omega^4}{16} \right)(\eta-\omega)^2 + \left(\frac1{12} -\frac{\omega^2}4 \right)(\eta-\omega)^4 \\
\qquad -\frac{(\omega^2-1)}2\eta^2(mr)^2 +\omega\hat\eta\Big[ \frac1{30}+\frac{\hat\eta}2 -\frac{4\,\hat\eta^2}3 +\hat\eta^3 -\frac{\hat\eta^4}5 \\
\qquad - \Big( \frac43 - 2\hat\eta+ \frac{4\,\hat\eta^2}3  \Big)(mr)^2 -(mr)^4 \Big] \,,\quad \delta\omega/2 \leq \eta \leq \delta\omega  \\
\\
\left(-\frac5{48} +\frac{5\,\omega^2}{24} + \frac{\omega^4}{16} \right)(\eta-\omega)^2 + \left(\frac1{12} -\frac{\omega^2}4 \right)(\eta-\omega)^4 \\
\qquad -\frac{(\omega^2-1)}2\eta^2(mr)^2  \,,\quad \eta \geq \delta\omega
\end{array} \right.
\eea

Lastly, we can assemble the total contribution of both branes to $\cV_{\rm branes} = \sum_b \cV_b$:
\bea
\mu\,\frac{\partial \cV_{\rm branes}}{\partial\mu} &=& \frac{1}{(4\pi r^2)^2\omega} \times \left\{
\begin{array}{l}
 \left(-\frac5{24} +\frac{5\,\omega^2}{12} + \frac{\omega^4}{8} \right)\eta^2 + \left(\frac1{6} -\frac{\omega^2}2 \right)\eta^4 \\
\qquad -(\omega^2-1)\eta^2(mr)^2 +\omega\eta\Big[ \frac2{15}-\frac{\eta^2}3 +\frac{\eta^4}5 \\
\qquad -\Big( \frac23 -\frac{2\,\eta^2}3 \Big)(mr)^2 + (mr)^4 \Big]  \,,\quad \eta \leq \delta\omega/2  \\
 \\
-\frac{5\,\omega^2}{96} + \frac{11\,\omega^4}{96} -\frac{\omega^2(\omega^2-1)}4(mr)^2 \\
\qquad -\left[ \frac5{24}-\frac{2\,\omega^2}3 +\frac{5\,\omega^4}8 +(\omega^2-1)(mr)^2\right] \left(\eta-\frac\omega2\right)^2 \\
\qquad-\left(\frac{\omega^2}2-\frac16\right)\left(\eta-\frac\omega2\right)^4 +\omega\hat\eta\Big[ \frac1{30}+\frac{\hat\eta}2 -\frac{4\,\hat\eta^2}3 +\hat\eta^3 -\frac{\hat\eta^4}5 \qquad\\
\qquad - \Big( \frac43 - 2\hat\eta+ \frac{4\,\hat\eta^2}3  \Big)(mr)^2 -(mr)^4 \Big] \,,\quad \delta\omega/2 \leq \eta \leq \delta\omega  \\
\\
-\frac{5\,\omega^2}{96} + \frac{11\,\omega^4}{96} -\frac{\omega^2(\omega^2-1)}4(mr)^2 \\
\qquad -\left[ \frac5{24}-\frac{2\,\omega^2}3 +\frac{5\,\omega^4}8 +(\omega^2-1)(mr)^2\right] \left(\eta-\frac\omega2\right)^2 \\
\qquad-\left(\frac{\omega^2}2-\frac16\right)\left(\eta-\frac\omega2\right)^4  \,,\quad \eta \geq \delta\omega \,.
\end{array} \right.
\eea
%


\end{document}